\documentclass[11pt]{article}
\usepackage{amssymb}
\usepackage{graphicx}
\usepackage{epsfig}
\usepackage{wrapfig}

\newcommand{\goes}{\rightarrow} 
 
\newcommand{\GeV}{\; \mathrm{GeV}} 
\newcommand{\TeV}{\; \mathrm{TeV}} 
\newcommand{\lapproxeq}{\lower .7ex\hbox{$\;\stackrel{\textstyle  
<}{\sim}\;$}} 
\newcommand{\gapproxeq}{\lower .7ex\hbox{$\;\stackrel{\textstyle  
>}{\sim}\;$}} 
\newcommand{\stackdown}[2]{\lower 1.4ex\hbox{$\;\stackrel{\textstyle{#1}}  
{\scriptstyle{#2}}\;$}}
\newcommand{\beq}{\begin{equation}} 
\newcommand{\eeq}{\end{equation}} 
\newcommand{\bea}{\begin{eqnarray}} 
\newcommand{\eea}{\end{eqnarray}}
\newcommand{\lsp}{\chi_1^0}
\newcommand{\mlsp}{m_{\lsp}}

\newcommand{\relic}{\Omega_{\lsp}\,h_0^2} 
 
\newcommand{\etal}{\textit{et. al.}}
\newcommand{\almuon}{\alpha_{\mu}^{\mathrm{SUSY}}}       
                     
\newcommand{\bsga}{b \goes s \, \gamma}

\makeatletter 
\def\slash{\@ifnextchar[{\fmsl@sh}{\fmsl@sh[0mu]}} 
\def\fmsl@sh[#1]#2{%
  \mathchoice 
    {\@fmsl@sh\displaystyle{#1}{#2}}%
    {\@fmsl@sh\textstyle{#1}{#2}}%
    {\@fmsl@sh\scriptstyle{#1}{#2}}%
    {\@fmsl@sh\scriptscriptstyle{#1}{#2}}} 
\def\@fmsl@sh#1#2#3{\m@th\ooalign{$\hfil#1\mkern#2/\hfil$\crcr$#1#3$}} 
\makeatother 

\def\beq{\begin{equation}}
\def\eeq{\end{equation}}

\catcode`\@=11 

\def\lsim{\mathrel{\mathpalette\@versim<}}
\def\gsim{\mathrel{\mathpalette\@versim>}}
\def\@versim#1#2{\vcenter{\offinterlineskip
    \ialign{$\m@th#1\hfil##\hfil$\crcr#2\crcr\sim\crcr } }}
\def\etal{{\em et. al.}}

\def\t1{{\tilde 1}}

\def\slash#1{#1\hskip-6pt/\hskip6pt}

\def\GeV{\,{\rm GeV}}
\def\TeV{\,{\rm TeV}}

\def\to{\rightarrow}

\begin{document}

\begin{titlepage}

\begin{flushright}
ACT-03-03 \\
FTUV-030826 \\
KCL-PH/TH 030826 \\
MIFP-03-19 \\
UA/NPPS 26-08-03
\end{flushright} 

\begin{centering} 

{\large {\bf WMAPing the Universe: Supersymmetry, Dark Matter, 
Dark Energy, Proton Decay and  Collider Physics}}
\vspace{0.3cm} 

{\bf A.B.~Lahanas$^a$}, {\bf N.E.~Mavromatos$^{b,c}$} and 
{\bf D.V.~Nanopoulos$^{d,e,f}$}

\vspace{0.3cm} 

$^a$ {\it University of Athens, Physics Department, 
Nuclear and Particle Physics section, Panepistimioupolis, 
Zografou Campus, Athens GR-15771, 
Greece.} 

$^b$ {\it King´s College London, University of London, Department of Physics, 
Strand WC2R 2LS, London, U.K.} 

$^c$ {\it Departamento de F\'isica T\'eorica, Universidad de Valencia,
E-46100, Burjassot, Valencia, Spain.}

$^d$ {\it George P. and Cynthia W. Mitchell 
Institute of fundamental  Physics, 
Texas A \& M University, College Station, TX 77843-4242,
USA.} 

$^e$ {\it Astroparticle Physics Group, Houston Advanced Research Center (HARC),
The Mitchell Campus, The Woodlands, TX 77381, USA} 

$^f$ {\it Chair of Theoretical Physics, Academy of Athens,
Division of Natural Sciences, 28 Panepistimiou Avenue, Athens GR-10679,
Greece.} 

\vspace{0.2cm}

{\bf Abstract} 

\vspace{0.1cm} 

\end{centering} 

{\small In this review we critically discuss constraints 
on minimal supersymmetric
models of particle physics as implied by the recent astrophysical
observations of WMAP satellite experiment. 
Although the prospects of detecting supersymmetry 
increase dramatically, at least within the context of 
the minimal models, and 90\% of the available parameter 
space can safely be reached by the sensitivity 
of future colliders, such as Tevatron, LHC and linear colliders, 
nevertheless we pay particular
emphasis on discussing 
regions of the appropriate phase diagrams, which -if realized in nature-
would imply that detection of supersymmetry, 
at least in the context of minimal models,  
could be out of colliders reach. 
We also discuss the importance of a precise determination of 
the radiative corrections to the muon anomalous magnetic moment, $g_\mu-2$,
both theoretically and experimentally, which could lead to elimination
of such ``out of reach'' 
regions in case of a confirmed    
discrepancy of $g_\mu -2$ from the standard model value.
Finally, we briefly commend upon recent
evidence,
supported by observations,
on a dark energy component of the Universe, 
of as yet unknown origin, covering 73\% of  
its energy content. To be specific, we 
discuss how supergravity quintessence (relaxation) models can be made
consistent with recent observations, which may lead to 
phenomenologically correct constrained supersymmetric models, 
accounting properly for this dark energy component.
We also outline their unresolved problems.}

\end{titlepage} 

\newpage

\section{INTRODUCTION}

Particle Physics and Astrophysics have been separate
for a number of years. Until a few years ago the accuracy 
with which astrophysical
measurements 
were made was much lower
than the corresponding one in particle physics  experiments,
thereby preventing a fruitful interaction between the 
two communities. 

However, in the past decade we have witnessed spectacular
progress in precision measurements in astrophysics as a result
of significant improvements in terrestrial and extraterrestrial 
instrumentation. The (second phase of the) 
Hubble telescope opened up novel paths
in our quest for understanding the Universe, by allowing observations
on distant corners of the observable Universe that were not accessible
before. 

From the point of view of interest to particle physics, 
the most spectacular claims from astrophysics came five years ago
from the study of distant supernovae (redshifts $z \sim 1$) 
by two independent groups~\cite{supernovae}.
These observations pointed towards
a current era acceleration of our Universe, something that 
could be explained
either by a non-zero cosmological {\it constant} in a Friedman-Robertson-Walker-Einstein Universe, or in general by a non-zero {\it dark energy} component,
which could even be relaxing to zero (the data are consistent
with this possibility). This claim, if true, could revolutionize our
understanding of the basic physics governing fundamental interactions
in Nature. Indeed, only a few years ago, particle theorists were trying to 
identify (alas in vain!) an exact symmetry of nature that could 
set the cosmological constant (or more generally the vacuum energy) to zero.
Now, astrophysical observations point to the contrary. 

The skeptics may question the accuracy of the supernovae 
observations, given that neither the nuclear physics associated with 
their evolution, nor the physics involved in the intergalactic and interstellar
matter are well understood to date so as to exclude the possibility 
that the observed effects on the dimering of the distant supernovae ($z \sim 1$) 
as compared to the nearby ones ($z \ll 1$) are due to conventional physics
and are not related to the geometry of the Universe. 
However, there is additional evidence from quite different in origin astrophysical observations, those associated with the measurement of 
the cosmic microwave background radiation (CMB), which point towards
the fact that $73$ \% of the Universe vacuum energy consists of 
a dark (unknown) energy substance, in agreement with the (preliminary) 
supernovae observations. Moreover, recently~\cite{recentsn} 
two more distant supernovae have been discovered 
($z > 1$), exhibiting  similar features as the previous measurements,
thereby supporting the geometric interpretation on the acceleration
of the Universe today, and arguing against the nuclear physics or
intergalactic dust effects.

Above all, however, there are the very recent data 
from a new probe of Cosmic Microwave Background Radiation Anisotropy 
(Wilkinson Microwave Anisotropy Probe (WMAP))~\cite{wmap}.
In its first year of running WMAP
measured CMB anisotropies to an unprecedented accuracy 
of billionth of a Kelvin degree, thereby correcting previous measurements
by the Cosmic Background Explorer (COBE) satellite~\cite{cobe} 
by several orders of magnitude.
This new satellite experiment, therefore, opened up a new era
for astroparticle physics, given that such accuracies allow 
for a determination (using best fit models of the Universe) 
of cosmological parameters~\cite{spergel}, and 
in particular cosmological densities, 
which, as we shall discuss in this review, 
is quite relevant for constraining models of particle physics
to a significant degree.

\begin{figure}[htb]
\centering
\epsfig{file=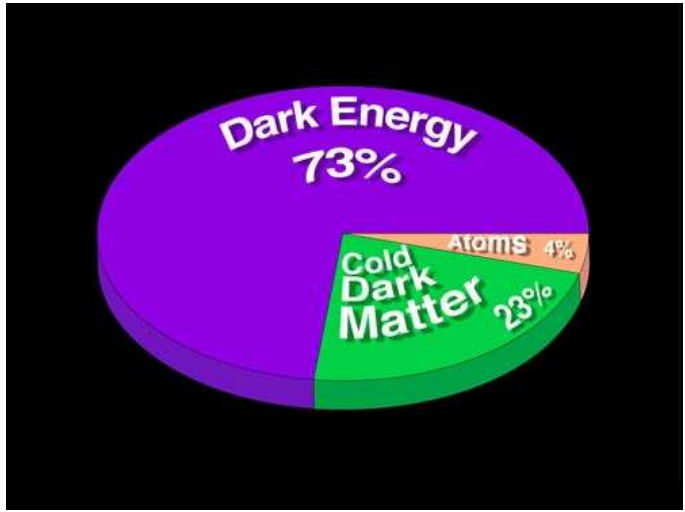, width=0.6\textwidth}
\caption{{\it The energy content of our Universe as obtained
by fitting data of WMAP satellite. The chart is in 
perfect agreement with earlier claims made by direct measurements
of a current era acceleration of the Universe from distant 
supernovae type Ia (courtesy of  
http://map.gsfc.nasa.gov/).}}
\label{budget}
\end{figure}

The WMAP satellite experiment determined the most important 
cosmological parameters that could be of relevance to 
particle physicists, namely~\cite{spergel}: the 
Hubble constant, and thus the 
age of the Universe,
the thickness of the last scattering surface (c.f. below), the 
dark energy and dark matter content of the Universe 
(to an unprecedented accuracy) (c.f. figure \ref{budget}), 
confirming the earlier claims from supernovae Ia data~\cite{supernovae},
and 
provided evidence for early reionization ($z \sim 20$),
which, at least from the point of view of large scale structure 
formation, excludes Warm Dark Matter particle theory models.

In this review we shall first describe 
briefly the above-mentioned measurements, and then use them
to constrain certain particle physics 
supersymmetric models (in particular, the minimally supersymmetric 
model, constrained by its embedding
in a minimal supergravity model (mSUGRA)).
We shall give a critical discussion on the derived constraints, 
and  discuss the capability of observing Supersymmetry 
in colliders after these latest WMAP data. We shall pay particular
attention to discussing regions of the parameter space of the models,
which if realized in nature, would imply impossibility of observing supersymmetry at LHC and/or linear colliders, even in the context of the mSUGRA models.
In this respect we shall also discuss  the importance of the $g_{\mu}-2$ experiments
of the muon gyromagnetic ratio, and how certain results,
pointing towards a discrepancy between the measured $g_{\mu}-2$ 
from the value calculated within the standard model,  could eliminate the 
above-mentioned precarious regions of parameter space.

An important comment we would like to make in this review 
will concern the dark energy component ($73$ \% ) of the Universe.
The WMAP measured equation of state 
for the Universe $p = w\rho $, 
with $p$ the pressure and $\rho$ the energy density, implies 
$-1 \le w < -0.78$ (assuming the lower bound for theoretical reasons, 
otherwise the upper limit may be larger~\cite{spergel}). 
For comparison we note that $w=-1$ characterizes a perfect fluid Universe
with non-zero, positive, cosmological constant.
As we shall remark, supergravity quintessence models do have this feature
of $w \to -1$, and it may well be that by exploiting further the data on  
this dark energy component of the Universe one may arrive at the physically correct supergravity model which could constrain the supersymmetric 
particle physics models.

The outline of this review is as follows: 
In section two we shall discuss 
briefly the WMAP experiment and its measurements, with emphasis 
on cosmological parameters relevant to particle physics models.
In particular, we shall describe the 
measurements of the equation of state of quintessence models, in an attempt
to shed light on the nature of the dark energy. 
In section three
we shall discuss the basics of inflationary models, which seem to 
be favoured by WMAP, but we shall be critical in our discussion, 
stressing the current inadequacies of the observations.     
In section four we shall embark on the main point of our discussion, 
namely the connection of the  
WMAP measurements and Supersymmetry. Specifically,  
we shall discuss various models of 
SUSY Dark Matter, and argue that only Cold Dark Matter (CDM) is favoured
by the data, given that Hot Dark Matter (neutrino) is excluded
in view of the strict upper bounds of neutrino masses imposed by the WMAP,
and Warm Dark Matter models with a low mass gravitino (less than $10$ KeV) 
are also excluded on account of evidence for early ($z \sim 20$) 
reionization of the Universe. 
By concentrating, for the purposes of this review, on 
CDM, made exclusively of 
neutralinos (viewed as the Lightest supersymmetric particle 
(LSP))~\cite{Hagelin}, 
we shall describe, in section five, 
the WMAP-derived constraints on the mSUGRA model,
and the  
prospects for detection of supersymmetry
within the context of this model in future colliders, as well as  
direct dark matter searches. Also we shall discuss briefly 
how such constraints, when combined with 
proton decay current 
lower limits, affect grand unified supersymmetric theories, 
specifically flipped SU(5) models. 
Finally, in section six  
we shall review critically the above constraints in light of 
the fact that mSUGRA and all the existing particle physics models
do not take proper account of the observed dark energy component.
In this respect we shall discuss supergravity/superstring/brane inspired
models with relaxing to zero ``vacuum energies´´ and argue how such models
may be used in the future to constrain particle physics supersymmetric 
models. The issue is of course still
unresolved, as it involves
the yet unsettled theoretical task of determining the precise mechanism
of low-energy supersymmetry breaking. 
Conclusions and outlook, outlining possible future directions,
will be presented in  
section seven. 

\section{WMAP  \& (ASTRO)PARTICLE PHYSICS} 

\subsection{What is WMAP}

\begin{figure}[htb]
\centering
  \epsfig{file=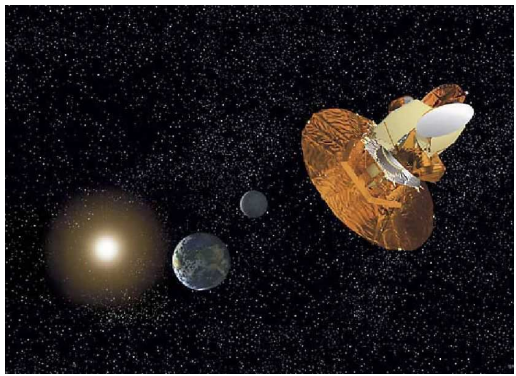, width=0.45\textwidth} \hfill 
\epsfig{file=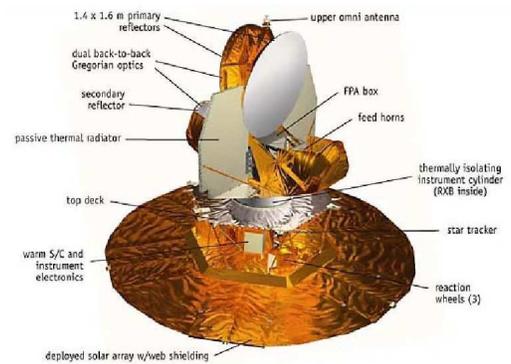, width=0.45\textwidth}
\caption{A graphical picture of the WMAP satellite (courtesy of  
http://map.gsfc.nasa.gov/).}
\label{wmapfigure}
\end{figure}

The Wilkinson Microwave Anisotropy Probe (WMAP)~\cite{wmap} 
is a satellite experiment (figure \ref{wmapfigure}) dedicated to measure 
temperature fluctuations of the cosmic microwave background (CMB),
with unprecedented accuracy, as compared to previous COBE measurements,
which reaches billionth of a Kelvin degree.

\begin{figure}[htb]
\centering
  \epsfig{file=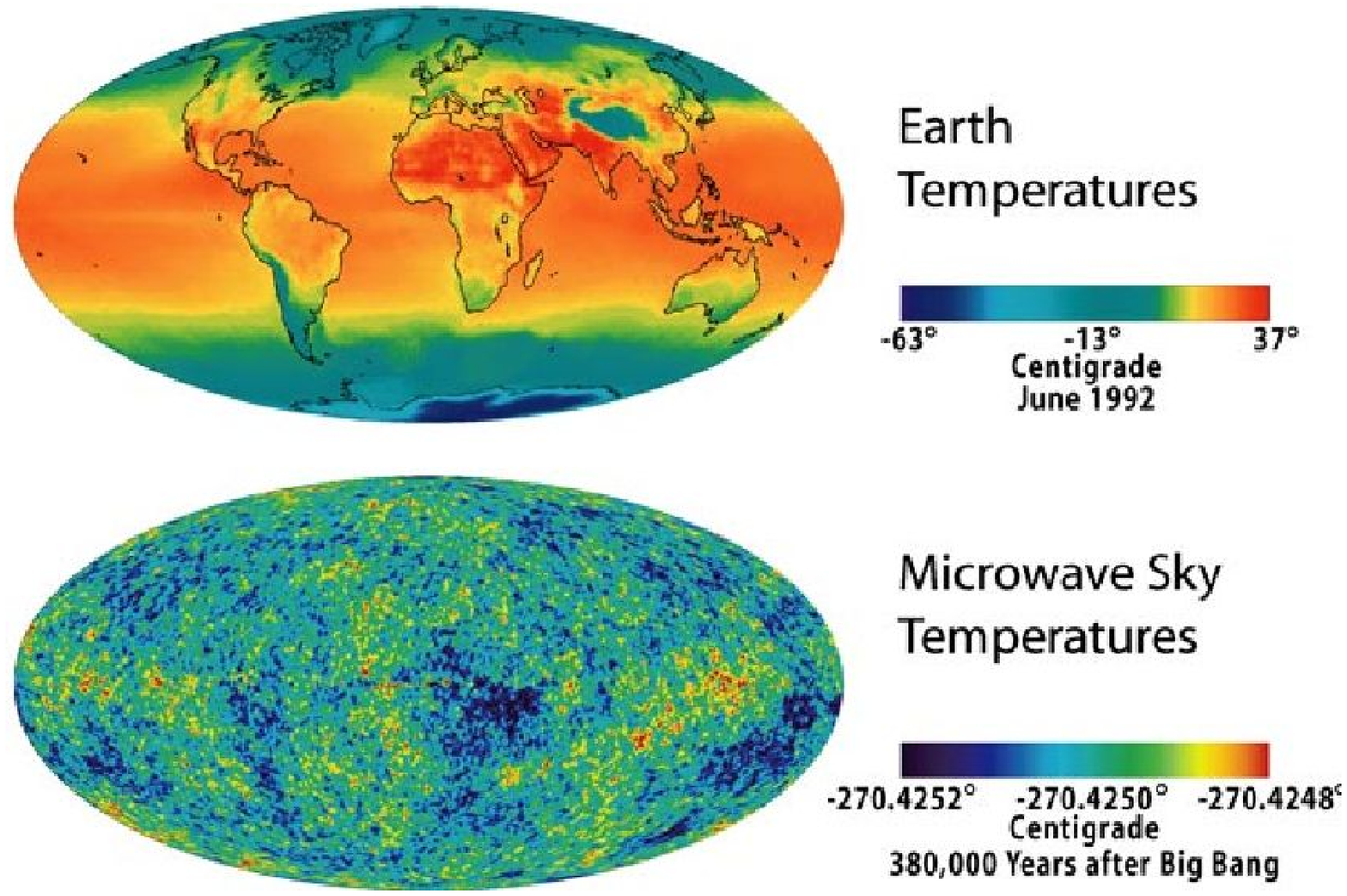, width=0.45\textwidth} \hfill 
\epsfig{file=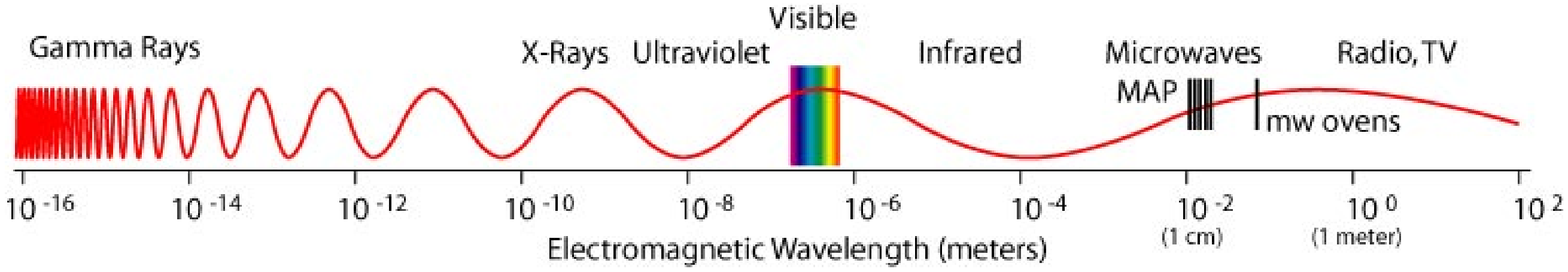, width=0.45\textwidth}
\caption{{\it  The expansion of the Universe causes 
stretching of the wavelengths of light, so 
light emitted at a gamma ray region of wavelengths reaches 
us today as microwave radiation (courtesy of  
http://map.gsfc.nasa.gov/).}}
\label{wmapexplan}
\end{figure}

The cosmic microwave background is 
the afterglow radiation left over from the hot Big Bang. 
Its temperature is extremely uniform all over the sky. 
However, tiny temperature variations or fluctuations 
(at the part 
per million level) can offer great insight into the 
origin, evolution, 
and content of the Universe.
The light that is reaching us has been stretched 
out as the Universe has stretched, so light that 
was once beyond gamma rays is now reaching 
us in the form of microwaves (longer wavelength)
(figure \ref{wmapexplan}). We can only observe
light that comes from the time of last scattering,
in a completely analogous way as in a cloudy sky we 
can only see light coming from the cloud surface 
that is facing the earth ground (c.f. figure 4).
We remind the reader that the existence of the last scattering surface 
is due to the fact that the Universe expands rapidly and, as a result,
there is significant dilution in the 
density of matter particles, including radiation. 
Hence, the scattering rate of matter and also of light is diminished,
and eventually stops (freezing out), implying 
a surface of last scattering.

\begin{figure}[htb]
\begin{center}
  \epsfig{file=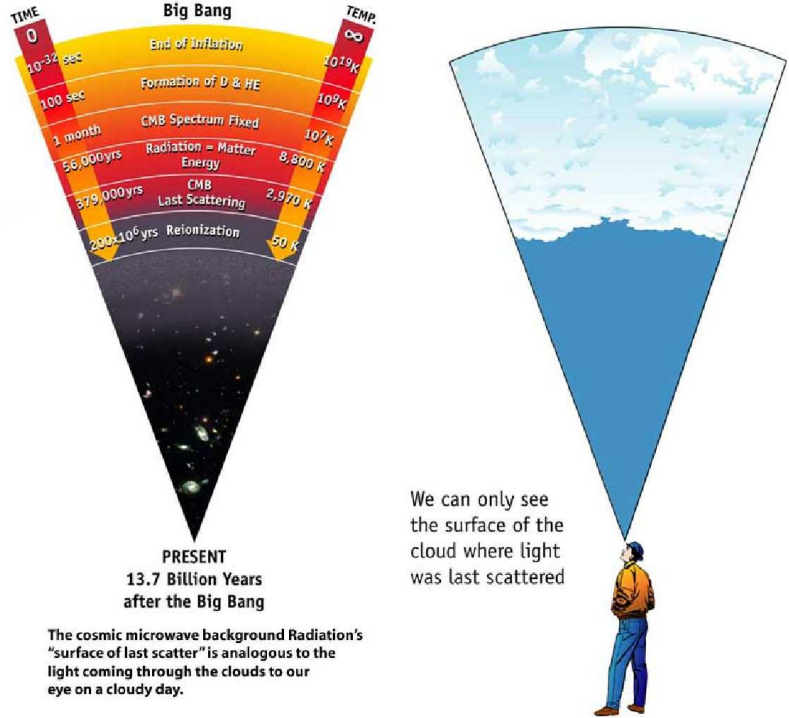,angle=0,width=0.7\linewidth} 
\caption{{\it We can only see the light emitted from the 
surface where it was last scattered. This happens when we look 
either at a cloudy sky, or at the cosmic
microwave background radiation (courtesy of  
http://map.gsfc.nasa.gov/).}}
\end{center}
\label{lastscatt}
\end{figure}

\begin{figure}
\centering
  \epsfig{file=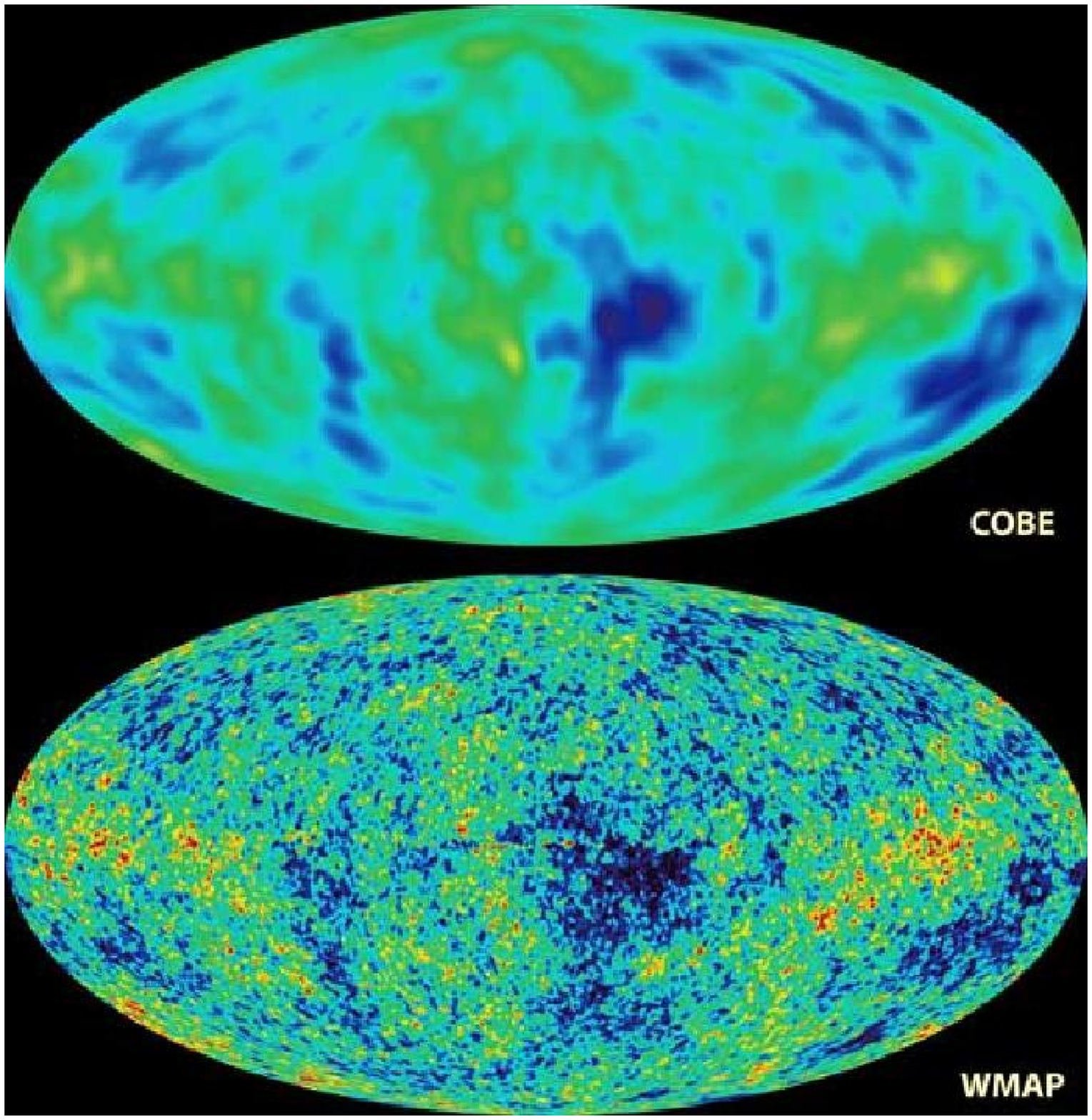, width=0.5\textwidth}
\hfill \epsfig{file=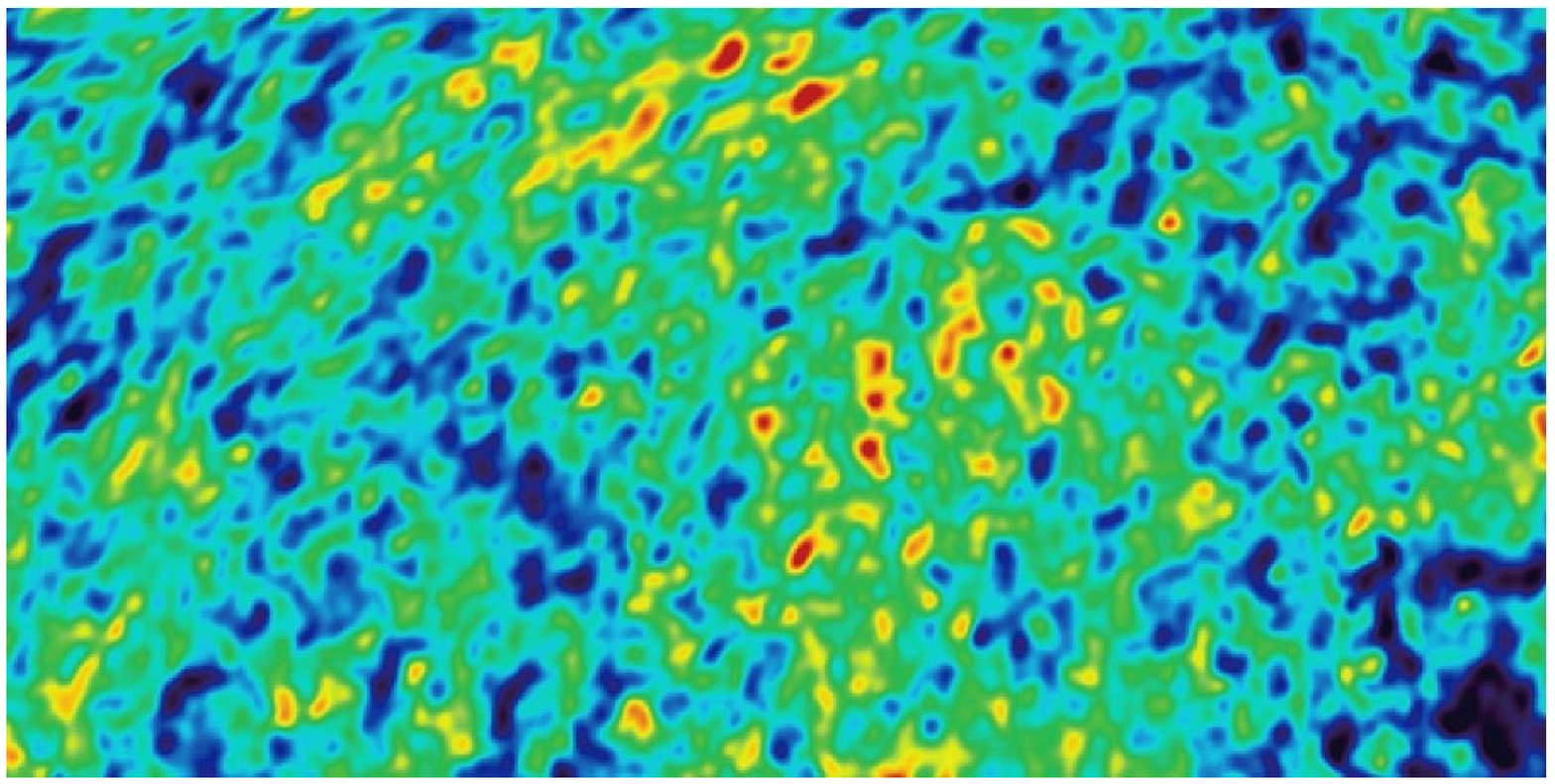, width=0.4\textwidth}
\caption{{\it CMB Temperature fluctuations 
provided by WMAP and comparison with COBE (left),
Importance of Detail (right) (courtesy of  
http://map.gsfc.nasa.gov/).}}
\label{cobewmap}
\end{figure}

In its first year of running, WMAP already  provides
a much more detailed picture of the temperature fluctuations 
than its COBE predecessor (figure \ref{cobewmap}), which can be analyzed to
provide best fit models for cosmology, leading to 
severe constraints on the energy content of 
various model Universes, useful for particle physics,
and in particular supersymmetric searches.

Theoretically~\cite{kolbturner}, 
the temperature fluctuations in the CMB radiation 
are attributed to: (i) our velocity w.r.t cosmic rest frame,  
(ii) gravitational 
potential fluctuations on the last scattering surface (Sachs-Wolf effect), 
(iii) Radiation field fluctuations on the last scattering surface, 
(iv) velocity of the last scattering surface, and 
(v) damping of anisotropies if Universe
reionizes after decoupling. 

The fluctuations due to 
(i) imply a dipole anisotropy, while (ii)are the dominant 
effects for large angular
scales $\theta >> 1^o$. The fluctuations 
(iii) - (v) on the other hand, are dominant for 
$\theta << 1^o$. The value of $\theta=1^o$
 is determined by 
the physics of decoupling. 

\begin{figure}
\centering
\includegraphics[angle=90,width=0.6\textwidth]{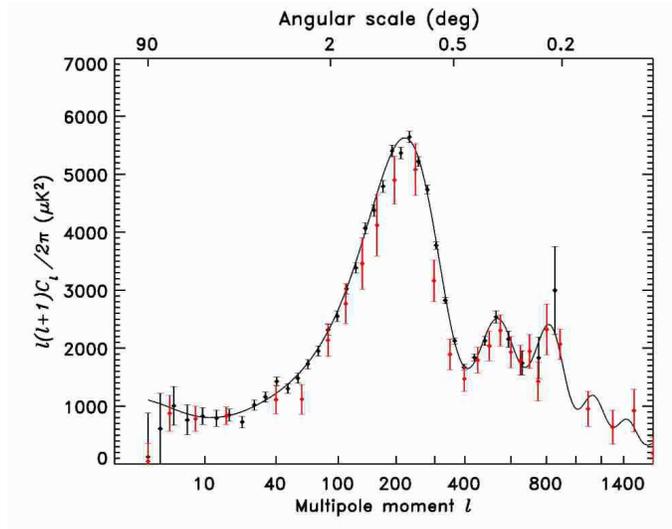}
\caption{{\it Red points (larger errors) are previous measurements. Black points (smaller errors) are WMAP measurements 
(Hinshaw, G. {\it et al.} arXiv:astro-ph/0302217).}}
\label{wmapgaussian}
\end{figure}

The CMB anisotropies in the Sky are expanded as: 
$$\frac{\delta T}{T}(\theta, \phi) = \sum_{\ell = 2}^{+\infty}\sum_{m=-\ell}^{+\ell} a_{\ell m}Y_{\ell m}(\theta, \phi) $$
where $Y_{\ell m} (\theta, \phi)$ are spherical harmonics;  
isotropy implies $<a_{\ell m} > = 0$.
Let us denote the variance of $a_{\ell m}$ 
by $C_\ell = < |a_{\ell m}|^2 > $. A non zero 
$C_\ell$  would imply temperature 
fluctuations. More accurately, within a Gaussian model  
with width $\sigma$, the above sum over $\ell$ 
in the expression for $\delta T/T$ is effectively
cutoff at $\ell \sim \sigma$: 
$$ C_\sigma = <\frac{\delta T ({\vec x_1})}{T}
\frac{\delta T ({\vec x_2})}{T}> = 
\frac{1}{4\pi} \sum_{\ell = 2}^{+\infty}(2\ell + 1)  
< |a_{\ell m}|^2 > P_\ell ({\vec x}_1, {\vec x}_2)e^{-(\frac{\ell + 1}{2})^2\sigma^2} $$
This Gaussian model of fluctuations 
is in very good agreement with the recent WMAP data
(see figure \ref{wmapgaussian}). The perfect fit of the first few peaks 
to the data allows a precise determination of the total density of the 
Universe, which as we shall discuss next implies its spatial flatness.

\subsection{WMAP Measurement of Cosmological Parameters}

\begin{figure}[htb]
\centering
  \epsfig{file=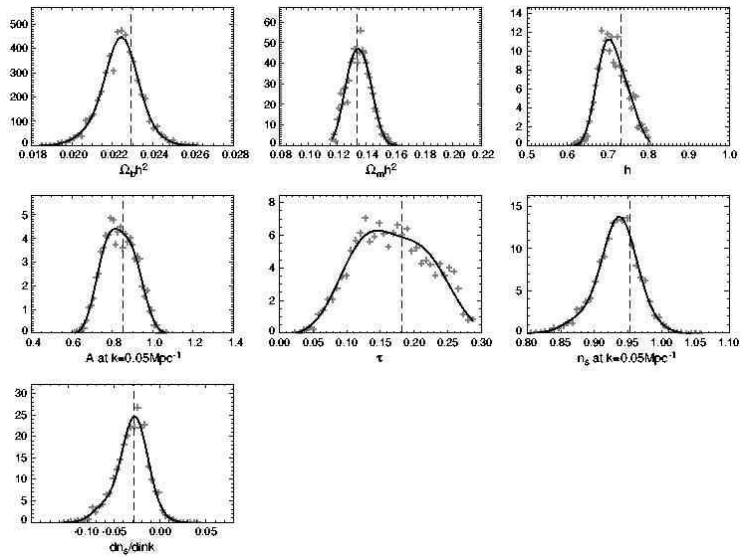, width=0.65\textwidth}
\caption{{\it Likelihood Functions for various cosmological 
parameters measured by WMAP~\cite{spergel}.}}
\label{spergel1} 

\end{figure}

\begin{figure}[htb]
\centering
  \epsfig{file=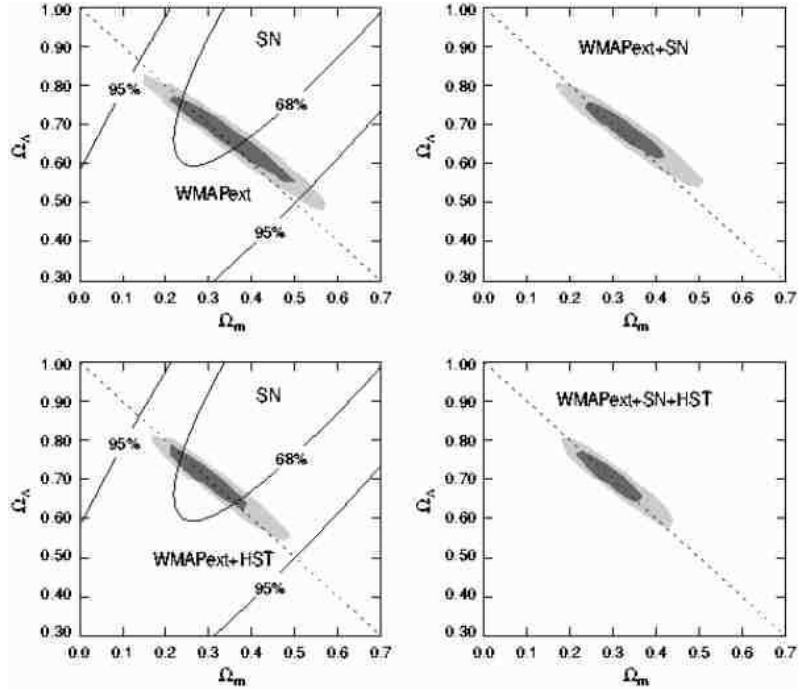, width=0.7\textwidth}
\caption{{\it Energy density and matter density content of the Universe as measured by the WMAP satellite alone, as well as in combination with other experiments~\cite{spergel}.
}}
\label{spergel2} 
\end{figure}

\begin{figure}[htb]
\centering
\epsfig{file=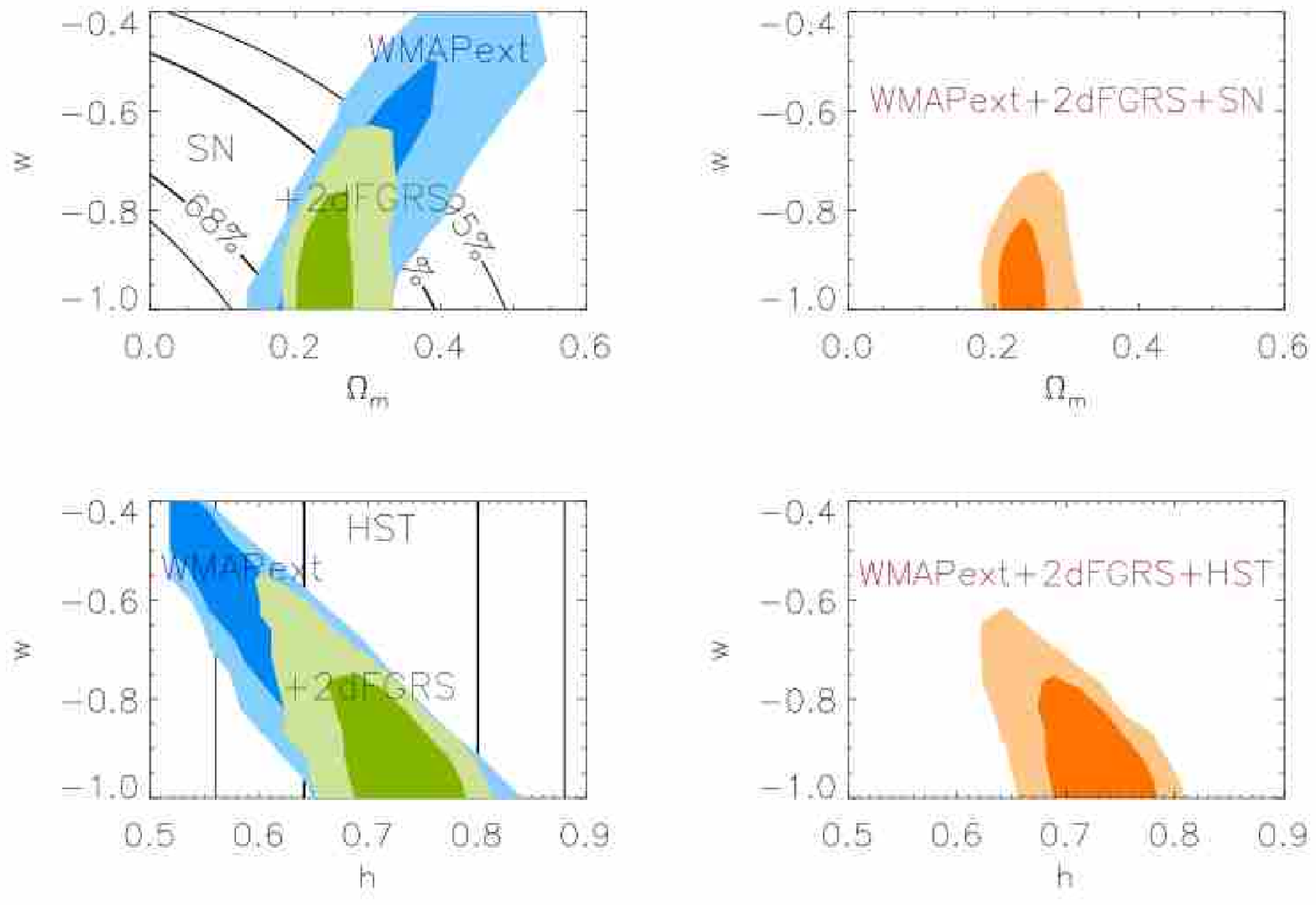, width=0.7\textwidth}
\caption{{\it Equation of State of the Dark Energy component of the Universe as measured by the
WMAP satellite alone, as well as in combination with other 
experiments~\cite{spergel}. 
}}
\label{spergel3} 
\end{figure}

The measurements of the WMAP~\cite{spergel} on the cosmological parameters
of interest to us here can be summarized in the
figures \ref{spergel1}-\ref{spergel3}, see Refs. \cite{spergel},
with $\Omega_i = \rho_i/\rho_c$. $\rho_i$ 
is the matter-energy density component,
$\rho_c=3H^2/8\pi G_N$ (with $G_N$ the Newton constant) the critical density for the closure of the Universe, 
determined within the Friedman-Robertson-Walker framework, 
and $h$ the rescaled Hubble parameter, defined through 
$H = 100 h~{\rm Km}~{\rm s}^{-1}~{\rm Mpc}^{-1}$. 
The WMAP satellite can measure directly $\Omega h^2$, which is Hubble 
parameter independent by construction, and can also separately measure $h$.  
For our purposes in this article we concentrate on the 
measurements of the density of matter and dark 
energy (see figure \ref{spergel2}), as well as on the equation of state  
of the dark energy component (c.f. figure \ref{spergel3}). 
 
We observe from figure \ref{spergel3}
that the WMAP results constrain severely the 
equation of state $p=w \rho $ ($p=$pressure), 
pointing towards $w < -0.78$, if one fits the data with the assumption 
$-1 \le w$  (we note for comparison that in the scenarios advocating the existence of a cosmological {\it constant}  
one has $w=-1$).  
Many quintessence models can easily satisfy
the criterion $-1 < w < -0.78$, especially the supersymmetric ones,
which we shall comment upon later in the article. 
Thus, at present, the available data are not sufficient to distinguish
the cosmological constant model from quintessence (or more generally
from relaxation models of the vacuum energy). 

\begin{figure}
\centering
\includegraphics[angle=270,width=0.7\textwidth]{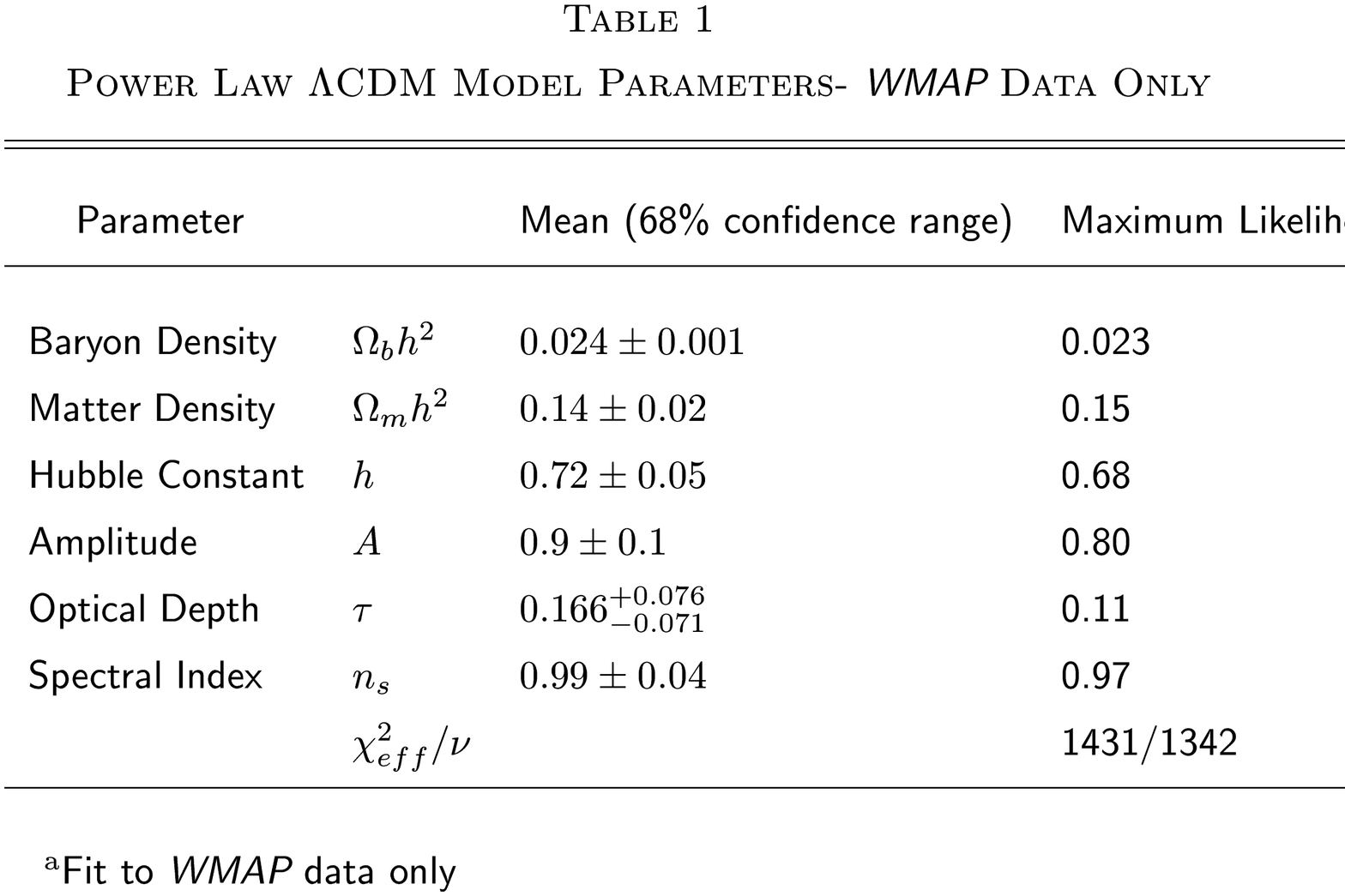}
\caption{{\it Cosmological parameters measured by WMAP (only):directly measurable quantities~\cite{spergel}.}}
\label{table1spergel}
\end{figure}

\begin{figure}
\centering
  \epsfig{file=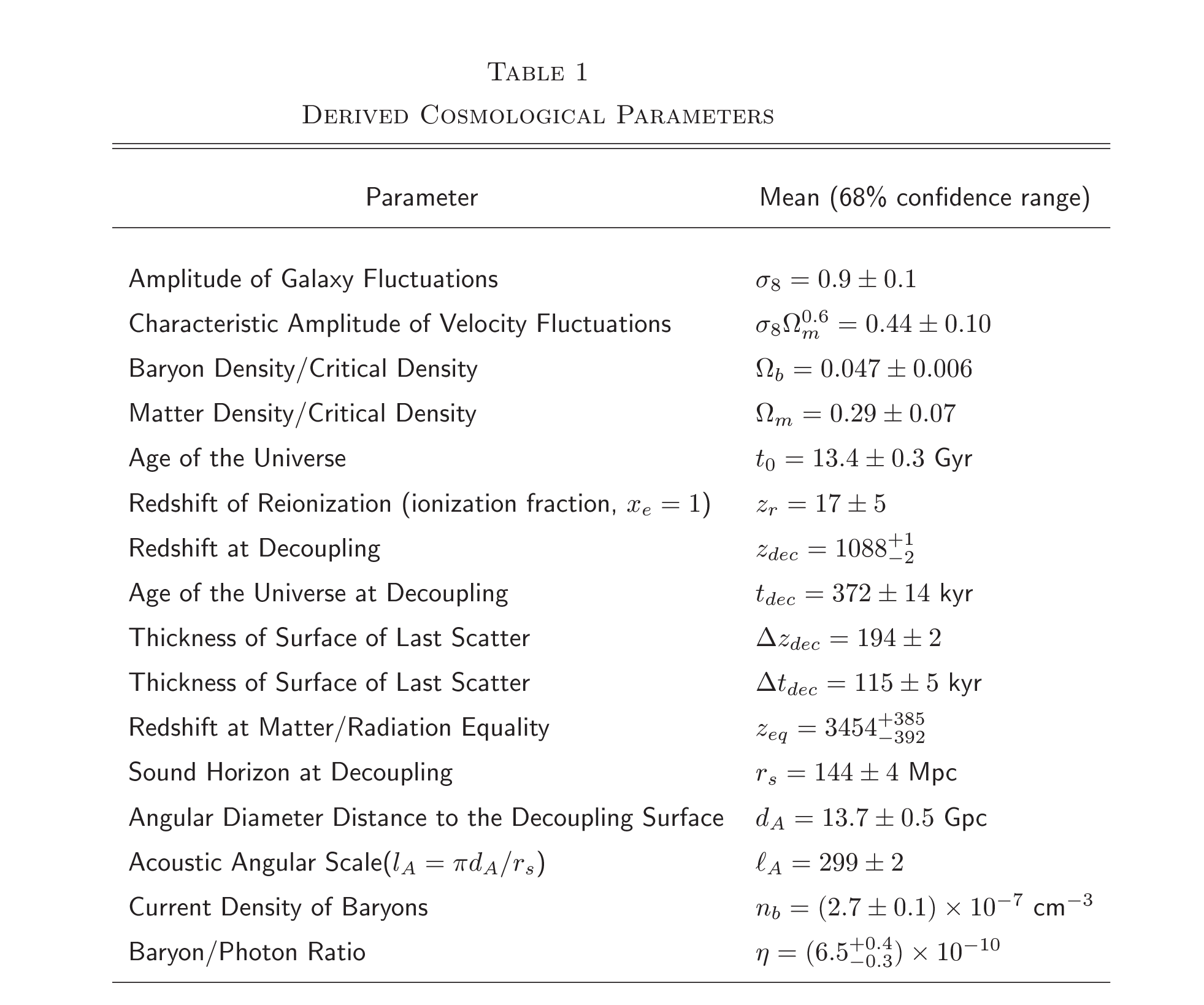, width=0.9\textwidth}
\caption{{\it Cosmological parameters measured by WMAP (only):derived 
quantities~\cite{spergel}.}}
\label{table2}
\end{figure}

The results of the WMAP analysis (alone),including directly measurable and
derived quantities,  are summarized 
in the tables appearing in figures \ref{table1spergel},\ref{table2}.

One therefore obtains the chart for the energy and 
matter content of our Universe depicted in figure \ref{budget}.
This chart is in perfect agreement  
with 
{\it direct} evidence on acceleration of the Universe 
(and hence cosmological constant) from Supernovae Ia Data~\cite{supernovae}
(c.f. figure 12).

\begin{figure}[htb]
\centering
\epsfig{file=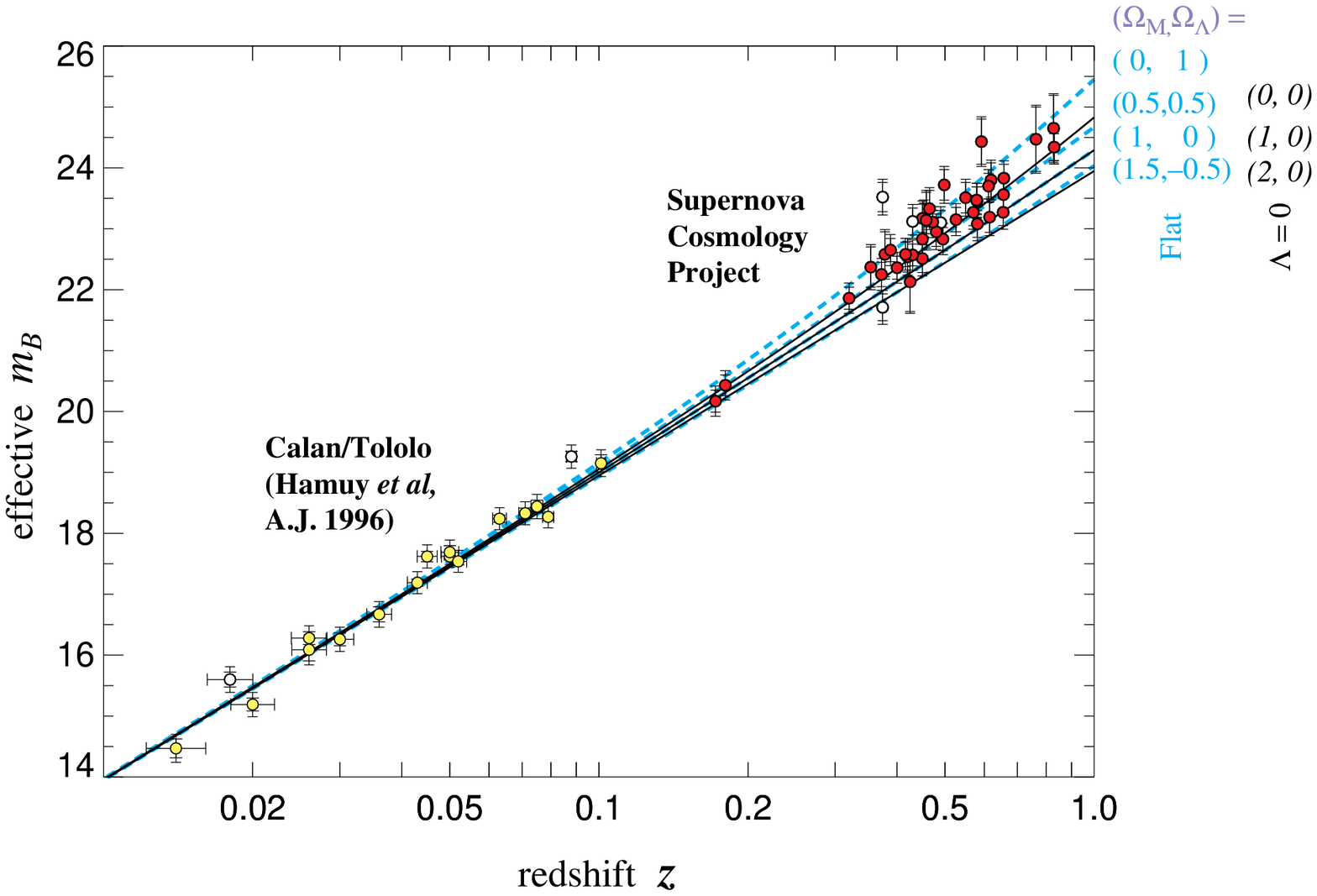, width=0.4\textwidth}
\hfill \epsfig{file=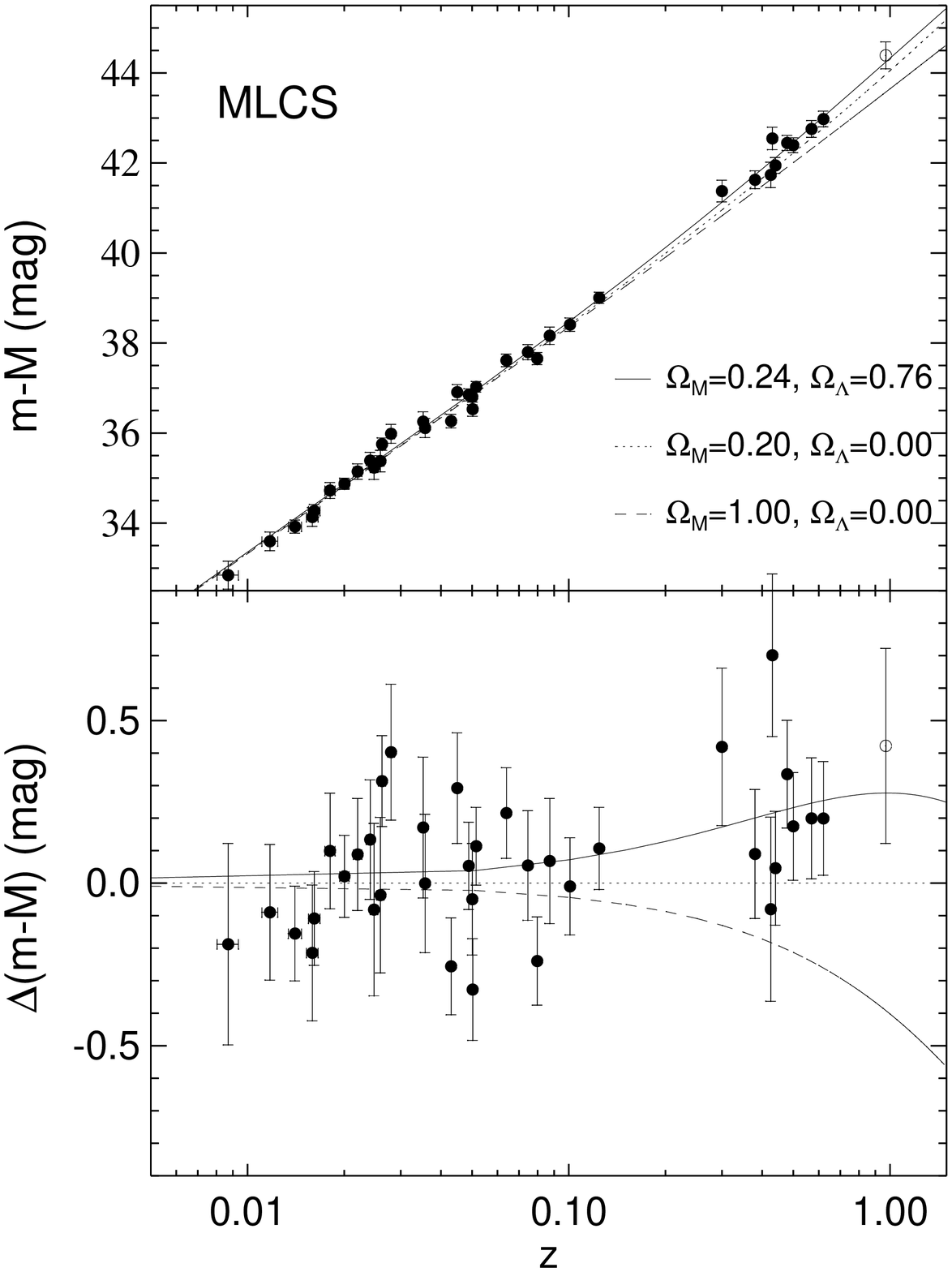, width=0.4\textwidth}
\caption{{\it The only Direct evidence for acceleration 
of the Universe (and hence Dark energy) so far is provided by 
high redshift  ($z \le 1$ ) 
supernovae measurements (SNIa)~\cite{supernovae}, which were prior to  
WMAP data.}}
\label{supnIa}
\end{figure}

It should be stressed that 
the interpretation 
of the supernovae data 
is based on a  {\it best fit} Friedmann-Robertson-Walker (FRW) 
Universe~\cite{supernovae}: 
\begin{equation} 
0.8\Omega_M - 0.6\Omega_\Lambda \simeq -0.2 \pm 0.1~,
\quad {\rm for}~\Omega_M \le 1.5
\label{o1}
\end{equation}
with $\Omega_{M,\Lambda}$ corresponding to the matter and
cosmological matter densities. 
Assuming a flat model (k=0), 
$\Omega_{\rm total} = 1$, as supported by the 
CMB data, the SNIa data alone imply: 
\begin{equation} 
\Omega_M^{Flat} =
0.28^{+0.09}_{-0.08}~(1\sigma~stat.)^{+0.05}_{-0.04}~(identified~syst.)
\label{o2}
\end{equation}
The deceleration parameter defined as
$q \equiv - \frac{{\ddot a} a}{ {\dot a}^2}$ , where $a$ is the
cosmic scale factor, receives the following form if we omit the
contribution of photons which is very small,
\begin{equation}
q =\frac{1}{2}\Omega_M - \Omega_\Lambda \simeq -0.57 < 0~, 
\qquad (\Omega_\Lambda \simeq 0.7)~.
\end{equation}
Hence (\ref{o1}) and (\ref{o2}) provide evidence for a  
{\it current era acceleration of the Universe. }
At this stage it should be 
stressed that the recent observation of two more supernovae 
at $z > 1$~\cite{recentsn} 
supports the geometrical interpretation on the existence
of a dark energy component of the Universe, and 
argues rather against the r\^ole 
of nuclear (evolution) or intergalactic dust effects. 

The recent data of WMAP satellite  
lead to a new determination of 
$\Omega_{\rm total} =
1.02 \pm 0.02 $, where $\Omega _{\rm total} = \rho_{\rm total}/\rho_c$, due 
to high precision measurements 
of secondary (two more) acoustic peaks 
as compared with previous CMB measurements 
(c.f. figure \ref{wmapgaussian}). Essentially the value of $\Omega$ 
is determined by the position of the first acoustic peak in a 
Gaussian model, whose reliability increases significantly 
by the 
discovery of secondary peaks and their excellent fit with the Gaussian 
model~\cite{spergel}.

\section{WMAP AND  INFLATION} 

\subsection{Basics of Inflationary Models}

The recent determination of the cosmological parameters by the WMAP
team~\cite{spergel} favours, by means of best fit procedure,
{\it spatially flat} inflationary
models of the Universe~\cite{wmapinfl}. In terms of Robertson-Walker metrics 
\begin{equation} 
ds^2 = -dt^2 + a^2(t) \left(\frac{dr^2}{1 - kr^2} + 
r^2 d\Omega^2 \right)~,
\label{rwmetric}
\end{equation} 
such models are described by an early phase in which the scale factor
$a(t)$ 
is exponentially expanding with the Robertson-Walker time $t$: 
$a(t) \sim e^{Ht}$, where 
$H$ is the Hubble parameter. Recent WMAP data confirm its value during 
inflation to be of order $H ={\rm const} = 10^{-5} M_{Pl}$,
where $M_{Pl}$ is the Planck energy scale ($\sim 10^{19}$ GeV). 
Spatially flat models are described by $k = 0$, since the spatial curvature
is proportional to this parameter, and they correspond to a total energy 
density $\Omega_{\rm total} = 1$.

\begin{figure}[htb] 
\centering
\epsfig{file=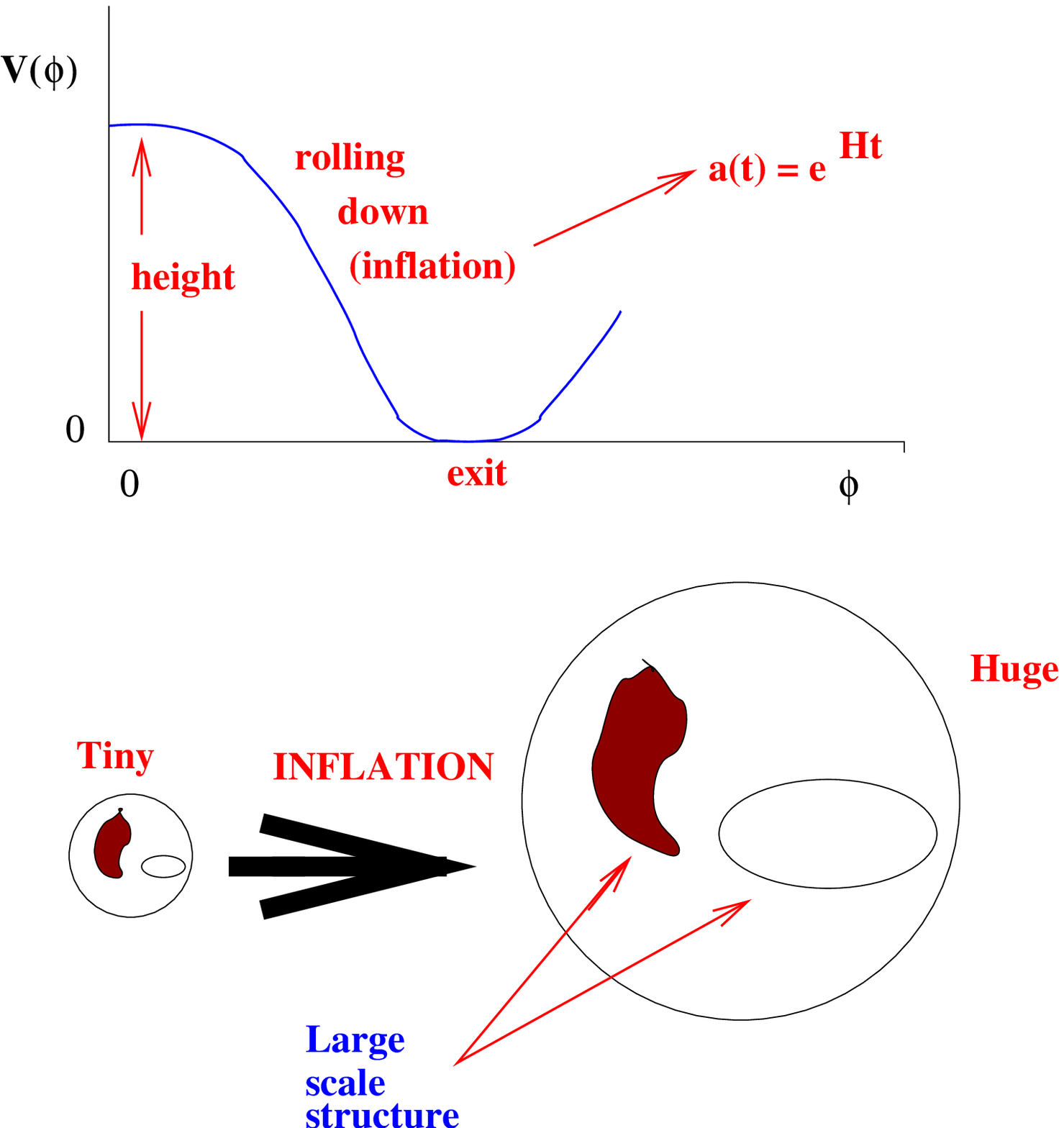, width=0.4\textwidth}
\caption{{\it Inflation Basics: the inflaton and its rolling down the hill phase which causes inflation (exponential expansion); due to this expansion small
quantum fluctuations may enhance and be responsible for large scale structures
of the Universe at present.}}
\label{inflation}
\end{figure}

Such a spatially flat Universe  is preferred in inflationary scenaria~\cite{inflation}. An important 
feature of conventional inflationary models is the presence of a 
scalar mode, the {\it inflaton} $\phi$~\cite{ONS}, 
which couples to a de Sitter 
Universe. The field's slow rolling down the hill of its potential 
(c.f. figure \ref{inflation}) causes the Universe to expand exponentially.
Its equation of motion in a de Sitter background has a frictional form,
with the friction provided by the Hubble parameter $H$ (assumed almost constant
during inflation): 
\begin{equation} 
{\ddot \phi} +  3H {\dot \phi} + 
\frac{\delta V(\phi)}{\delta \phi} = 0~.
\label{infleq}
\end{equation} 
where the overdot denotes derivative w.r.t Robertson-Walker time $t$.

The quantum fluctuations of a Planck-size ($10^{-35}$ meters) Universe, 
which is assumed to be 
the state of the Universe right after the big bang in inflationary scenaria
are enormously enhanced (c.f. figure \ref{inflation}) 
by the exponential expansion and result 
to cosmic microwave background radiation anisotropies. They may also
provide the explanation of the presently observed large scale
structure of the Universe. 

\begin{figure}
  \centering 
\epsfig{file=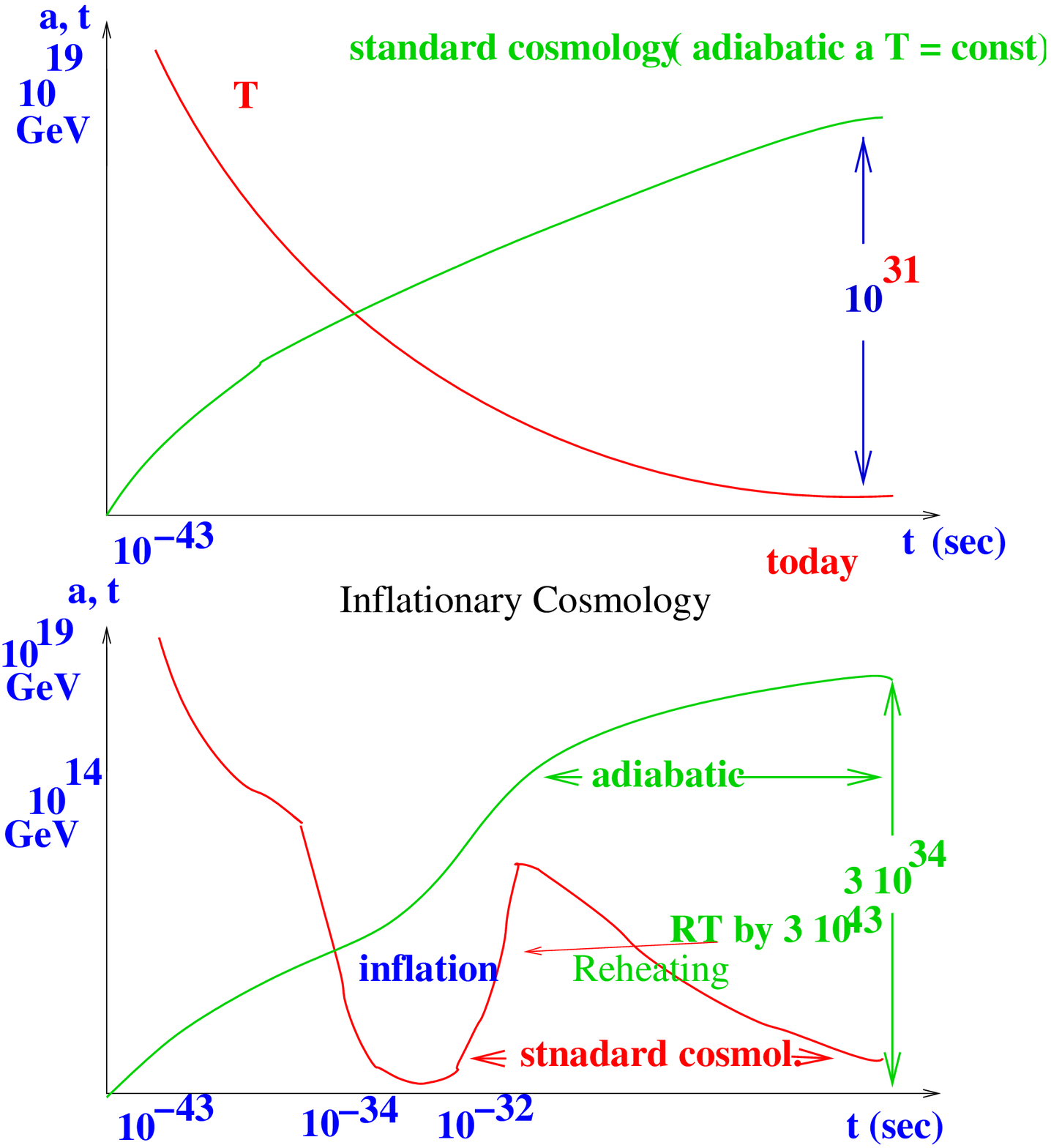,angle=0,width=0.4\linewidth}
\caption{{\it The thermal history of the inflationary Universe as 
compared to the standard cosmology: the exponential expansion causes 
the Universe to cool down significantly, before is reheated again.
The physics of reheating is not well understood at present. The inflationary 
phase results in an enormous entropy production at the end of inflation.}}
\label{inflationtemp}
\end{figure}

The (thermal) history of the inflationary Universe~\cite{kolbturner} 
is quite different from the 
one expected in standard cosmology (c.f. figure \ref{inflationtemp}).
During the exponential expansion the Universe cools down significantly,
and then is reheated after the end of inflation. The physics of reheating 
is not well understood at present, and many scenaria exist in the literature
on this issue, some of which involve fundamental strings. 
In the present article we would not like to  discuss such issues, as they
lie outside our scope. An important characteristic of inflation is the enormous
entropy production at the end of inflation.

\subsection{WMAP-Measured Inflationary Parameters}

The measurements of the WMAP satellite 
which support inflation are associated with the so called  
{\it Running Spectral Index}, a quantity defined 
as follows~\cite{runningspect}. 

Let one consider density fluctuations in the Universe, 
$$
\delta ({\vec x}) = \frac{\rho ({\vec x}) - {\overline \rho}}
{{\overline \rho}} $$ 
where barred quantities denote average over the Universe volume $V$. 
Taking the Fourier transform: 
$\delta ({\vec x}) = \frac{V}{(2\pi)^3}\int d^3 k \delta_{{\vec k}}
e^{i {\vec k} \cdot {\vec x}}~,$
one can define the power spectrum of the fluctuations as~\cite{kolbturner}
\begin{equation} 
<\delta ({\vec x} + {\vec r})\delta ({\vec x})> 
\equiv \xi = \frac{1}{(2\pi)^3}V \int d^3k |\delta _{\vec k}|^2 
e^{-i {\vec k} \cdot {\vec r}}
\label{powrspect}
\end{equation} 
Then, in terms of a comoving scale $k$, one can define 
the primordial power spectrum of scalar density
fluctuations~\cite{runningspect}: 
$$P(k) \equiv |\delta _{\vec k} |^2 \; . $$
The running scalar spectral index $n_s (k)$ is
$$ n_s (k) = \frac{d{\rm ln}P(k)}{d{\rm ln}k}~.$$
One may fix 
$n_s$ and its slope w.r.t. $k$ at $k_0 = 0.05 Mpc^{-1}$ 
and it is customary to assume~\cite{runningspect} that 
$\frac{d^2n_s}{d{\rm ln}k^2}=0 $, from which it follows that 
\begin{equation} 
n_s (k) = n_s (k_0) + \frac{dn_s}{d{\rm ln}k}{\rm ln}(\frac{k}{k_0}) \; .
\label{runningindex} 
\end{equation}
Then the power spectrum $P(k)$ above is written as 
$$
P(k_0) = 
P(k_0) \left(\frac{k}{k_0}\right)^{n_s(k_0)
+ \frac{1}{2}\frac{dn_s}{d{\rm ln}k}{\rm ln}(\frac{k}{k_0})}  \; .
$$
The quantity 
$n_s(k)$ carries information on the field derivatives of the inflaton
potential $V(\phi)$ in best fit models:
$V',~ V'',~ V''' $ ($ V' \equiv \frac{d}{d\phi}V(\phi)$), and can be measured
by WMAP data. In fact, standard inflation implies a constant $n_s$, i.e.
scale invariance of the fluctuation spectrum. The current value of the
running of the
spectral index by the available set of WMAP data at present 
is very small but non-zero, 
however there is not enough statistics as yet to 
arrive at definite conclusions on the shape of the inflaton potential.

Below we give a list of  the 
basic WMAP measurements that favour inflation~\cite{spergel}.
 For definiteness we give first the 
conventional slow-roll inflationary 
parameters~\cite{kolbturner,inflation}
\begin{equation}
\epsilon \equiv \frac{1}{2}M_P^2\left(\frac{V'}{V}\right)^2,~ 
\eta \equiv M_P^2\left(\frac{V''}{V}\right),~\xi \equiv M_P^4\left(\frac{VV'''}{V^2}\right) 
\label{inflparam}
\end{equation}
where $M_P =\left(\frac{1}{8\pi G_N}\right)^{1/2} \simeq 2.4 \times 10^{18}$~GeV is the four-dimensional Planck mass, wth $G_N$ the Newton constant, 
$V(\phi)$ is the inflaton ($\phi$) potential, and the prime indicates
derivative with respect to $\phi$. 
WMAP has measured the following (c.f. figs. \ref{table1spergel},\ref{table2})~\cite{wmap,spergel}: 

\begin{itemize} 

\item{} Flatness of the Universe ($k=0$)
\item{} Density fluctuations $\Delta_R^2$, which are expressed in terms
of (\ref{inflparam}) as: 
$\Delta_R^2 = \frac{V}{24\pi^2M_P^2\epsilon}=2.95 \times 10^{-9}A$, 
\qquad WMAP result: $A=0.77 \pm 0.07$ 

\item{} WMAP results alone are {\it too crude}~\cite{inflatonpot} to
determine the shape and height of inflaton potential. When, 
and only when, the WMAP data are combined with all other data, 
then they seem to {\it exclude}, at $3-\sigma$ level,
a $V \sim \phi^4$ potential~\cite{spergel}.
 
\item{} The ratio of tensor to scalar perturbations 
(at quadrupole scale) $r = \frac{A_T}{A_S}=16 \epsilon,$ \qquad  
WMAP result : $r \simeq 0.16$

\item{} The Spectral Index for (i) scalar density 
perturbations $n_s = 1 - 6\epsilon + 2\eta \;$ \qquad WMAP result:
$n_s \simeq 0.96$, 
and (ii) for tensor perturbations 
$n_T = -2\epsilon~.$

\item{} The running of Scalar Spectral Index,  
a quite important WMAP result,
$\frac{dn_s}{d{\rm ln}k} = \frac{2}{3}[(n_s - 1)^2 - 4\eta^2 ] + 2\xi,$
\qquad WMAP result: $\frac{dn_s}{d{\rm ln}k} \simeq 8 \times 10^{-4}~.$
\end{itemize}

\subsection{WMAP and the Inflaton} 

\begin{figure}[htb]
  \centering 
\epsfig{file=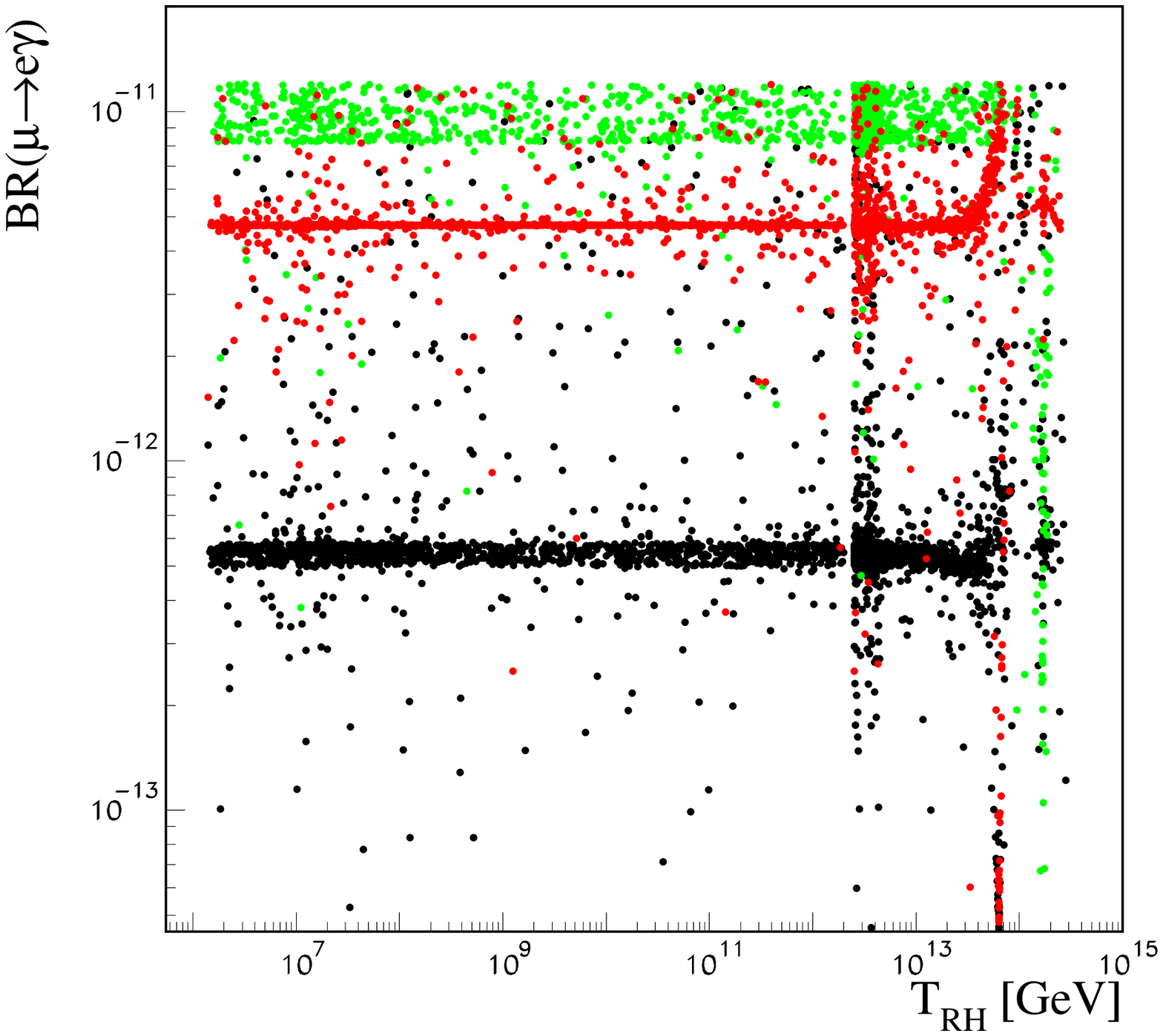,angle=0,width=0.4\linewidth}
\hfill \epsfig{file=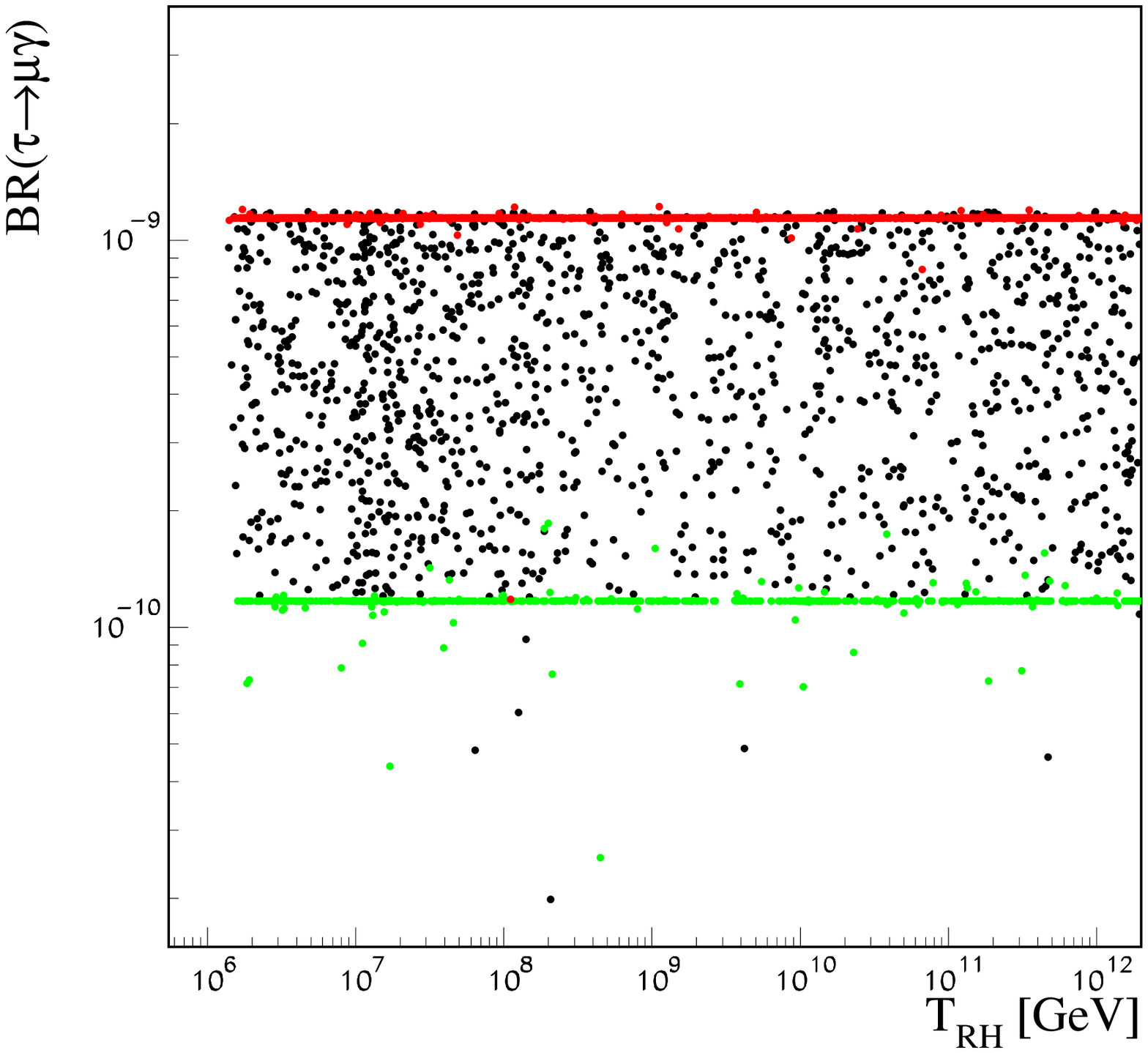,angle=0,width=0.4\linewidth}
\caption{{\it Branching ratios 
of various Lepton-Flavour-Violation processes in scenaria 
viewing the Sneutrino as the Inflaton~\cite{sneutrino}.}}
\end{figure}

Despite the agreement, via best fit scenaria, of the WMAP data with 
inflationary models, unfortunately no information can be obtained
at this stage on the nature of the inflaton field. In fact, all the WMAP data 
imply that at an early stage of our Universe there was an era of 
exponential expansion,
but cannot shed light on what caused this 
expansion. Although the most popular and successful 
model of inflation to date involves the inflaton~\cite{inflation},
however this is not the only possibility. 
An exponential expansion of the Universe 
does not have to be 
produced by an extra fundamental scalar field. 
The inflationary phase may be dynamical
in nature, without the need of inflaton fields. Such issues can only 
be understood 
when the full non-equilibrium dynamics of inflationary Universe 
is at hand. 

We mention  in passing at this point that within the context
of string theory~\cite{dilaton}, the dilaton can play 
the r\^ole of the inflaton 
field, and that this mode may arise dynamically 
in some non-critical string scenaria by incorporating 
properly non-equilibrium dynamics via the so-called Liouville 
mode~\cite{emninfl}. 
The Liouville mode is a world-sheet field~\cite{ddk}, which arises 
from the need for restoration of the conformal invariance 
of the underlying stringy $\sigma$-model theory, which is broken 
as a result of a cosmically catastrophic event, such as an initial
quantum fluctuation or, in a  modern context, the collision  between brane Universes, one of which represents our world~\cite{diamandis,gm,mav}.
In such models, the Liouville field appears with a time-like signature, and hence may be viewed~\cite{emnliouv,kogantime} as an (irreversible) target 
time variable in the system. The irreversibility is guaranteed by 
world-sheet renormalization-group arguments, given that the zero mode of the 
Liouville field plays the r\^ole of a world-sheet scale~\cite{emnliouv}. 
 
Due to its specific coupling with the two-dimensional curvature
of the world-sheet of the stringy $\sigma$-model, the Liouville 
mode contributes linear (in target time) terms in the dilaton excitation
of the string spectrum. Linear-in-target-time dilaton cosmology 
has been discussed in 
\cite{aben}, where it was shown that it implies linearly expanding 
Universes in the Robertson-Walker time~\footnote{In stringy cosmology, the 
Robertson-Walker frame is defined as the one in which 
the time in the metric has unit coefficient, but also 
the Einstein 
scalar curvature 
term in the low-energy effective field-theory action has no exponential
dilaton terms in front. 
This requires some redefinition, not only
of the time variable, but also of the target space metric
background by appropriate exponential dilaton factors~\cite{aben}.
Under such redefinitions, an originally linear in target time dilaton, 
becomes logarithmic in Robertson-Walker time.} 
To obtain inflation in such non-critical
string scenaria requires more complicated string 
backgrounds~\cite{emninfl,diamandis,gm,mav}, and such models have been 
discussed in the literature, also 
in connection with a current-era acceleration 
of the Universe. We shall come back to such issues in section 6.

Quite recently, 
other interesting particle physics scenaria have been 
suggested~\cite{sneutrino}, in which 
the inflaton may be the supersymmetric partner of the neutrino
(sneutrino). The model is argued to be consistent with the 
WMAP data, corresponding to a quadratic inflaton potential 
$V \sim \phi^2$. In such scenaria Leptogenesis leads to a prediction for  
the Lepton Flavour Violation (LFV) of the model 
which appears consistent with 
the WMAP constraint on the baryon density 
at a 3-$\sigma$ level: $7.8 \times 10^{-11} < Y_B < 1.0 
\times 10^{-10}$. It must be noted however that 
the model is not consistent with 
the WMAP non-zero value of the running of the spectral index $dn_s/d{\rm ln}k$.
However at present the data are too crude for an unambiguous 
determination of such quantities, and so the authors of \cite{sneutrino}
used only the $1\sigma$ level result of that particular measurement
to fit their model.

One can calculate in this model LFV 
branching ratios $BR(\tau \to \mu \gamma ) \lsim 10^{-9}$, 
$BR(\mu \to e \gamma ) \lsim 10^{-11}, BR(\tau \to e \gamma) < 10^{-7} $,
as well as electron and muon dipole moments, which have been argued~\cite{sneutrino} to 
be  
consistent with present experimental limits. The most sensitive
experiments to measure such BR´s are those involving 
neutrino oscillations. 
Note that such BR cannot be measured in the 
ATLAS/LHC experiment, whose sensitivity to such LFV processes is 
much poorer ($BR(\tau \to \mu \gamma)_{\rm ATLAS} \sim 10^{-6}$).

\section{WMAP AND DARK MATTER}

\subsection{Hot and Warm Dark Matter Excluded}

\begin{figure}[htb]
  \centering 
\epsfig{file=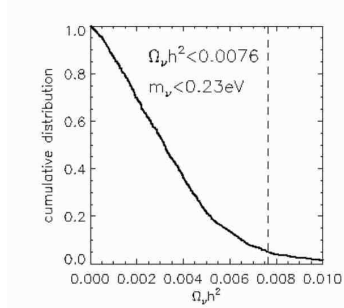,angle=0,width=0.3\linewidth}
\caption{{\it Hot Dark Matter models are excluded in view of the WMAP 
determination of (low) upper bounds of neutrino masses~\cite{wmap}.}}
\label{neutrino}
\end{figure}

The WMAP results on the cosmological parameters
discussed previously 
disfavor strongly Hot Dark Matter (neutrinos), as a result of the 
new determination of the upper bound on neutrino masses.

The contribution of neutrinos to the energy density of the 
Universe depends upon the sum of the mass of the light neutrino 
species~\cite{kolbturner,spergel}:
\begin{equation} 
\Omega_\nu h^2 = \frac{\sum_{i} m_i }{94.0~{\rm eV}}
\label{neutrinodens}
\end{equation} 
where the sum includes neutrino species
that are light enough to decouple while still relativistic. 

The combined results from WMAP and other experiments~\cite{spergel}
on the cumulative likelihood of data as a function of the energy 
density in neutrinos is given in figure \ref{neutrino}. Based on this analysis 
one may conclude that $\Omega_\nu h^2 < 0.0067$ (at 95\% confidence limit).
Adding the Lyman $\alpha$ data, the limit weakens slightly~\cite{spergel}: 
\begin{equation} 
\Omega_\nu h^2 < 0.0076
\label{neutrinolimit} 
\end{equation} 
or equivalently (from (\ref{neutrinodens})): 
$\sum_i m_{\nu_i} <0.69$ eV, where, we repeat again, the sum includes 
light species of neutrinos. 
This may then imply 
an average upper limit on electron neutrino mass 
$<m_\nu>_e < 0.23$ eV. The 
upper bound on the relevant densities (\ref{neutrinolimit})
strongly disfavours 
Hot Dark Matter scenarios. 

\begin{figure}[htb]
  \centering 
\epsfig{file=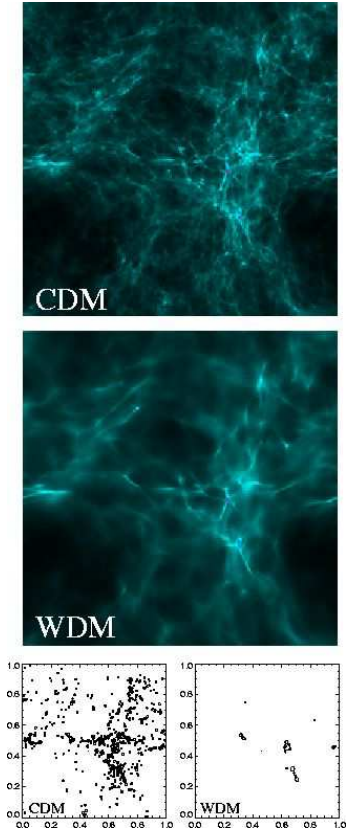,angle=0,width=0.3\linewidth}
\caption{{\it Numerical simulations for structure formation 
in Cold Dark Matter (CDM)  and Warm Dark Matter (WDM) 
models with mass $m_X = 10$ KeV~~\cite{warm}. 
\underline{Upper two figures}: The projected gas distribution in the CDM (top)
and the WDM (middle) simulations at $z=20$.\underline{Lower bottom figures}:  
the distribution of dark halos with mass greater than $10^5 M_{\odot}$
for the CDM (left) and for the WDM (right) model.}}
\label{warmdm}
\end{figure}

Another important result of WMAP is the evidence for
early re-ionization of the Universe at redshifts $z \simeq 20$.
If one assumes that structure formation is responsible for
re-ionization, then such early reionization periods 
are compatible only for high values of the masses $m_X$ of {\it 
Warm Dark Matter }. Specifically, one can exclude models 
with $m_X \le 10$ KeV~\cite{spergel,warm}
based on numerical simulations of structure formation for such models. 
Such simulations (c.f. figure \ref{warmdm}) imply that dominant structure
formation responsible for re-ionization, for Warm Dark Matter candidates
with $m_X \le 10$ KeV, occurs at much smaller $z$ than those observed
by WMAP. In view of this, one can therefore exclude popular 
particle physics models  
employing 
light gravitinos ($m_X \simeq 0.5 $ KeV) as the Warm Dark Matter candidate.

It should be noted at this stage that such structure formation arguments
can only place a lower bound on the mass of the Warm Dark Matter candidate.
The reader should bear in mind that 
Warm Dark Matter with masses $m_X \ge 100$ KeV becomes indistinguishable 
from Cold Dark Matter, as far as structure formation is concerned.

\subsection{Cold Dark Matter in Supersymmetric Models: Neutralino}

After the exclusion of Hot and Warm Dark Matter, the only type of 
Dark matter that remains consistent with the recent WMAP results~\cite{spergel}
is the {\it Cold Dark Matter }, which in general may consist of 
axions, superheavy particles (with masses 
$\sim 10^{14 \pm 5}$ GeV)~\cite{kryptons,chung}
and stable supersymmetric partners. 
 
Indeed, one of the major and rather
unexpected predictions of Supersymmetry (SUSY), broken
at low energies $M_{SUSY} \thickapprox \mathcal{O}(1 \TeV) $, 
while $R$-parity is conserved, is the existence of a stable, neutral
particle, the lightest neutralino  ($\lsp$), 
referred to as the lightest supersymmetric particle (LSP)~\cite{Hagelin}.
Such particle is an ideal candidate 
for the Cold Dark Matter in the Universe \cite{Hagelin}.
Such a prediction fits well with the fact that 
SUSY is not only  indispensable in 
constructing consistent string
theories, but
it also seems unavoidable at low energies ($\sim 1 \TeV$)
if the gauge hierarchy problem is to
be resolved.
Such a resolution provides a measure of the SUSY
breaking scale $M_{SUSY} \thickapprox \mathcal{O}(1 \TeV) $. 
There is  indirect evidence for such a
low-energy supersymmetry breaking scale, from the unification
of the gauge couplings \cite{Kelley} and from the apparent lightness
of the Higgs boson as determined from precise electroweak measurements,
mainly at LEP \cite{EW}. 

This type of Cold Dark Matter will be our focus from now on,
in association with the recent results from WMAP on relic 
densities~\cite{wmap,spergel}. The WMAP results for the 
baryon and matter densities (including dark matter)
are given in the table of figure \ref{table1spergel}. After combining 
the WMAP data with other existing data, one obtains
the following final values:   
\begin{eqnarray}
&&\Omega_m h^2 = 0.135^{+0.008}_{-0.009}~ {\rm (matter)}~, \nonumber \\
&&\Omega_b h^2 = 0.0224 \pm 0.0009~ {\rm (baryons)}~.
\label{baryonmatter}
\end{eqnarray}
Hence, assuming that CDM is given by the difference of these two,
one has 
\begin{equation}
\Omega_{CDM} h^2 = 0.1126^{+0.0161}_{-0.0181}~, \qquad (2-\sigma \quad {\rm level})~.
\label{cdmlimits}
\end{equation} 

As mentioned already, in supersymmetric (SUSY) theories the 
favorite candidate for CDM is
the lightest of the Neutralinos $\lsp$
(SUSY CDM), which is stable, as being the Lightest SUSY particle (LSP)
{\footnote{There are cases where the stau or the sneutrino can be
the lightest supersymmetric particles. These cases are not favoured 
\cite{nickref}  and hence are not considered. }}. 
From the above-described WMAP results, then,  assuming $\Omega_{CDM} 
\simeq \Omega_\chi$,  we 
can infer stringent limits for the neutralino $\chi$ relic density:
\begin{equation}
0.094 < \Omega_\chi h^2 < 0.129~,\qquad (2-\sigma \quad {\rm level})
\label{relicdensity}
\end{equation}
It is important to notice that in this 
inequality  only the upper limit is {\it rigorous}. 
The lower Limit is {\it optional}, given  
that there might (and probably do) exist other contributions 
to the overall (dark) matter density.

The constraints in SUSY models we shall review in this 
work come precisely from 
the use of this constraint,  
together with
other results mainly from
the anomalous magnetic moment of the muon, $g_\mu - 2$, 
the decay $b \to s\gamma $, as well as 
LEP-induced  Higgs mass lower limits. 
Specifically, we shall constrain the minimal SUSY Standard Model 
(MSSM) with universal values for the masses of scalars $m_0$
the gauginos
$m_{1/2}$ and trilinear soft couplings $A_0$ at the GUT scale. Furthemore
Electroweak Symmetry Breaking (EWSB) is assumed. 
This is the so-called
Constrained MSSM (CMSSM).  
For the purposes of this review we shall 
concentrate on plots $(m_0, m_{1/2})$ for 
the minimal supergravity model (mSUGRA)~\cite{msugra,nilles}, 
in which such CMSSM are embedded. 

It is imperative to notice that all the constraints we shall discuss
in this review are {\it highly model dependent} and the results 
cannot be extrapolated
to non-minimal models. But most of the analysis can be extrapolated to such models, with possibly different results. 
In what follows, therefore, for definiteness
we shall concentrate exclusively to mSUGRA models. We shall 
not discuss non-minimal models in this article.

\section{WMAP AND SUSY CDM}

\subsection{CMSSM/mSUGRA: Basic Features}

Before embarking into a detailed analysis of the constraints
of the minimal supersymmetric standard model
embedded in a minimal
supergravity model (CMSSM)~\cite{msugra}, we consider it
useful to
outline the basic features of these models, which will be used 
in this review. 

The embedding of SUSY models into the minimal supergravity 
(mSUGRA) model implies that there are five independent parameters: 
Three of them, the scalar and gaugino masses $m_0, m_{1/2}$ as well as the trilinear soft coupling 
$A_0=$, at the unification scale, set the size of the Supersymmetry breaking scale. In addition one can consider as input parameter ${\rm tan}\beta =\frac{<H_2>}{<H_1>}$, the ratio  of the v.e.v's of the Higgses $H_2$ and $H_1$ giving masses to up and down quarks respectively. The sign 
( signature) of the Higgsino mixing parameter $\mu$ is also an input but not its size which is determined from the Higgs potential minimization condition
\beq
\mu^2+\frac{{ M}_Z^2}{2} = 
\frac{m^2_{H_1}-m^2_{H_2}\; {\tan \beta}^2}  {{\tan \beta}^2-1}+
\frac{{\Sigma_1}-{\Sigma_2}\; {\tan \beta}^2}{{\tan \beta}^2-1} \; .
\label{mineq}
\eeq
In (\ref{mineq}) ${M}_Z$ is actually the running Z - boson mass and $m_{H_{1,2}}$ the soft masses of the Higgses $H_2$ and $H_1$ respectively. $\Sigma_{1,2}$ on the r.h.s of this equation are contributions arising from loop corrections of the effective potential \cite{arnonath,hbbranch}. 

The parameter space of mSUGRA can be effectively described in terms of two branches as can be seen by exploring the minimization condition (\ref{mineq}). 
Following the discussion in~\cite{chatto} we can demonstrate this analytically ignoring,
for simplicity,  the coupling of the $b$ quark. In this case (\ref{mineq}) can be approximated by
\cite{hbbranch} :
\begin{equation} 
C_1 m_0^2 + C_3{m'}_{1/2}^2 + C_2'A_0^2 + \Delta \mu_{\rm loop}^2 = 
\mu^2 + \frac{1}{2}M_Z^2
\label{constraintsresb} 
\end{equation} 
where
\begin{equation} 
{m'}_{1/2} \equiv m_{1/2} + \frac{1}{2}A_0\frac{C_4}{C_3}~, \qquad
C_2' = C_2 - \frac{1}{4}\frac{C_4^2}{C_3} 
\label{defmprime}
\end{equation}
and 
\begin{eqnarray} 
&& C_1 = \frac{1}{t^2 - 1}(1 - \frac{3D_0 -1}{2}t^2)~, \quad 
C_2 = \frac{t^2}{t^2 - 1}k~, \nonumber \\
&& C_3 = \frac{1}{t^2 - 1} (g - t^2 e)~, \quad 
C_4 = -\frac{t^2}{t^2 -1}f~, \Delta \mu_{\rm loop}^2 = \frac{\Sigma_1 - t^2 \Sigma_2}{t^2 - 1}
\; .
\label{C1}
\end{eqnarray}
$\Delta \mu_{\rm loop}$ denote the loop correction to the parameter $\mu$, $t={\rm tan}\beta$,
$\Sigma_{1,2}$ were defined before,  
the functions $e,f,g,k$ are as defined in~\cite{ibanez}, and 
$D_0 = 1 - (m_t/m_f)^2$ where $m_f \simeq 200 \; {\rm sin}\beta$ GeV. 
\begin{itemize} 

\item{Ellipsoidal Branch (EB) of Radiative Symmetry Breaking :} 

For small to moderate values of tan$\beta \lsim 7$, 
the loop corrections are typically small, and the renormalization 
group analysis shows that both $C'_2 > 0$ and $C_3 > 0$. In this case one 
finds $C_1 >0$ independently of any scale choice $Q$ for the radiative 
electroweak symmetry breaking. Then, one finds that the 
radiative symmetry breaking constraint demands that the allowed set of 
soft parameters $m_0$ and ${m'}_{1/2}$ lie, for a given value of $\mu$,  
on the surface of an {\bf{Ellipsoid}}. This places   
upper bounds on the sparticle masses for a given value of 
$\Phi \equiv \mu^2/M_Z^2 + 1/4$. This is the ellipsoidal branch (EB) of the 
Electroweak Radiative Symmetry Breaking (EWRSB). 

Although this is always true for values tan$\beta \lsim 7$,
it {\it does not necessarily hold} for larger values tan$\beta \gsim 7$. In the latter case $m_0$ and ${m'}_{1/2}$ may lie on the surface of a
{\bf{Hyperboloid}}. This situation, which peresents phenomenological interest, is described in more detail below \cite{hbbranch}.

\item{Hyperbolic Branch (HB) of Radiative Symmetry Breaking}: 
 
For large values of tan$\beta \gsim 7$,
the loop corrections to $\mu$ are significant and $\mu$ also exhibits a significant variation with varying the scale $Q$. If we choose the running scale at $Q_0$ where the loop corrections are minimized, or even vanish, the loop corrections to the r.h.s of (\ref{mineq}) can be ommited and all quantities appearing in eq. (\ref{mineq}) should be considered at this scale. 
Usually the scale $Q_0$ is the geometric average of the stop masses which give the largest contributions to $\Sigma_{1,2}$. In some regions of the parameter space it may occur that the 
sign of $C_1$ is flipped, i.e. ${\rm sign}(C_1(Q_0))=-1$, so that the minimization condition receives the form 
\begin{equation} 
\frac{{m'_{1/2}}^2}{\alpha^2(Q_0)} - \frac{{m_0}^2}{\beta^2(Q_0)} 
\simeq \pm 1 \; . 
\label{hbrancheq}
\end{equation} 
In (\ref{hbrancheq}) $\alpha = \frac{|(\Phi_0 + \frac{1}{4})M_Z^2 - C_2'A_0^2|}{|C_3|}~, 
\beta = \frac{|(\Phi_0 + \frac{1}{4})M_Z^2 - C_2'A_0^2|}{|C_1|}$, and the 
$\pm$ sign is determined by the condition ${\rm sign}\left((\Phi + \frac{1}{4})M_Z^2 - C_2'A_0^2\right)= \pm 1~.$ It is apparent from (\ref{hbrancheq}) that unlike the previous case 
$(m_0, m'_{1/2})$ lie now on the surface of a {\bf{hyperboloid}} and hence the name Hyperbolic Branch (HB).
\end{itemize}

What is interesting in the HB case it the fact that $m_0$ and or  ${m}_{1/2}$ can become very large  while much smaller values for $\mu$. A subset of HB is 
the so called {\bf{high zone}}. In this case EWSB can occur in regions where $m_0$ and ${m}_{1/2}$ can be in the several TeV range, with much smaller values for the parameter $\mu$ which however is much larger than $M_Z$. This has important consequences for phenomenology as we shall see. In this zone the lightest of the neutralinos, $\chi_1$, is almost a Higgsino having mass $\mu$. This is called {\it{inversion phenomenon}} since the LSP is a Higgsino rather a Bino. 

Except the high zone where the inversion 
phenomenon takes place the HB includes the so called {\bf{Focus Point}} 
(FP) region~\cite{focusfirst,focus}, 
which is defined as a region in which some renormalization group (RG) 
trajectories 
intersect (FP region would be only a point, 
were it not for threshold effects which smear it out). 
We stress that the FP is {\it not} a fixed point of the RG. 
The FP region is a subset of the HB limited to relatively low values of ${m}_{1/2}$ and values of $\mu$ close to the electroweak scale, $M_Z$, while $m_0$ can be a few TeV but not as large as in the high zone due to the constraints imposed by the EWSB condition. The LSP neutralino in this region is a mixture of Bino and Higgsino and the Higgsino impurity  allowes for rapid s-channel LSP annihilations, resulting to low neutralino relic densities at experimentally acceptable levels. 

From the previous discussion concerning the HB region it is obvious that the crucial property characterizing this region is the inversion of the sign of $C_1$. For large  tan$\beta$ 
( tan$\beta \gsim 7$ ), the sign of $C_1$ is mainly driven by that of the quantity $3 D_0 - 1$ as is seen from eq. (\ref{C1}). In order for $C_1$ to be negative $3 D_0 - 1$  should be positive which with the approximations adopted so far entails to 
$$
\sin \beta > {\sqrt{\frac{3}{2}}}\; \frac{m_t}{200 \;GeV} \approx \frac{m_t}{163.3 \;GeV} 
$$
Even from this crude approximation it is apparent the central role the value
of the running top  mass $m_t$ it plays for the location of the Hyperbolic Branch.
If it happens $m_t > 163.3 \; GeV$, at the scale $Q_0$,   the HB is never
reached. Thus the preference towards smaller values of the top quark mass,
close to the lowest experimental limit, for the location of the HB is
manifest from this discussion. Therefore the HB region is sensitive  to
variations of the parameters of the theory which may induce a shift of the
theoretically determined $m_t$ to higher values by 1-2 $\%$.
It then becomes evident that two loop RGE effects to Yukawa couplings,
radiative corrections  to the top mass and supersymmetric corrections to
the bottom Yukawa coupling, which are important for large values of
tan$\beta$, as well as low energy threshold effects to the strong coupling
constant, which greatly affect the RGE running of the parameters of the
theory and hence low energy predictions, must be duly taken into account
for a correct and unambiguous phenomenological description of this region.
These subtleties may drastically affect the conclusions reached in various analyses and
this is the reason this phenomenologically interesting region is sometimes
considered as being fine tuned.
\subsection{Focusing The Cosmologically Relevant Regions}

Having described the two principal regions, EB and HB, with different
phenomenolological characteristics each, we now pass to briefly outline the
regions which posses special interest as far as CDM is concerned.

In the EB region the EWSB condition drives the parameter $\mu$ to be of the
same order of magnitude with $m_0, \; m_{1/2}$. In the bulk of this region
the lightest of the neutralinos, $\chi_1^0$, is mostly a bino except a region
with small $m_{1/2}$ in which $\chi_1^0$ has a sizeable Higgsino component.
This region is however experimentally ruled out by $b \to s\gamma$ data and
light chargino searches. Excluding this region we focus our attention to the
following subregions which possess special features as far as Cosmology and
particle physics is concerned. The first is the stau coannihilation region in
which $\chi_1^0$ is almost degenerate in mass, but heavier, with the
light $\tau$-slepton. When this happens the calculation of the neutralino
CDM requires the coannihilations $\chi_1^0 - \tilde{\tau}  $
and $\tilde{\tau} - \tilde{\tau} $
\cite{efo2,lazarides,lezsek}. In this region the mass of $\chi_1^0$ is allowed to
be extended up to 500 GeV assuming that the WMAP data are respected. The
second region is characterized by large tan $\beta \gsim 45$, and concerns
the region where $2 \; m_{\chi_1^0} \simeq m_A \;$ where $m_A$ is the
pseudoscalar Higgs mass. In this region s-channel pseudoscalar annihilation
to a $b - \bar b$ or a 
$\tau - \bar \tau$ pair becomes important in opening up extended regions
compatible with WMAP data. Interestingly enough this large tan
$\beta $ region yields high values, $\simeq 10^{-8,9}$ pb for the elastic
$\chi_1^0$ - nucleon cross section close to the sensitivity limits of on
going and future direct dark matter experiments. We postpone a more detailed
discussion of this issue later on in this review.

Moving to the HB, 
in general the scalar $m_0$ and gaugino $m_{1/2}$ 
masses can get very large, which implies that the squarks 
and sleptons can be very heavy, with masses that lie in the several 
TeV range. The feature of the large $m_0$ mass is also shared by 
the focus point region of mSUGRA models~\cite{focusfirst,focus}.  
Consider the high zone of the HB, in which 
$m_{1/2} \gg \mu \gg M_Z$. In this scenario one finds~\cite{hbbranch,chatto} 
that the two lightest neutralino states, $\chi_1^0$, $\chi_2^0$ and the light 
chargino state $\chi_1^\pm$ are essentially degenerate with mass $\sim \mu$. 
Theis masses range from several hundreds GeV to above TeV while their mass
differences are much smaller in the range 1-10 GeV.To first order they are
degenerate all having a mass $\mu$ but this degeneracy is lifted by small
corrections ${\cal{O}} ( \frac{M_Z}{\mu} )$ entailing to a mass spectrum
$$
m_{\chi_1^0} < m_{\chi_1^{\pm}} < m_{\chi_2^0} 
$$
with ${\chi_1^{\pm}}$ the lightest chargino states and $\chi_2^0$ the next
to lightest neutralino.
As mentioned earlier in this region the {\it inversion phenomenon} takes
place given that the lightest neutralino 
switches from being mostly a Bino to being a higgsino.

The inversion phenomenon has dramatic effects on the nature 
of the particle spectrum and 
SUSY phenomenology in this HB. 
Indeed, in mSUGRA one naturally has coannihilation 
with the sleptons when the neutralino mass extends to masses 
beyond 150-200 GeV with processes of the type
$\chi \tilde \ell_R^a \rightarrow \ell^a \gamma, \ell^a Z, \ell^a h$,
$\tilde \ell_R^a \tilde \ell_R^b \rightarrow \ell^a \ell^b$,
 and $\tilde \ell_R^a \tilde \ell_R^{b*} \rightarrow \ell^a\bar \ell^b,
\gamma \gamma, \gamma Z, ZZ, W^+W^-, hh$ where $\tilde{\it l}$ is 
essentially a $\tilde \tau$. 
Remarkably the relic density constraints can
be satisfied on the hyperbolic branch also by coannihilation. 
However, on the HB the  coannihilation is of
an entirely different nature as compared with the stau coannihilations
discussed previously:
instead of a neutralino-stau co-annihilation, and stau - stau 
in the HB one has co-annihilation processes 
involving the second lightest neutralino and chargino states~\cite{gondolo}, 
$\chi_1^0-\chi_1^{\pm}$, followed by 
$\chi_1^0-\chi_2^{0}$,$\chi_1^+-\chi_1^{-}$,$\chi_1^{\pm}-\chi_2^{0}~.$
Some of the dominant processes that contribute
to the above coannihilation processes are \cite{gondolo}
\begin{eqnarray}
\chi_1^0 \chi_1^{+}, \chi_2^0 \chi_1^{+}\rightarrow
u_i\bar d_i, \bar e_i\nu_i, AW^+,Z W^+,
W^+h \nonumber\\
\chi_1^{+} \chi_1^{-}, \chi_1^0 \chi_2^{0}\rightarrow
u_i\bar u_i, d_i \bar d_i,
W^+W^-
\label{charginocoani}
\end{eqnarray}
Since the mass difference between the states 
$\chi_1^+$ and $\chi_1^{0}$ is the smallest the $\chi_1^0 \chi_1^{+}$
coannihilation dominates. 
In such cases, the masses $m_0$ $m_{1/2}$ 
may be pushed beyond $10$ TeV, so that squarks and sleptons 
can get masses up to several TeV, i.e.  
beyond 
detectability limits of immediate future accelerators such as LHC.

Except the high zone the other interesting region belonging to the HB is
the Focus Point (FP).
As already mentioned in the previous chapter unlike the high zone this is
characterized by $m_0$ in the few TeV range, low values of $m_1{1/2}<< m_0$
and rather small values of $\mu$ close to
$M_Z$. 
The LSP neutralino in this case is a mixture of Bino and Higgsino and its
Higgsino impurity is adequate to give rize to rapid s-channel LSP
annihilations so that the neutralino relic density is kept low at
experimentally acceptable values. Since $\mu$  is small the lightest
chargino may be lighter than 500 GeV and the FP region may be accessible to
future TeV scale colliders. Also due to the relative smallnes of ${m}_{1/2}$
in this region gluino pair production may occur at a high rate making the
FP region accessible at LHC energies.

As we remarked before although the HB may be  viewed as fine tuned,
nevertheless recent studies~\cite{baer}, based on a $\chi^2$ analysis, have
indicated that the WMAP data,
when combined with data on $b \to s\gamma$ and $g_\mu -2$, 
seem to favour the Focus Point HB region and the large tan $\beta$
neutralino resonance annihilation of mSUGRA.

\subsection{SUSY CDM in the pre-WMAP era and the importance of 
the muon's anomaly}

Before embarking onto a discussion on the constraints on 
SUSY CDM implied by the recent WMAP data on relic densities, 
we consider it as instructive to review phenomenological  
works before the publication of these data.  

The issue of the Dark Matter has attracted the interest of many physicists
and in the literature
there are numerous theoretical works dealing with this issue
\cite{x1,x2,x3,x4,x5,x6,x7,x8,allofthem,Hagelin,gondolo}.
Departures from the minimal mSUGRA models and studies including
non-universalities \cite{nonuniv}, CP - violating effects \cite{cpviol}
as well as efforts related to Yukawa unification \cite{lazarides,unific}
have been pursued.
Phenomenological studies including the effects of variations of the WIMP
velocities on the detection rate have also been considered \cite{variation} 
and regions where coannihilation phenomena are important have been analyzed
\cite{mizuta,gondolo,efo2,lazarides,lezsek,santoso,klapdorcoan}

The importance of the smallness of the 
Dark Matter (DM) relic density in constraining supersymmetric predictions 
had been pointed out in the past~\cite{LNS1,LNSd,lns,lnsmuon}, 
paying special attention at the 
to the large $\; \tan \beta \; $ regime.
In this region the neutralino ($\lsp$) pair annihilation 
through $s$-channel
pseudo-scalar Higgs boson ($A$) 
exchange, leads to an enhanced annihilation cross sections
reducing significantly the relic 
density \cite{Drees}. 
The importance of this mechanism, in conjunction with the 
cosmological data which favour small values of the DM
relic density,
has been stressed in \cite{LNS1,LNSd}.

\begin{figure}[t]
\centering 
\epsfig{file=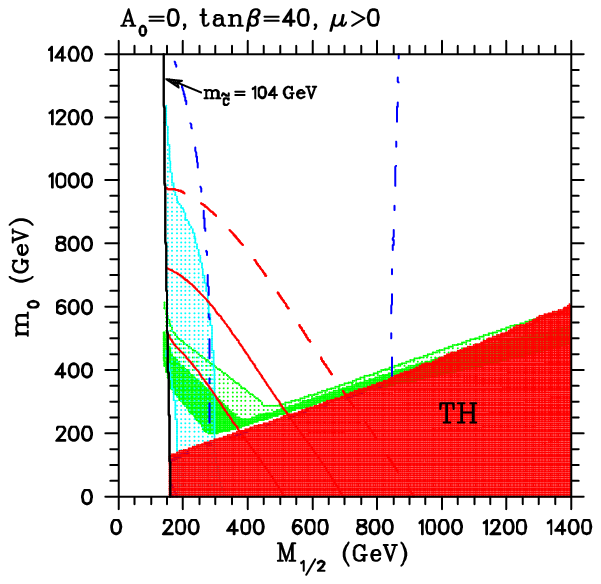,angle=0,width=0.30\linewidth}
\hfill \epsfig{file=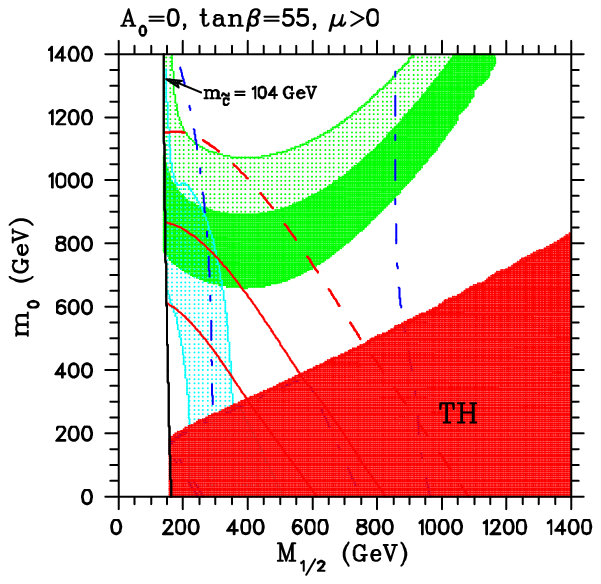,angle=0,width=0.30\linewidth}
\caption{{\it mSUGRA/CMSSM  constraints before WMAP~\cite{LNS1,lns}. 
The Mass of the top quark is taken 175 GeV. Dark Green 
shaded areas correspond to neutralino relic densities in the range: 
$0.08 < \Omega_\chi h^2 < 0.18,$ whilst  
light green areas 
correspond to $ 0.18 < \Omega_\chi h^2 < 0.30$. The Solid
Red Line indicates constraints
implied by the anomalous magnetic moment of the muon 
$a_\mu ^{\rm SUSY} = (306 \pm 106) \times 10^{-11}$. The 
Dashed red line indicates the boundary of the region for which 
the lower bound is moved to 2$\sigma$ 
limit. The Dashed Blue lines correspond 
to constraints implied by the Higgs mass $113.5 {\rm GeV} < m_{\rm Higgs} < 
117.0 {\rm GeV} $. Finally Cyan shaded region on the right 
are excluded
due to $ b \to s \gamma $ constraint.}}
\label{early}
\end{figure}

In previous analyses regarding 
DM direct searches \cite{LNSd}, we had stressed 
that  the contribution of the $CP$-even Higgs bosons exchange
to the LSP-nucleon scattering cross sections increases with $\tan \beta$.
Therefore in the large $\tan \beta$ 
region one obtains the highest possible rates
for the direct DM searches and the smallest LSP relic densities. 
Similar results are presented in Ref.~\cite{Kim}. 
Early studies on small  
$\Omega_\chi h^2 \lsim 0.18$,
taking into account constraints from the anomalous
magnetic moment of the muon,  
have been carried out in \cite{lnsmuon},  
based on the mSUGRA model. The results from such analyses 
are depicted in figure \ref{early}. 

\begin{figure}[t]
\centering 
\epsfig{file= 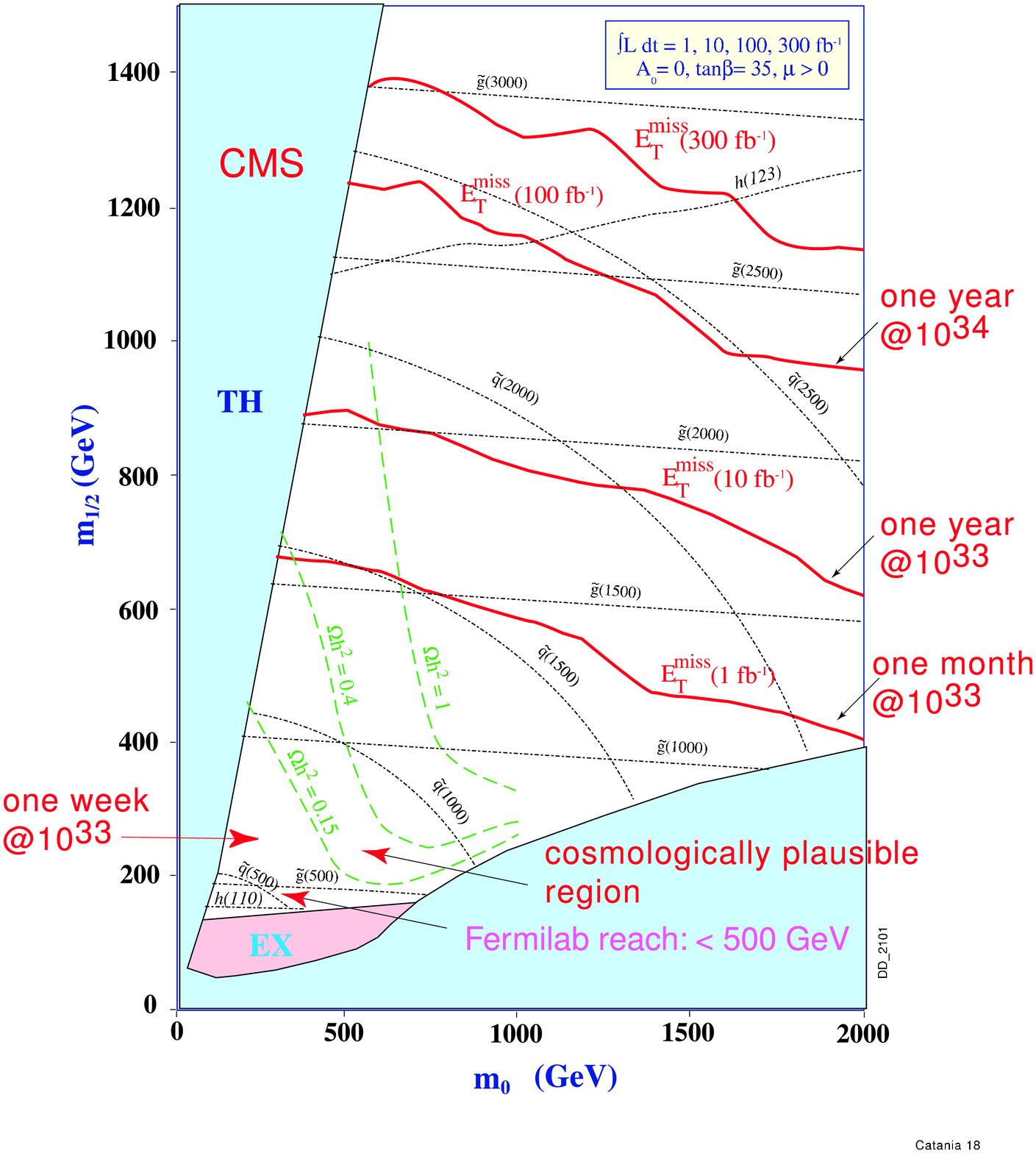,angle=0,width=0.44\linewidth,clip=}
\hfill \epsfig{file= 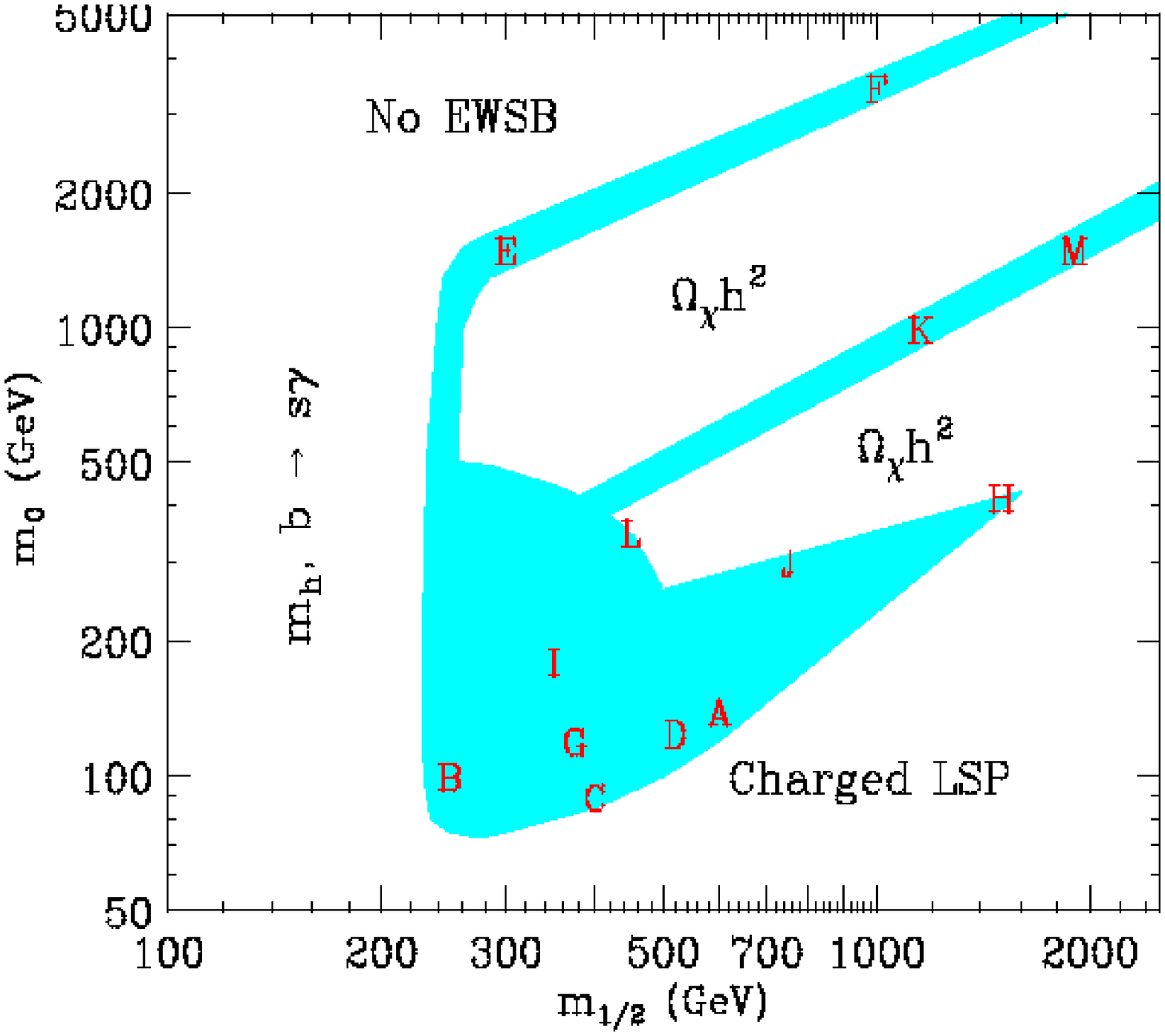,angle=0,width=0.51\linewidth,clip=}
\caption{{\it \underline{Left figure:} SUSY detection prospects 
before WMAP data~\cite{Ellisrev},  
explorable at LHC using missing energy and jet signature. 
\underline{Right Figure:} Post-LEP 
Benchmark SUSY scenaria~\cite{benchmark}. Benchmark points have 
$\Omega_\chi h^2$ larger than the bounds implies by the WMAP data. }} 
\label{bench}
\end{figure}

The pre-WMAP situation~\cite{Ellisrev}, involving the efforts of other
groups, concerning prospects for
detecting SUSY is summarized in figure \ref{bench},
where we also give some benchmark scenaria~\cite{benchmark}.

Undoubtedly one of the most significant experimental results of the last 
years is the measurement of the anomalous magnetic moment of the
muon~\cite{E821Expt}.
Deviation of its measured value from the Standard Model (SM) predictions is
evidence for new physics with Supersymmetry being the prominent candidate to
play that role.
Adopting Supersymmetry as the most natural extension of the SM, such
deviations may be explained and  impose at the same time severe constraints
on the predictions of the available SUSY models by putting upper bounds on
sparticle masses. Therefore knowledge of the value of $g_\mu - 2$  is of
paramount importance for Supersymmetry and in particular for the fate of
models including heavy sparticles in their mass spectrum, as for
instance those belonging to the Hyperbolic Branch.

Unfortunately the situation concerning the anomalous magnetric moment is
not clear as some theoretical uncertainties remain unsettled as yet. 
At present, there are two theoretical estimates for the difference of the 
experimentally measured~\cite{E821Expt} value of 
$a_\mu = (g_\mu -2)/2$ from the theoretically 
calculated one within the SM~\cite{narison},
\begin{itemize} 

\item
{Estimate (I)} $\; \;a_\mu^{\rm exp} - a_\mu^{\rm SM} = 1.7(14.2) \times 
10^{-10} \;  \quad [ 0.4(15.5) \times 10^{-10} ]  $

\item
{Estimate (II)} $a_\mu^{\rm exp} - a_\mu^{\rm SM} = 24.1(14.0) \times 
10^{-10}  \quad [ 22.8(15.3) \times 10^{-10} ] $

\end{itemize} 
In (I) the $\tau$-decay data are used in conjuction with Current Algebra
while in (II) the ${e^{-}}{e^{+}} \rightarrow$ Hadrons data are used in
order to extract the photon vacuum polarization which enters into the
calculation of $g_\mu - 2$.
Within square brackets are the most updated values of Ref.~\cite{narison}
{\footnote{Due to the rapid updates concerning $g_\mu - 2$ the values of     
$a_\mu^{\rm exp} - a_\mu^{\rm SM}$ used in previous works quoted in this
article may differ from those appearing above.}}. 
Estimate (I) is considered less reliable since it carries additional
systematic uncertainties and for this reason in many studies only the
Estimate (II) is adopted. Estimate (II) includes the contributions of
additional scalar mesons not taken into account in previous calculations. 

In order to get an idea of how important the data on the muon anomaly might
be we quote Ref.\cite{chatto} where both estimates have been used. 
If Estimate (II) is used at a 1.5$\sigma$ range 
much of the HB  and all of the inversion region 
can be eliminated. In that case the usually explored region of 
SUSY in the EB is the only one that survives, which, as we shall discuss 
below, can be severely constrained by means of the recent
WMAP data.     

On the other hand, Estimate (I), essentially implies no difference
from the SM value, 
and hence, if adopted, leaves 
the HB, and hence its high zone (inversion
region), {\it intact}.  In such a case, SUSY may not be detectable
at colliders, at least in the context of the mSUGRA model, but may be 
detectable in some direct dark matter searches, to which we shall turn to 
later in the article. 

For the above reasons, it is therefore {\it imperative} 
to determine unambiguously the muon anomalous magnetic moment 
$g_\mu -2$ by reducing the errors
in the leading order hadronic contribution, experimentally, and improving
the theoretical computations within the standard model. 
In view of its importance for SUSY searches, it should also be necessary
to have further experiments in the future, that could provide
independent checks of the measured muon magnetic moment by the 
E821 experiment~\cite{E821Expt}. 

\subsection{WMAP and SUSY constraints}
\subsubsection{Constraints in the EB}

After the first year of running of the WMAP 
satellite~\cite{wmap}, the situation 
concerning mSUGRA model constraints in the EB 
of radiative symmetry breaking parameter space,  
has changed dramatically~\cite{olive,ln}.
The constraints on the neutralino relic density 
(\ref{cdmlimits}, \ref{relicdensity}), implied by the best fit 
Universe model of the WMAP data combined with other 
experiments~\cite{wmap,spergel}, 
and with other constraints such as the post-LEP Higgs mass limits, 
$b \to s\gamma$ and $g_\mu - 2$,  
have restricted severely the available
parameter space of the EB. 

Some important remarks are in order at this stage: 
  
In Ref.~\cite{ln} in calculating the $\lsp$
relic abundance, the authors solve the Boltzmann equation  
numerically using the machinery
outlined in Ref.~\cite{LNS1}. In this calculation the coannihilation effects, 
in regions where $\tilde{\tau}_R$ approaches in mass the LSP, which is a high
purity Bino, are properly taken into account.

The  details of the 
procedure in calculating the spectrum of the CMSSM can be
found elsewhere \cite{LS,lnsmuon,lns}. Here 
we shall only briefly 
refer to some subtleties which turn out to be
essential for a correct determination of the 
Higgsino, $\mu$, and Higgs, $m_3^2$, mixing parameters, which 
greatly affect the Neutralino, Chargino and Higgs  mass spectra, especially in the 
region of large tan$\beta$. This region is of particular 
phenomenological interest as we discussed in previous sections.
$\mu$ and $m_3^2$ are determined by minimizing 
the  one-loop corrected effective potential \cite{arnonath}.
For large $\tan \beta$ the derivatives of the effective potential
with respect the Higgs fields, which enter into the minimization conditions, 
are plagued by terms which are large and hence potentially dangerous, making
the perturbative treatment untrustworthy.
In order to minimize the large $\tan \beta$
corrections we had better calculate the effective potential using as
reference scale the average stop scale
$Q_{\tilde t}\simeq\sqrt{m_{{\tilde t}_1} m_{{\tilde t}_2} }$~\cite{scale}. 
At this scale these terms are small and hence perturbatively trusted. 

The proceedure in solving the renormalization group equations for masses 
and couplings follows essentially that of Ref. \cite{BMPZ}. The low energy 
threshold corrections to the strong coupling constant are duly taken into
account
and the unification scale $M_{GUT}$ is defined at the point where the gauge 
couplings $\alpha_{1,2}$ 
meet. No unification of the gauge coupling constant $\alpha_{3}$ is enforced with 
$\alpha_{1,2}$ at $M_{GUT}$ as this usually results to too high values for the 
strong coupling constant $\alpha_s $ at $M_Z$. Therefore to comply with the 
experimentally value for $\alpha_s(M_Z)$ we take it as input and thus the value 
of $\alpha_{3}$ at the unification scale is an output rather than an input.  

A significant correction, which 
drastically affects the numerical proceedure 
arises from the gluino--sbottom and chargino--stop corrections to the bottom
quark Yukawa coupling  $h_b$~\cite{mbcor,wagner,BMPZ,arno}.
The proper resummation of these corrections
is important for a correct determination of $h_b$ \cite{eberl,car2} and 
this has been taken into account. 

Seeking a precise determination  of the Higgs boson mass
the dominant two-loop corrections to this have been included \cite{zwirner}.
Concerning the calculation of the $\bsga$ branching ratio,
the important contributions beyond the leading order, 
especially for large $\tan \beta$,  
have been taken into account \cite{gamb}. 

There have been two independent groups working on this update of the 
CMSSM in light of the WMA{P data, 
with similar results~\cite{olive,ln}
and below we summarize 
the results of their analysis in  
figures \ref{olivefig1},\ref{lnfig1},
for some typical values of the 
parameters ${\rm tan}\beta$ and signature of $\mu$.

As mentioned in previous sections the region designated as large tan$\beta$ 
has particular phenomenological interest. In this region the dominant 
neutralino annihilation mechanism is through s - channel annihilation 
of the pseudoscalar Higgs boson (A), 
$\lsp \, \lsp \stackrel{A}{\goes} b \, \bar{b}$ or $\tau \, \bar{\tau}$. 
The reason is that 
by increasing $\tan \beta$ the mass $m_A$ decreases, while the
neutralino mass remains almost constant, if the other parameters are kept
fixed. Thus the pseudoscalar Higgs mass $m_A$ is expected eventually to enter 
into the pole regime in which $m_A\,=\, 2 m_{\lsp}$.
At the same time the pseudoscalar Higgs couplings with $\tau \bar {\tau}$,
$b \bar{b}$ are large as being proportional to tan$\beta$. Thus although 
the $\chi_1^0$ is a high purity bino, which does not couple to A, its small
Higgsino component, which does couple, can give sizeable effects due to the
largeness of the $A b \bar{b}$ and $A \tau \bar{\tau}$ couplings. 
Both effects result to
enhanced annihilation cross sections dominating the process and hence  
to small neutralino relic densities~\cite{ln} over
a broad region of the parameter space characterized by large values 
tan$\beta$ values.
From this discussion it becomes obvious that 
for a correct determination of the relic density,  
in the large $\tan\beta$ region, an unambiguous and
reliable determination of the $A$-mass, $m_A$, is required.
In the CMSSM, 
$m_A$ is not a free parameter but 
is determined once the other parameters  are given.
$m_A$  depends sensitively on the Higgs
mixing parameter, $m_3^2$, which is determined from the 
minimization conditions of the effective potential, 
and for its calculation all 1 - loop corrections must be taken 
into account. 
It is found \cite{KLNS} that the handy calculation of the 
pseudoscalar mass through 
the second derivatives of the effective potential is scale 
independent and approaches the pole mass to better than 
$2 \%$ provided that one includes the corrections of all particles, 
especially those of Charginos and Neutralinos, 
in addition to the leading corrections of 
the third generation sfermions.  

\begin{figure}[htb]
\centering
\epsfig{file= 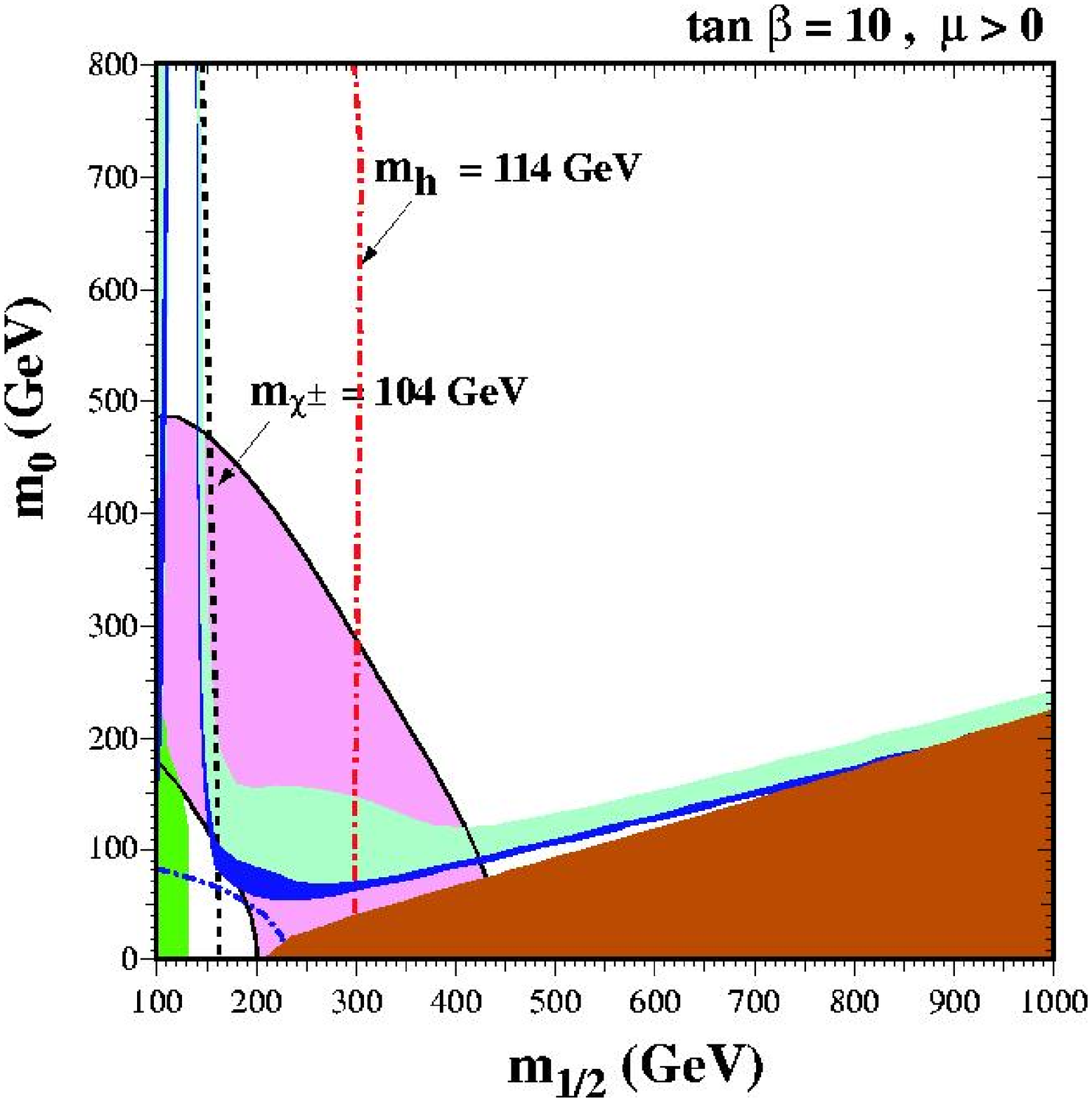,angle=0,width=0.35\linewidth}
\hfill \epsfig{file= 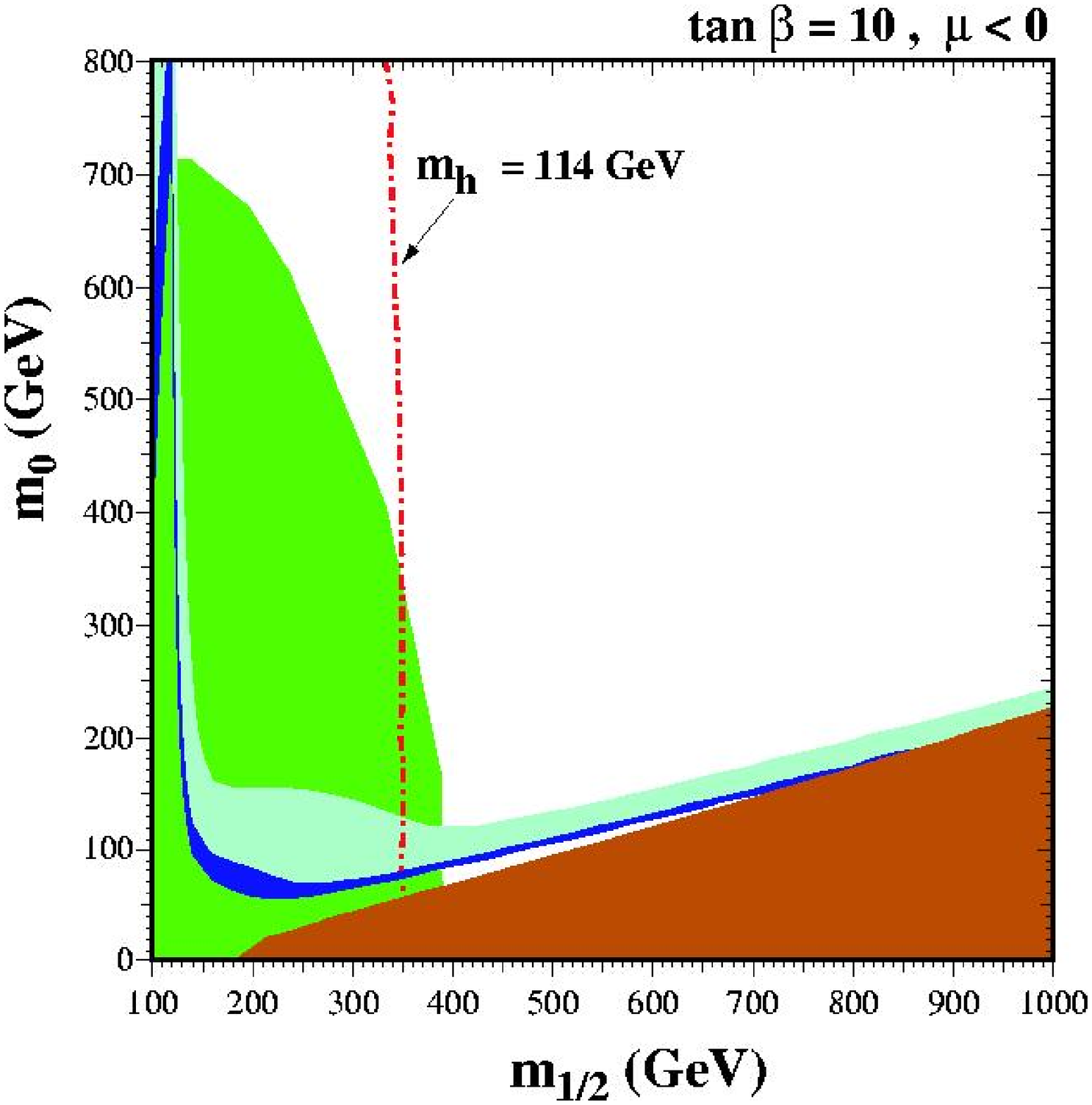,angle=0,width=0.35\linewidth}
\epsfig{file= 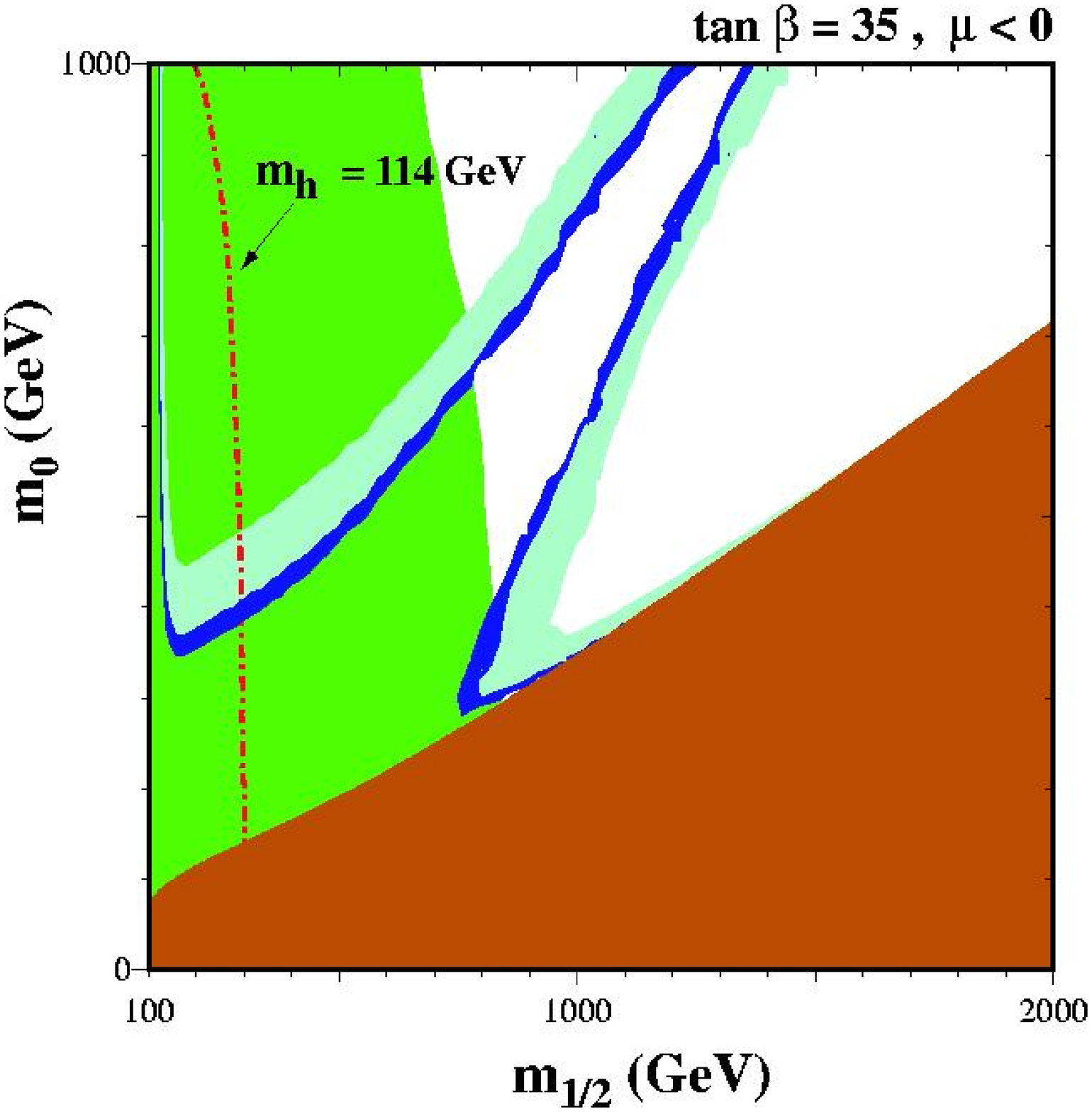,angle=0,width=0.35\linewidth}
\hfill \epsfig{file=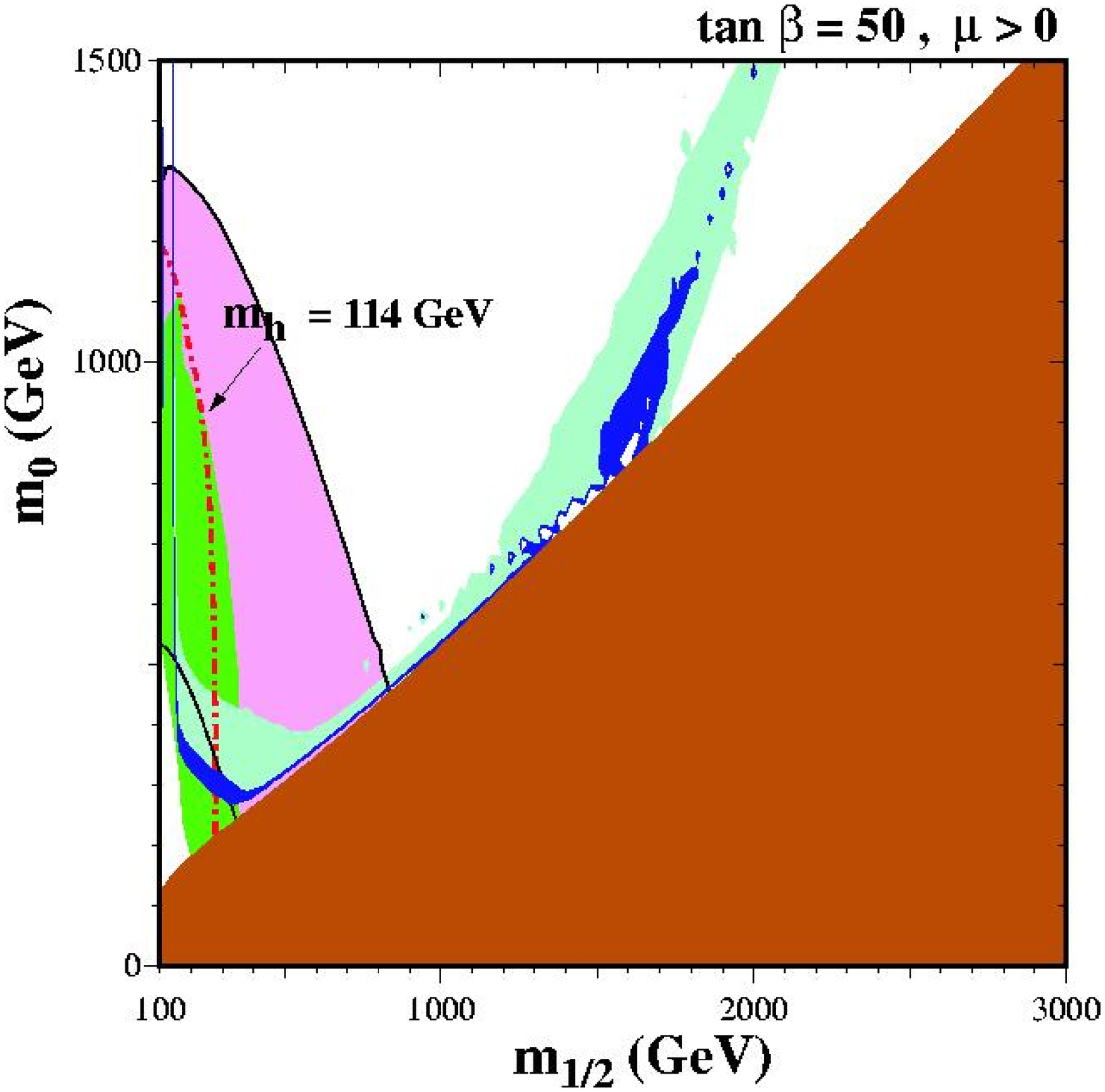,angle=0,width=0.35\linewidth}
\caption
{{\it mSUGRA/CMSSM constraints after WMAP~\cite{olive}.
Dark Blue shaded region favoured by WMAP
( $0.094 \le \Omega_\chi h^2 \le 0.129$ ). 
Turquoise shaded regions have $0.1 \le \Omega_\chi h^2 \le 0.3$.
Brick red shaded regions are excluded because LSP is charged.
Dark green regions are excluded by $b \to s\gamma $.
The Pink shaded region includes $2-\sigma$ effects of $g_\mu - 2$. 
Finally, the dash-dotted line represents the 
LEP constraint on ${\tilde e}$ mass.}}
\label{olivefig1}
\end{figure}

The analysis of \cite{ln} is concentrated on large values 
of tan$\beta$ for reasons mentioned above. 
It is found (see the right panel of the lower part of figure~\ref{lnfig1})
that for such values,
a region opens up within which the relic density
is cosmologically allowed. 
This is due to the pair annihilation of the
neutralinos through the pseudo-scalar Higgs exchange in the $s$-channel.
In this region 
the ratio $m_A / 2 m_{\lsp}$
approaches unity and the pseudo-scalar exchange dominates as explained above.
It is for this 
reason that we give special emphasis to this particular mechanism which opens 
up for large $\; \tan \beta \;$ and delineates cosmologically allowed 
domains of small relic densities and large elastic neutralino - nucleon 
cross sections. 
In this case the lower bound put by the $( g_\mu -2 )$ data 
cuts the cosmologically allowed
region which would otherwise allow for very large values of $m_0, m_{1/2}$.

\begin{figure}[htb] \centering 
\epsfig{file= 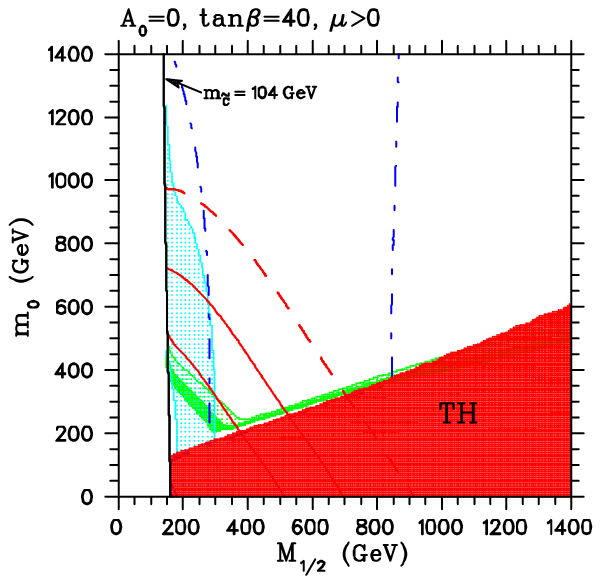,angle=0,width=0.35\linewidth}
\hfill \epsfig{file=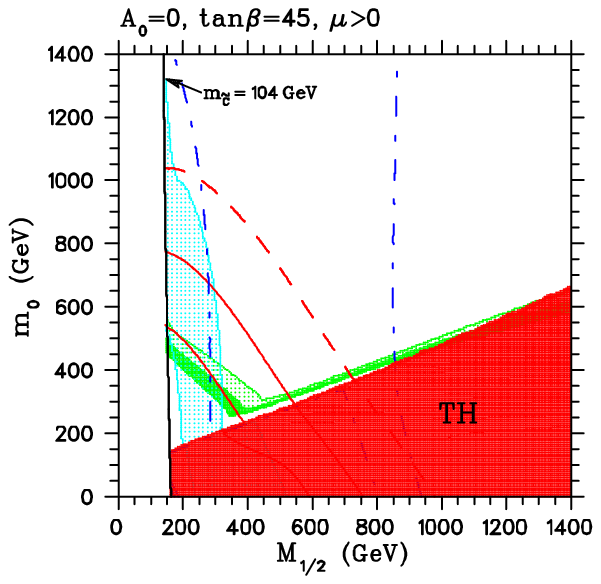,angle=0,width=0.35\linewidth}
\epsfig{file= 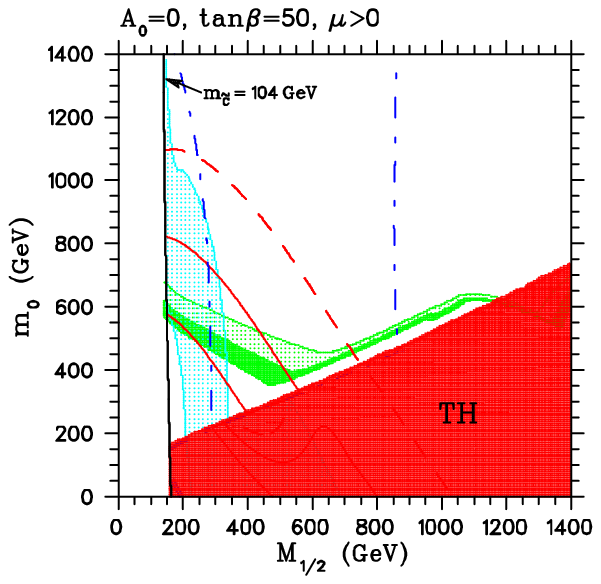,angle=0,width=0.35\linewidth}
\hfill \epsfig{file=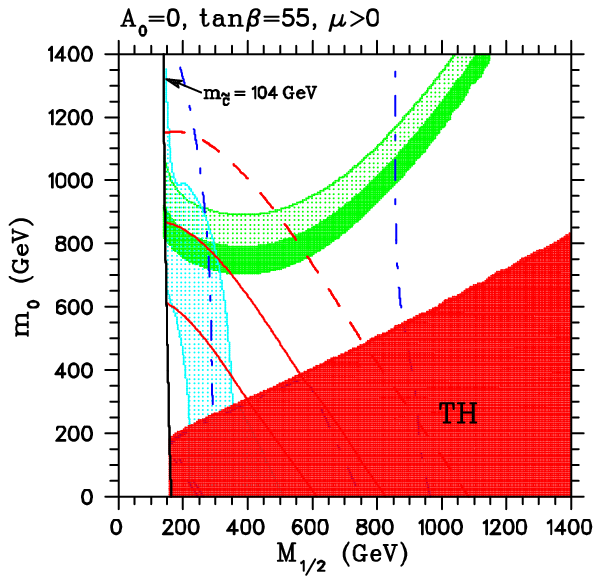,angle=0,width=0.35\linewidth}
\caption{{\it {\small Cosmologically allowed regions of the relic density for 
in the $(m_{1/2},m_0)$ plane for large tan$\beta$ values~\cite{ln}.  
The mass of the top is taken $175\GeV$. In the dark
green shaded area $0.094<\relic<0.129$. In the light green shaded
area $0.129<\relic<0.180\;$. The solid red lines mark the region within which
the supersymmetric contribution to the anomalous magnetic moment of the
muon is $\alpha^{SUSY}_{\mu} = (361 \pm 106) \times 10^{-11}$.
The dashed red line
is the boundary of the region for which the lower bound is moved to
its $2 \sigma$ limit. 
The dashed-dotted blue lines are
the boundaries of the region $113.5 \GeV \leq m_{Higgs} \leq 117.0 \GeV$.
The cyan shaded region is excluded due to
$\bsga$ constraint. 
}}}
\label{lnfig1}  
\end{figure}
\clearpage

In figure~\ref{lnfig1} we see that for the $\tan \beta = 55$ case, 
close to the highest possible value, and considering the 
$2 \, \sigma$ lower bound on the muon's anomalous magnetic moment
$\alpha_{\mu}^{SUSY} \geq 149 \times 10^{-11}$ and values of
$\relic$ in the range $\;0.1126^{+ 0.0161}_{-0.0181}  \;$, 
allowed points are within a narrow stripe.  
The point with the highest value for $m_0$ is (in GeV) 
at $(m_0, m_{1/2}) \;=\; ( 850 , 550 ) \;$ and that with the highest 
$\; M_{1/2}\;$ at $(m_0, m_{1/2}) \;=\; ( 750 , 600 \;) \;$. The latter 
marks the lower end of the line segment of the boundary 
$\; 149 < 10^{-11}\;\alpha_{\mu}^{SUSY}\;$ which amputates the 
cosmologically allowed stripe.

It should be noted that within $1 \sigma$ 
of the E821 data only a few points survive which lie in a small region 
centered at $(m_0, m_{1/2}) \;=\; ( 725 , 300 ) \;$.
The bounds on $m_0, m_{1/2}$ displayed in (the lower part of) 
figure~\ref{lnfig1} refer to the 
$\; A_0=0 \;$ case.
Allowing for $A_0 \neq 0$ values, the upper bounds put on $m_0, M_{1/2}$
increase a little and so do the corresponding bounds on sparticle
masses.
For the LSP, the lightest of the charginos, stops, staus and Higgses the upper 
bounds on their masses are displayed in Table \ref{table1} for various values 
of the parameter $\; \tan \beta \;$, if the new WMAP determination~\cite{wmap,spergel} of the
Cold Dark Matter (\ref{relicdensity}) and the $2 \sigma$ bound
$\; 149 < 10^{-11}\;\alpha_{\mu}^{SUSY}<573\;$ of E821 is respected.  
We have also taken into account the limits arising from Higgs boson searches 
as well as from $\bsga$ experimental constraints. 
In extracting these values we used~\cite{ln} a random sample  
of $40,000$ points in the region $|A_0| < 1 \; \mathrm{TeV}$, $\tan \beta < 55 $, 
$m_0,M_{1/2} < 1.5 \; \mathrm{TeV}$ and $\; \mu > 0 \;$. 
The lightest of the charginos has a mass 
whose upper bound is $\; \approx 550 \; \mathrm{GeV} \;$, 
and this is smaller than the upper bounds put on the masses of the 
lightest of the other charged sparticles, namely the stau and stop, as is 
evident from Table \ref{table1}. 
Hence the prospects of discovering CMSSM 
at a $e^{+} e^{-}$ collider with center of mass energy $\sqrt s = 800 \GeV$,
are  {\em not} guaranteed. Thus a center of mass energy of at least 
$\sqrt s \approx 1.1 \; \mathrm{TeV}\;$
is required to discover SUSY through chargino pair production.  
Note that in the allowed regions 
the next to the lightest neutralino, ${\tilde{\chi}^{\prime}}$, has a mass
very close to the lightest of the charginos and hence the process
$e^{+} e^{-} \goes {\tilde{\chi}} {\tilde{\chi}^{\prime}}$, with 
${\tilde{\chi}^{\prime}}$ subsequently decaying to
$ {\tilde{\chi}} + {l^{+}} {l^{-}}$ or 
$ {\tilde{\chi}}+\mathrm{2\,jets}$,  
is kinematically allowed for such large $\tan \beta$, provided
the energy is increased to at least $\sqrt{s} = 860 \GeV$. It should
be noted however that this channel proceeds via the $t$-channel exchange
of a selectron and it is suppressed due to the heaviness of the exchanged
sfermion.
Therefore only if the center of mass energy is increased to  
$\; \sqrt{s} = 1.1 \; \mathrm{TeV} \;$ supersymmetry can be 
discovered in a $e^{+} e^{-}$ 
collider provided it is based on the Constrained scenario~\cite{ln}.

\begin{figure}
\centering
\epsfig{file=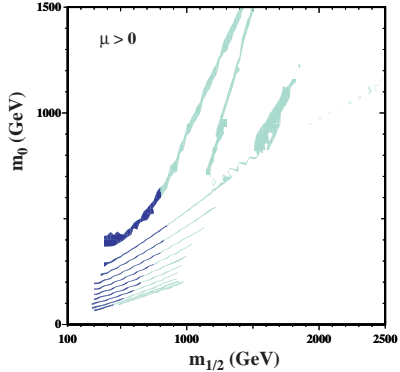,angle=0, width=0.35\linewidth}
\caption{{\it Strips correspond to tan$\beta = 5, 10, 15, 20, 25, 30, 35, 40, 45, 50, 55$. Light Green Regions of the graph 
$(m_0,m_{1/2})$ are those compatible with the WMAP
result on the neutralino relic density $0.094 \le \Omega_\chi h^2 \le 0.129$.
Darker Blue shaded parts of strips 
are compatible with $g_\mu - 2$ results (at a $2-\sigma$ level)~\cite{olive}.}}
\label{severe}
\end{figure}

An important conclusion, therefore, which can be inferred by inspecting  
the figures \ref{olivefig1}, \ref{lnfig1}, is 
that the constraints implied by a possible discrepancy of $g_\mu - 2$ 
from the SM value 
( $\alpha_{\mu}^{SUSY} \gsim 15.0 \times 10^{-10}$ ), 
when combined with the 
WMAP restrictions on CDM (neutralino) relic densities (\ref{relicdensity}),  
imply severe restrictions on the available parameter space of the EB
and lower significantly the upper bounds on the allowed neutralino masses
$\mlsp$. This is clearly seen in  figure \ref{severe}. 
In this the width of the strips is
much smaller than the spacing between them.
For tan$\beta < 40$  one obtains~\cite{olive} 
$\mlsp \lsim 500$ GeV (see figure \ref{neutrmass}), 
which is significantly lower than the 
bounds obtained in pre-WMAP 
cosmological models, $\mlsp \lsim 650$ GeV.

\begin{figure}[htb]
\centering  
 \epsfig{file= 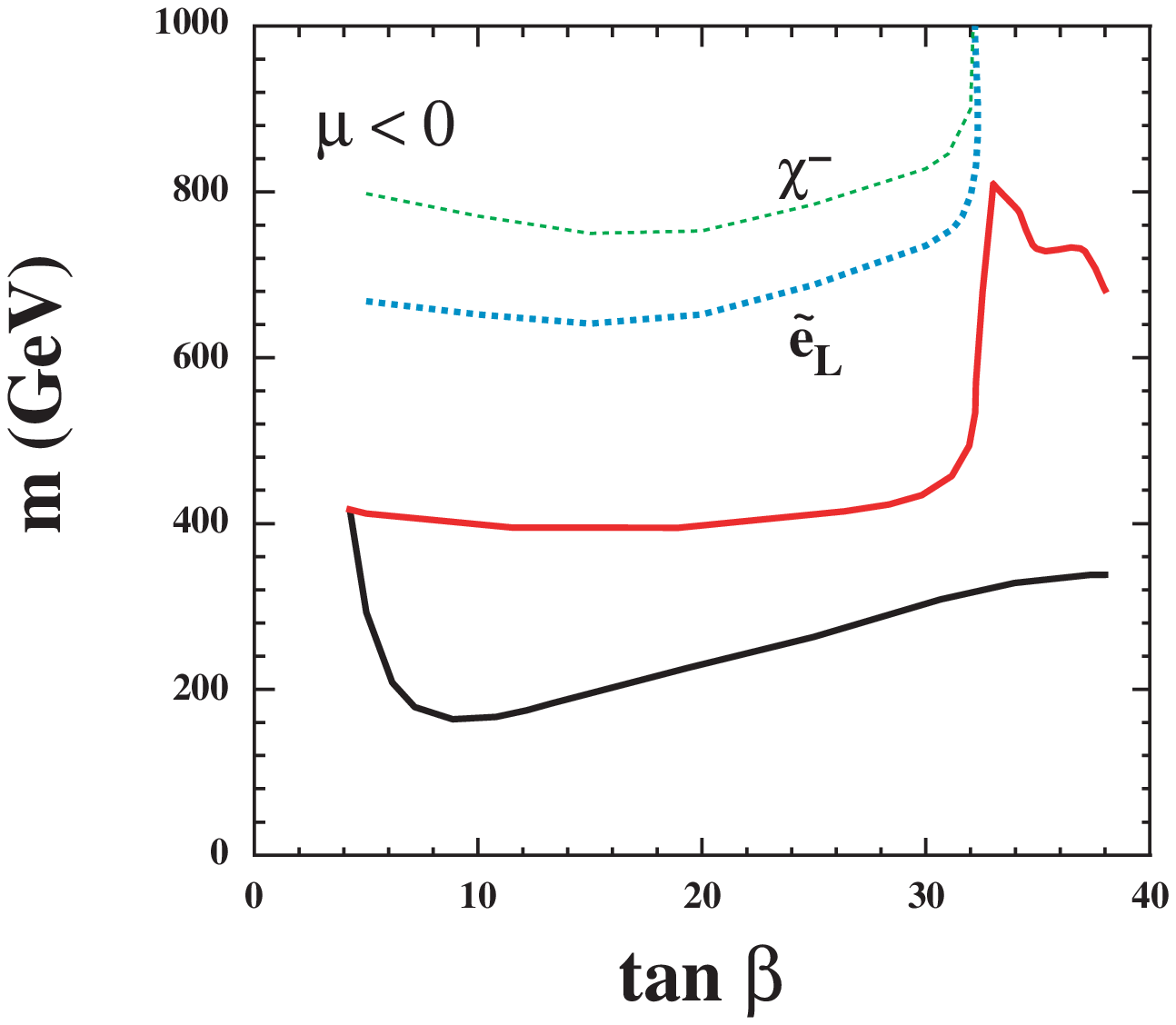,angle=0,width=0.4\linewidth}
\hfill \epsfig{file=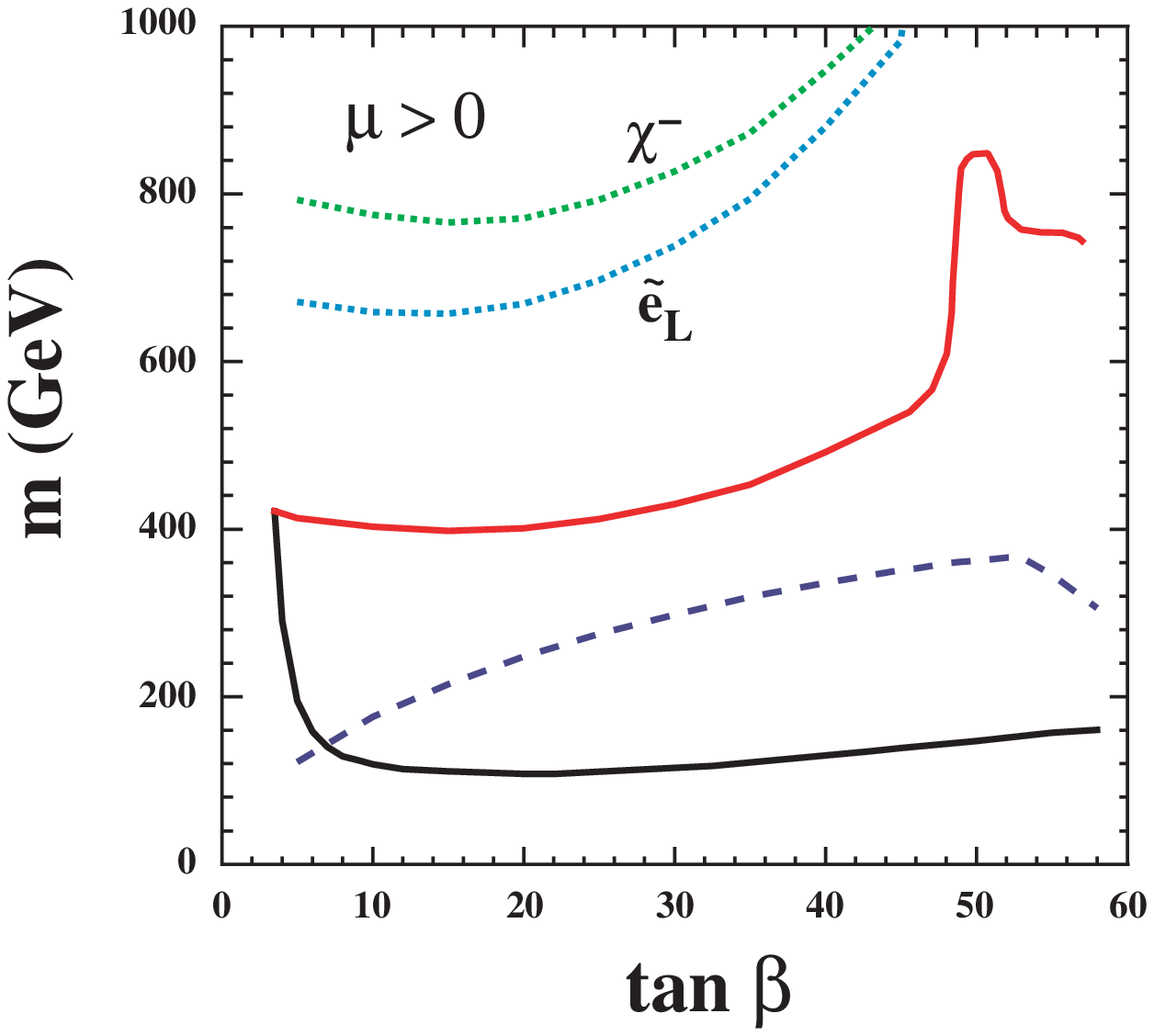,angle=0,width=0.4\linewidth}
\caption{{\it {\small The Neutralino Mass $m_\chi$ Range in the EB of
radiative symmetry breaking parameter space, 
implied by the WMAP data and other constraints
($b \to s\gamma$, $g_\mu - 2$ {\it etc}),
in the context of the mSUGRA
model~\cite{olive}.
\underline{Left:} $\mu < 0$, {\underline Right:}  
$\mu > 0$. Red solid lines: 
Upper limits without $g_\mu - 2$.
Dashed Blue lines: Upper limits with $g_\mu - 2$. Dotted lines: 
Green: Masses of $\chi^{\pm}$, Blue masses of ${\tilde e}_L$,  
at the tip of co-annihilation tails.}}}
\label{neutrmass}
\end{figure} 

\begin{table}[htb]
\begin{center}
\begin{tabular}{|c|c|c|c|c|c|} \hline \hline
 $\tan\beta$ & ${\lsp}$ & $\chi_1^+$ & $\tilde{\tau}$ & $\tilde{t}$ & $h$ 
                                                                    \\ \hline
  10  &  155  & 280  & 170 & 580  &  116 \\
  15  &  168  & 300  & 185 & 640  &  116 \\
  20  &  220  & 400  & 236 & 812  &  118 \\
  30  &  260  & 470  & 280 & 990  &  118 \\
  40  &  290  & 520  & 310 & 1080 &  119 \\
  50  &  305  & 553  & 355 & 1120 &  119 \\
  55  &  250  & 450  & 585 & 970  &  117 \\ 
 \hline \hline
\end{tabular}
\end{center}
\caption{{\it {\small Upper bounds, in GeV, 
on the masses of the lightest of the neutralinos,
charginos, staus, stops and  Higgs bosons for various values of
$\tan\beta$ if the new WMAP value \cite{wmap} for $\Omega_{CDM} h^2$ 
and the $2 \sigma$  E821 bound, 
$149 \times 10^{-11}<\alpha_{\mu}^{SUSY}<573 \times 10^{-11}$,  
is imposed~\cite{ln}.
}}}
\label{table1}
\end{table}

To summarize, therefore, 
in the EB 
the recent WMAP data~\cite{wmap} can easily be adapted through by 
most of the benchmark points 
(in bulk and co-annihilation regions) of figure \ref{bench}
by reducing $m_0$. 
In the post-WMAP benchmark scenaria~\cite{benchmarksnew} 
the only exception is the point H (tip 
of co-annihilation tail): WMAP data will lower $m_{1/2}$ 
thereby facilitating detection prospects  
of SUSY at LHC or future linear $e^+e^-$ colliders
a center of mass energy $\sqrt{s}=1.1. $ TeV~\cite{ln,olive},
provided the HB region, and in particular its high zone (inversion), 
is not realized in nature. 

\subsubsection{Constraints in the HB}

Despite the above-mentioned good prospects of 
discovering minimal SUSY models at future colliders, if the EB is realized, 
however, things may not be that simple in Nature. 
A recent $\chi^2$ study~\cite{baer} of mSUGRA in light of the recent 
WMAP data has indicated that the HB/focus point region
of the model's
parameter space seems to be favoured along with the neutralino resonance 
annihilation region for $\mu>0$ and large tan$\beta$ values. 
The favoured focus point region corresponds to 
moderate to large values of the Higgs parameter $\mu^2$, and large 
scalar masses $m_0$ in the several TeV range. The situation is summarized briefly 
in figure \ref{focus}.

\noindent 
\begin{figure}[htb]
\centering 
\epsfig{file=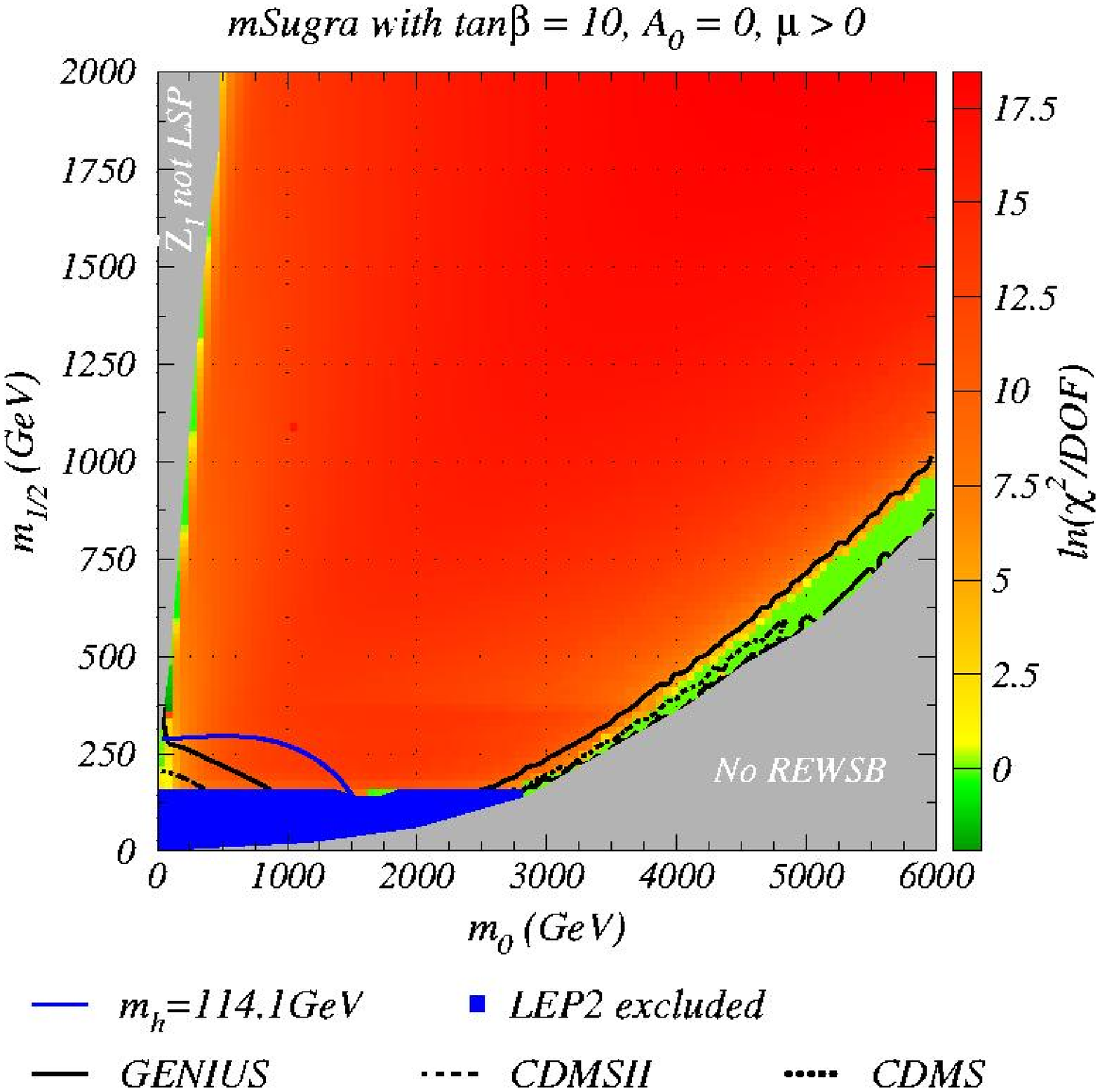,angle=0,width=0.4\linewidth}
\hfill \epsfig{file=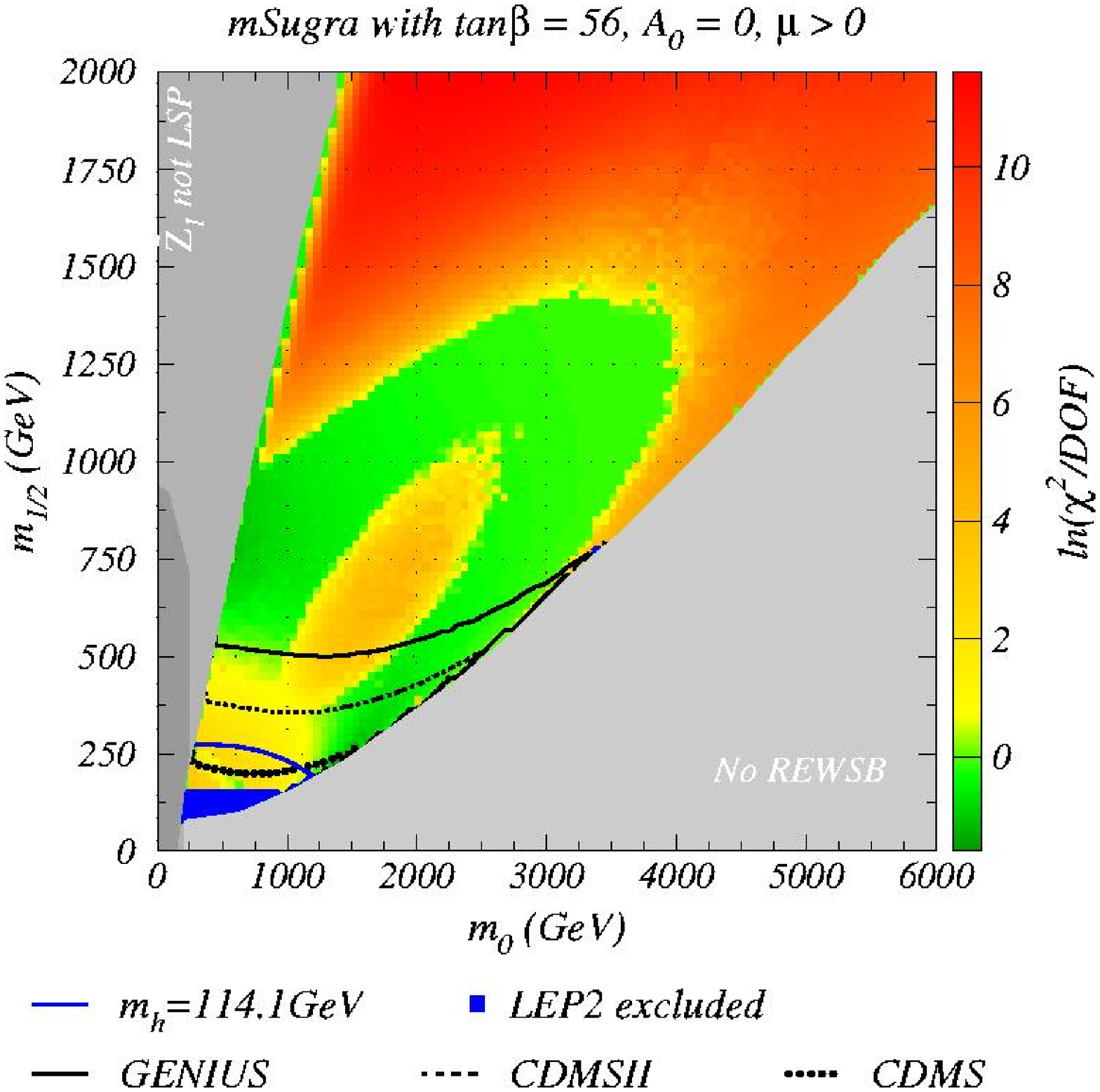,angle=0,width=0.4\linewidth}
\caption{{
\it WMAP data seem to favour ($\frac{\chi^2}{dof}<4/3$)
(green) the HB/focus point
region (moderate to large values of $\mu$, large $m_0$ scalar masses)
for almost all tan$\beta$  (Left), as well as s - channel 
Higgs resonance annihilation (Right) for $\mu>0$ and large 
tan$\beta$~\cite{baer}. 
}}
\label{focus}
\end{figure}

The situation in case the HB is included in the analysis 
is depicted in figure \ref{chattofigure}~\cite{chatto},
where we plot the $m_0-m_{1/2}$ graphs,
as well as graphs of $m_0$, $m_{1/2}$ vs the neutralino LSP mass.
The neutralino density is that of the WMAP data.
We stress again that, in case the high zone (inversion) region 
of the HB is realized,
then the detection prospects of SUSY at LHC are diminished significantly,
in view of the fact that in such regions slepton masses may lie
in the several TeV range (see figure \ref{chattofigure}). 

\begin{figure}[htb]
\centering 
\epsfig{file=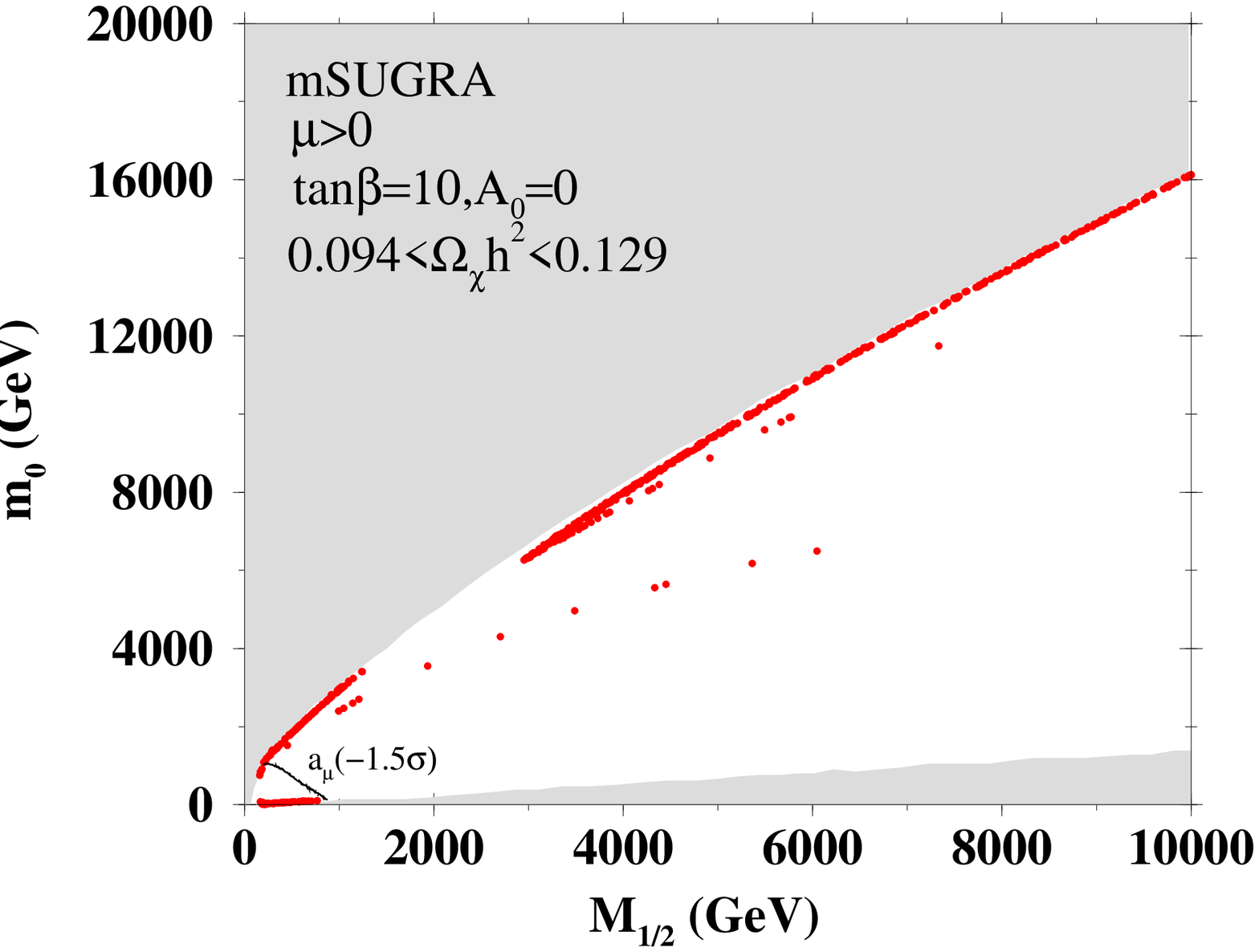,angle=0,width=0.41\linewidth,clip=}

\vspace*{-0.5cm}

\centering 
\epsfig{file=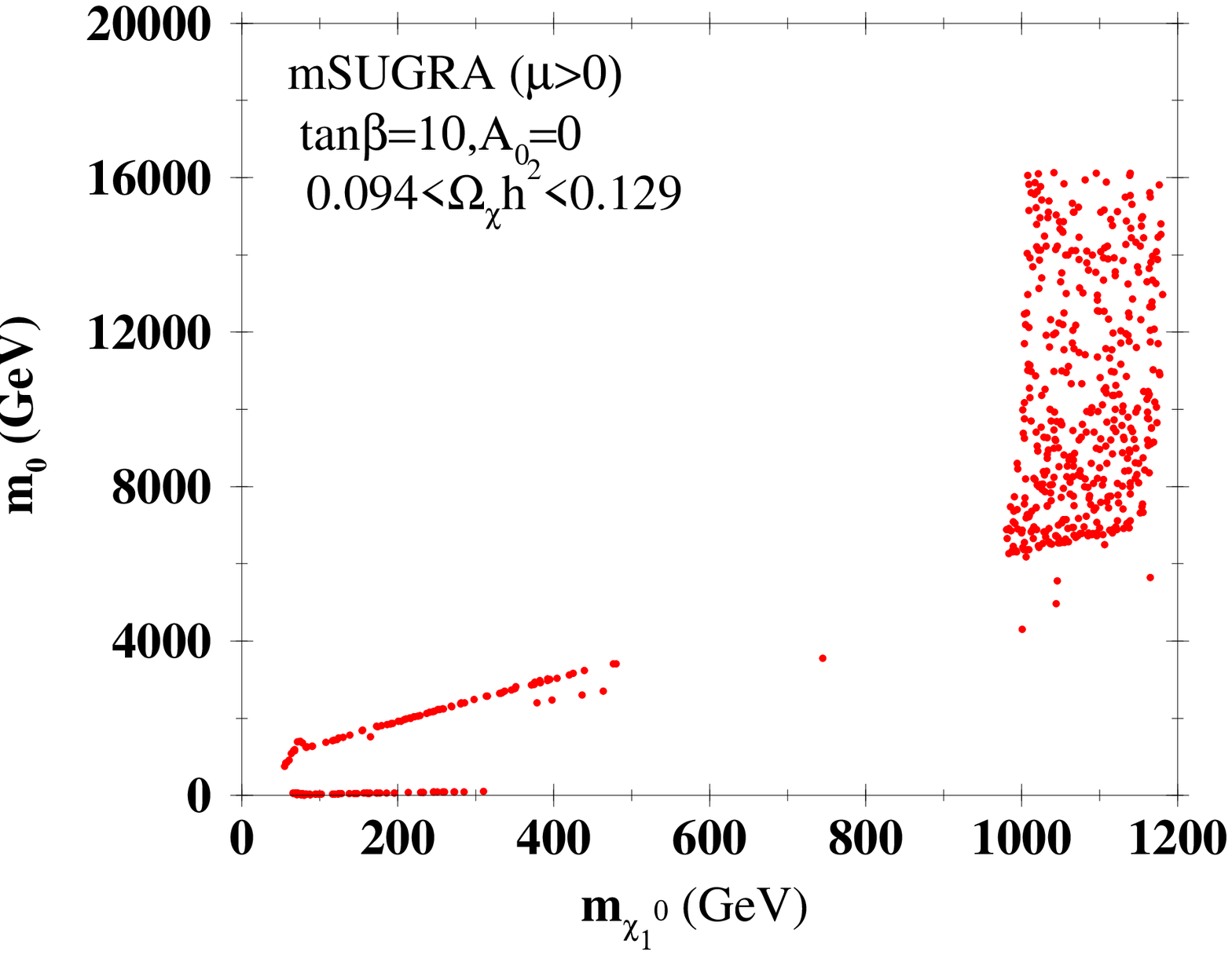,angle=0,width=0.41\linewidth,clip=}
\hfill\epsfig{file=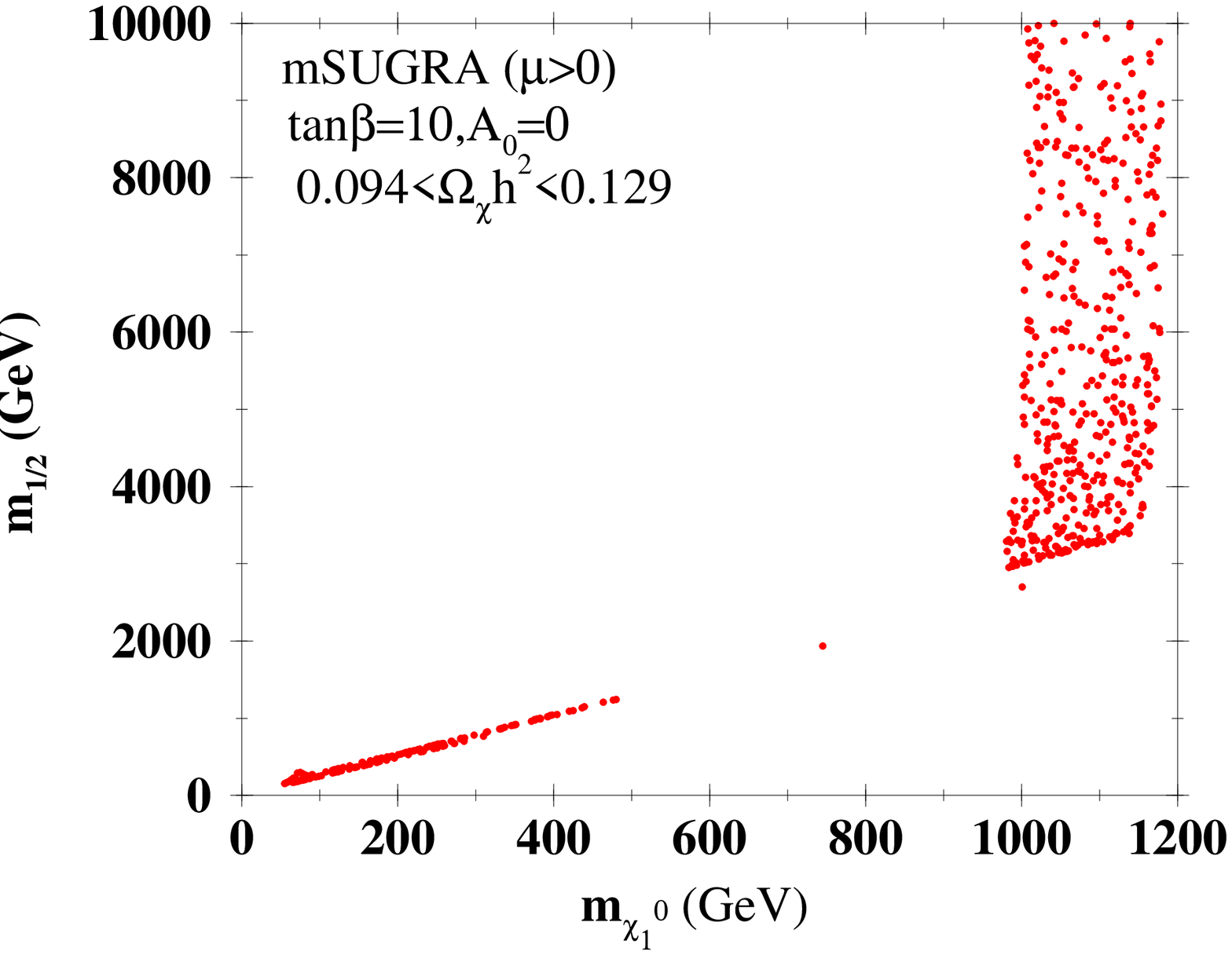,angle=0,width=0.41\linewidth,clip=}
\caption{{\it $m_0-m_{1/2}$ graph, and $m_0$ and $m_{1/2}$ vs. 
$m_\chi$ graphs, including the HB of mSUGRA~\cite{chatto}. 
Such regions are favoured by the WMAP data~\cite{baer}.}}
\label{chattofigure}
\end{figure}

In view of the above results, 
an updated reach of LHC in view of the recent WMAP 
and other constraints discussed above
(see figure \ref{reach})
has been performed in \cite{reachlhc}, showing that  
a major part of the HB, but certainly not its high zone, 
can be accessible at LHC. 
\noindent 
\begin{figure}[htb]
\centering 
\epsfig{file=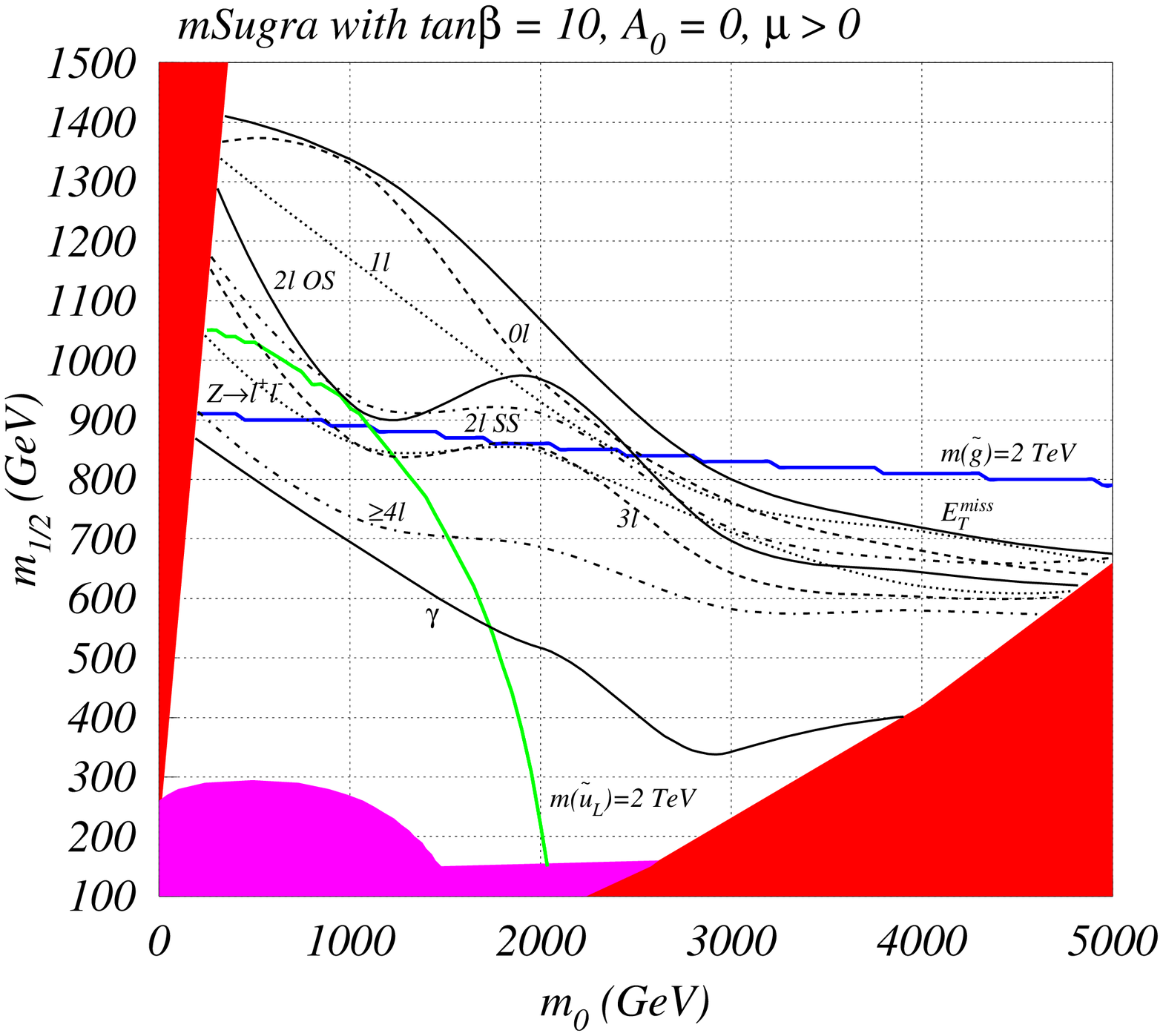,angle=0,width=0.35\linewidth}
\hfill \epsfig{file=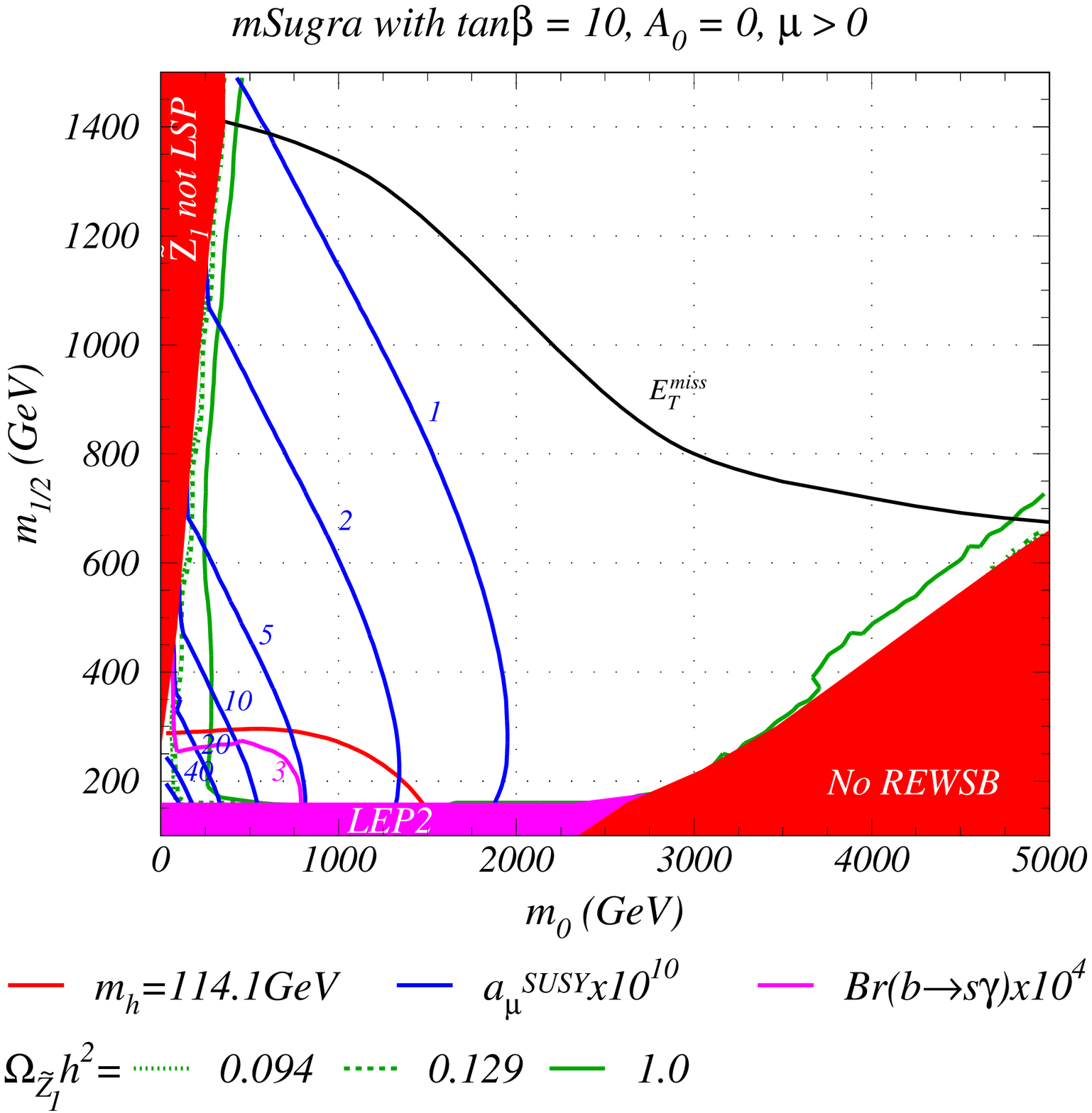,angle=0,width=0.35\linewidth}
\caption{{\it \underline{Left}: 
The updated Reach in $(m_0, m_{1/2})$ parameter plane of mSUGRA
assuming 100 $fb^{-1}$ integrated luminosity. Red (magenta) regions
are excluded by theoretical (experimental) constraints~\cite{reachlhc}.
\underline{Right}: Contours 
(in view of the uncertainties) of several
low energy observables : CDM relic density (green), 
contour of 
$m_h=114.1$ GeV (red), contours of $a_\mu 10^{10}$ (blue) 
and contours of $b \to s \gamma $ BF ($\times 10^{4}$)(magenta).}}  
\label{reach}
\end{figure}
The conclusion from this study~\cite{reachlhc} is that for an 
integrated luminosity 
of $100~fb^{-1}$ values of $m_{1/2} \sim 1400$ GeV can be probed 
for small scalar masses $m_0$, corresponding to gluino masses 
$m_{\tilde g} \sim 3$ TeV. For large $m_0$, in the hyperbolic branch/focus point region, $m_{1/2} \sim 700$ GeV can be probed, corresponding to 
$m_{\tilde g} \sim 1800$ GeV. It is also concluded that 
the LHC (CERN) can probe the entire stau co-annihilation
region and most of the heavy Higgs annihilation funnel
allowed by WMAP data, except for some range of $m_0,m_{1/2}$ in the case
tan$\beta \gsim 50$. 
 
A similar updated reach study in light of the new WMAP data 
has also been done for the 
Tevatron~\cite{reachtev}, extending previous analyses to large $m_0$ masses 
up to 3.5 TeV, in order to probe the HB/focus region favoured by the 
WMAP data~\cite{baer}. Such a study 
indicated that for a $5\sigma$ (3$\sigma$) signal 
with 10 (25) $fb^{-1}$ of integrated luminosity, the Tevatron reach in the 
trilepton channel extends up to $m_{1/2} \sim 190$ $(270)$ GeV independent of 
tan$\beta$, which corresponds to a reach in terms of gluino 
mass of $m_{\rm g} \sim 575 (750)$ GeV.

\subsection{WMAP and Direct Dark Matter Searches}

We turn now to study the impact of the new WMAP data on Dark Matter, 
the constraints from $\; ( g_{\mu}-2 )\;$ and $\bsga$, 
and the Higgs mass bounds on direct 
Dark Matter searches. 
A quantity of great interest is the spin-independent ( scalar ) neutralino-proton
cross section on which experimental limits exist 
from the current dark matter experiments 
$\sigma_{\chi^0p}(SI) < 10^{-6} $ picobarns.

\begin{figure}[t]
\centering 
\epsfig{file=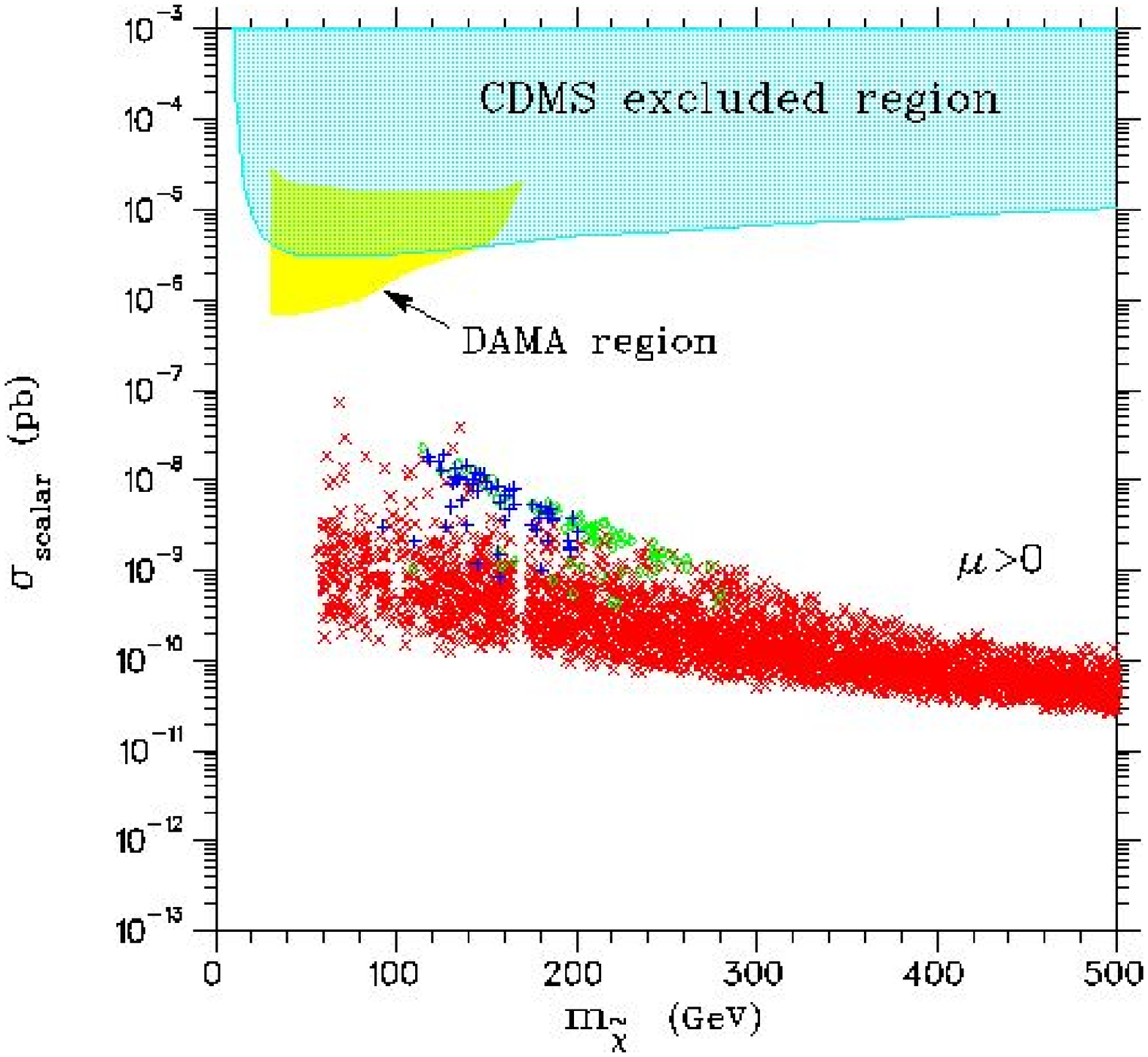,angle=0,width=0.4\linewidth}
\caption{\it Neutralino-nucleon scalar cross section vs. 
$m_\chi$ in the EB~\cite{ln}.} 
\label{fig3}
\end{figure}

In figure~\ref{fig3} we plot the scalar $\lsp$-nucleon   
cross section as function of the LSP mass, $\mlsp$~\cite{ln}.
On the top of it the shaded region (in cyan colour) is
excluded by the CDMS experiment \cite{cdms}.
The DAMA sensitivity region is
plotted in yellow \cite{dama}. 
Pluses ($+$) (in blue colour) represent points 
which are both compatible with the E821 data 
$\almuon = (361 \pm 106)\times 10^{-11}$
and the WMAP cosmological bounds $\Omega_{CDM} h^2 = 0.1126^{+ 0.0161}_{-0.0181}$.  
Diamonds ($\diamond$) (in green colour) represent points 
which are  compatible with the $2 \, \sigma$ E821 data  
$149 \times 10^{-11}  < \almuon  < 573 \times 10^{-11} $
and the cosmological bounds.
The crosses ($\times$) (in red colour) represent the rest of the
points of our random sample which consists of 40,000 points in the range
$|A_0| < 1 \; \mathrm{TeV}, m_0, m_{1/2} < 1.5 \; \mathrm{TeV}$,
tan$\beta < 55$ and $\mu > 0$ belonging to the EB region.   
Here the Higgs boson
mass, $m_h > 113.5 \GeV$ and the
$\bsga$ bounds have been properly taken into account~\cite{ln}.
From this figure it is seen that the     
points which are compatible both with the ($g_{\mu}-2 )$ E821
and the cosmological data can yield cross sections
slightly above $10^{-8}$ pb when $\mlsp$
is about $120 \GeV$. The maximum value of $\mlsp$
is around $200 \GeV$ but in this case the scalar cross section drops 
by almost an order of magnitude $10^{-9}$ pb.   
Accepting the $2\, \sigma$ $( g_{\mu}-2 )$ bound the maximum value of the 
scalar cross section is again $\approx 10^{-8}$ pb,  
for $\mlsp \approx 120 \;GeV$, but the 
$\mlsp$ bound is increased to about $280\;GeV$ at the expense of having 
cross sections slightly smaller than $10^{-9}$ pb. 
Considering the $\mu>0$ case,  
it is very important that using all available data~\cite{ln}, 
one 
can put a lower bound $\approx 10^{-9}$ pbarns on the scalar cross section
which is very encouraging for 
future DM direct detection experiments \cite{Klapdor}. Such 
a lower bound cannot 
be imposed when $\mu<0$, since the scalar cross section can become very small due 
to accidental cancellations between the sfermion and 
Higgs exchange processes. 
However, this case is not favoured by $( g_{\mu}-2 )$ and $b \to s\gamma$ data.  

\begin{figure}[t]
\centering 
\epsfig{file=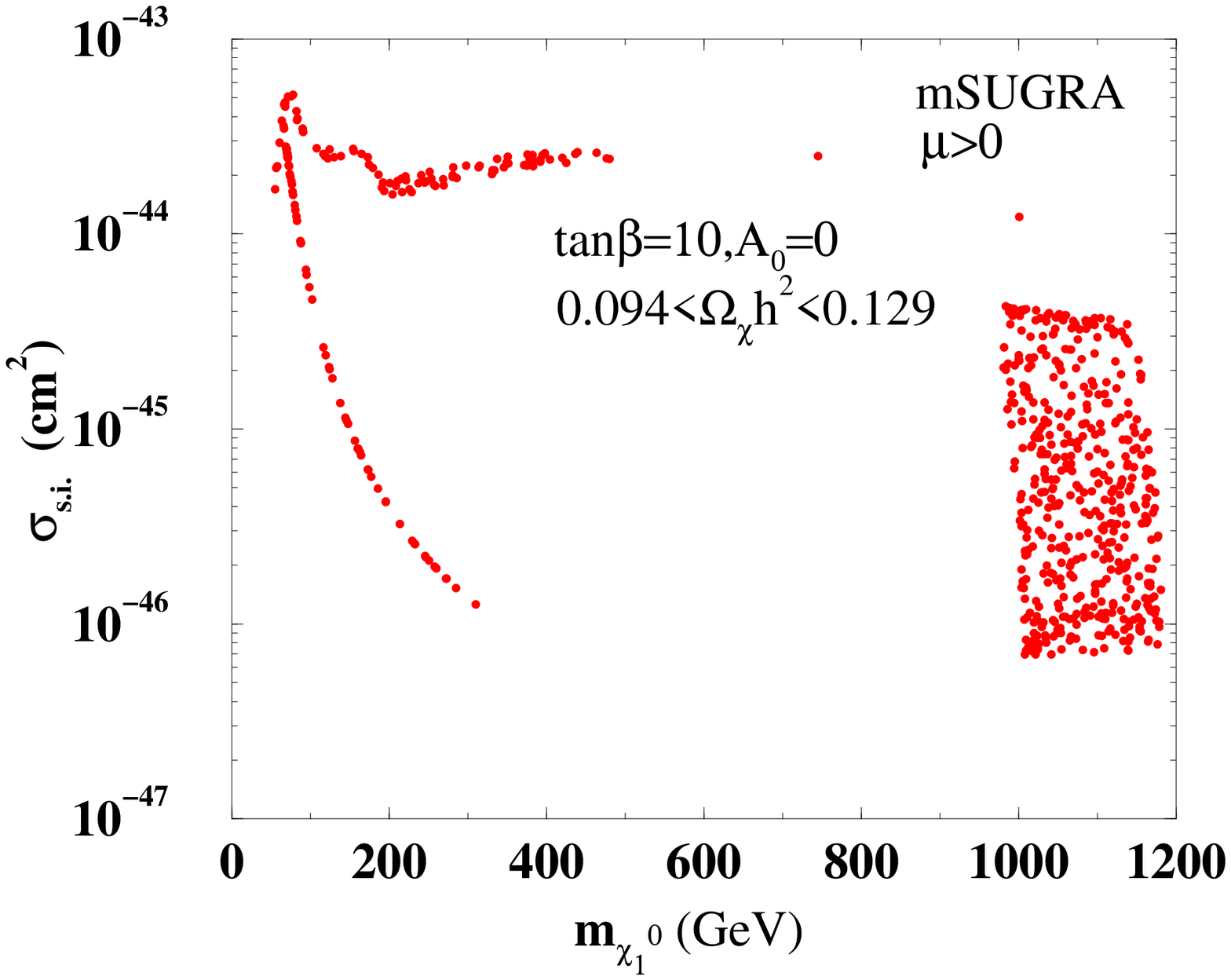,angle=0,width=0.4\linewidth}
\caption{\it Spin independent part of the 
neutralino-proton cross section vs. $m_{\chi^0}$ including the 
HB~\cite{chatto}.} 
\label{scatthb}
\end{figure}

Portions of the high zone (inversion) region of the HB may be detected by 
such direct Dark Matter Search Experiments, which are thus complementary
to collider physics. As in the EB case 
such scattered plots can be produced also for the HB~\cite{chatto}, 
which seems to be 
favoured by the WMAP data~\cite{baer}, and which has regions that lie beyond 
the reach of LHC. 
As stressed in \cite{chatto} it is important to distinguish the spin-dependent
from the spin-independent (scalar) cross sections in this branch.

\begin{figure}[ht]
\centering   
\epsfig{file=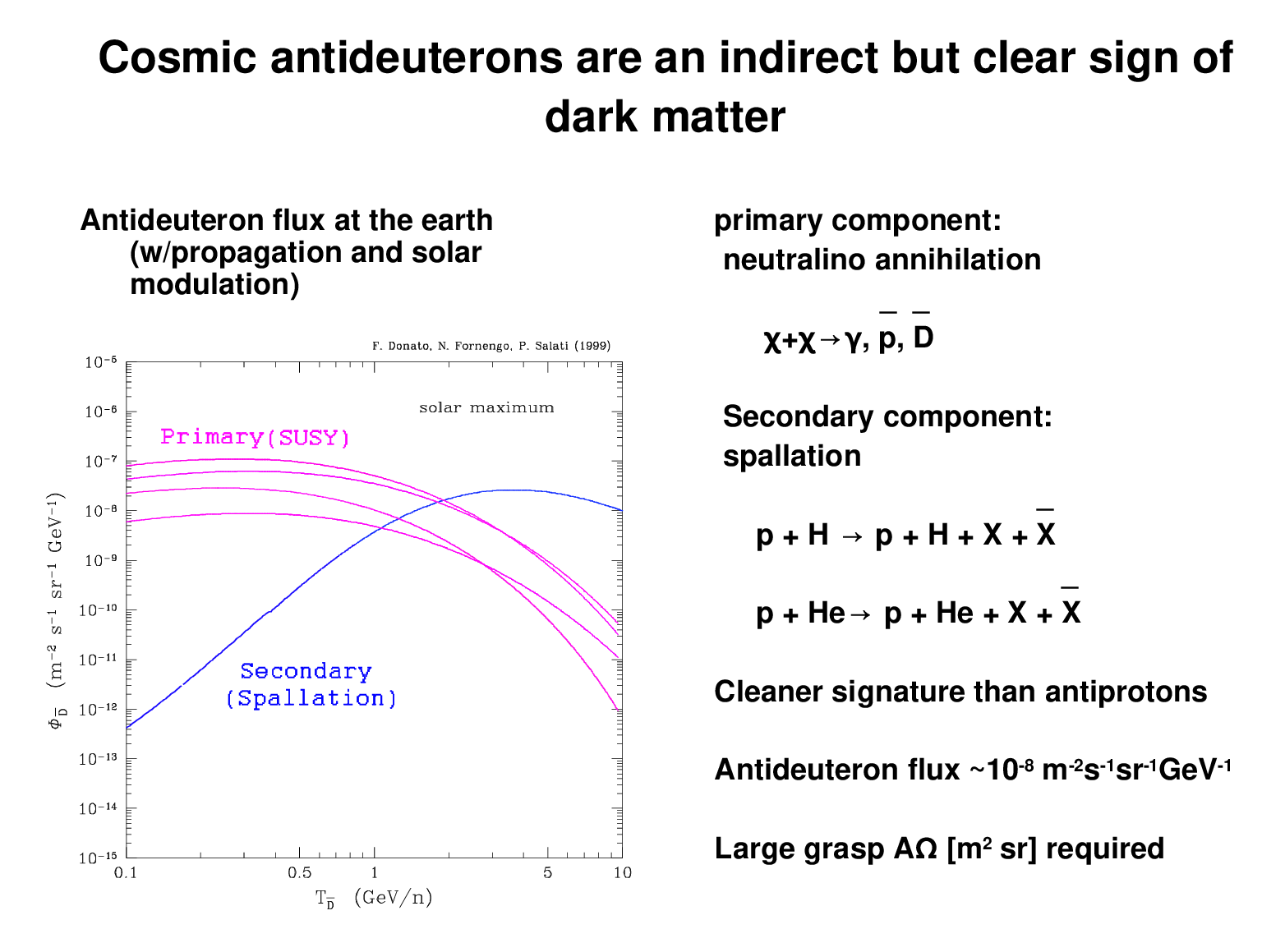,angle=0,width=0.49\linewidth,clip=}
\hfill \epsfig{file=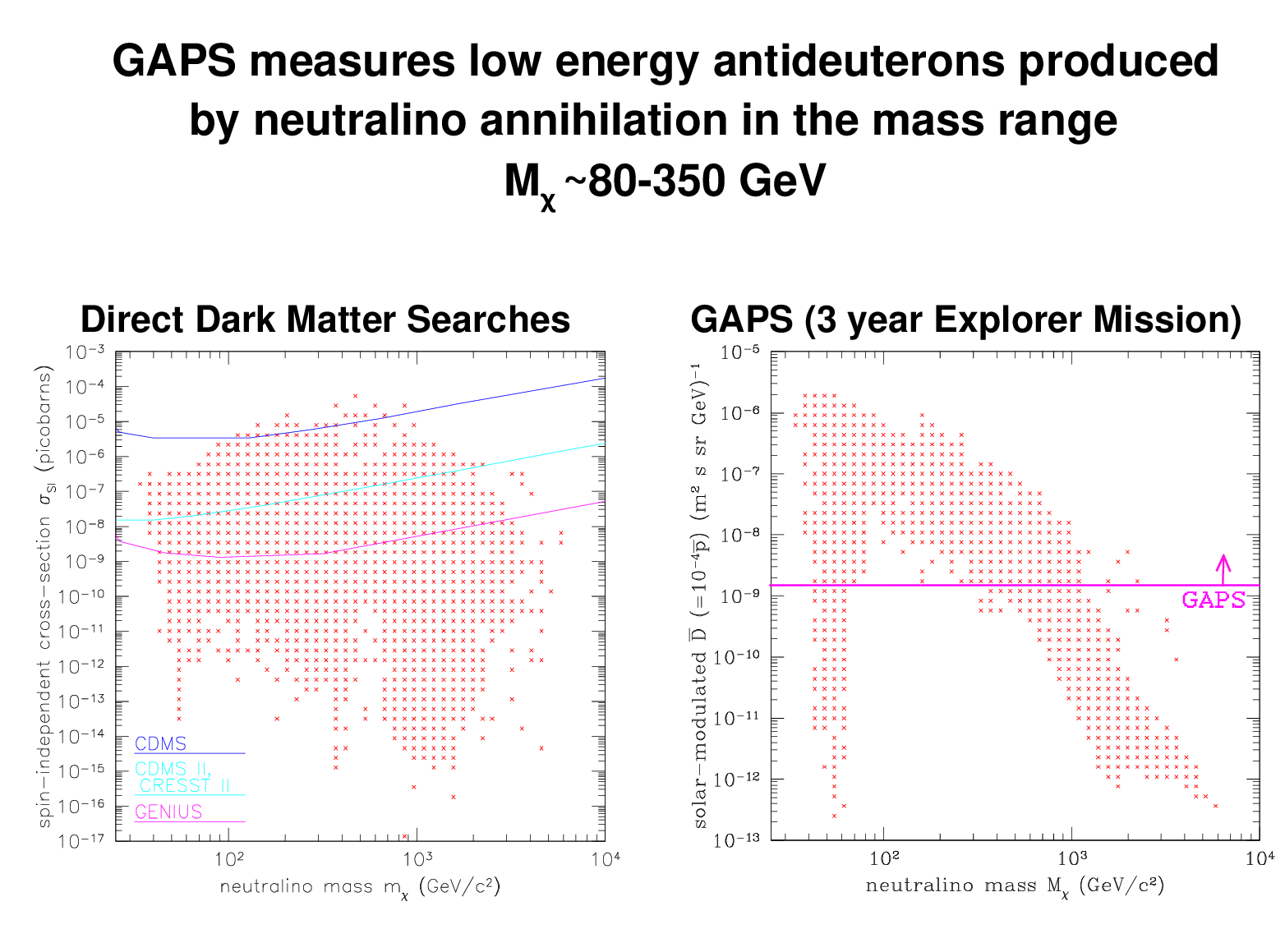,angle=0,width=0.49\linewidth,clip=}
\caption{{\it GAPS Detector of Indirect SUSY Dark Matter Searches, 
through the detection of cosmic antideuteron, 
may cover a neutralino mass range that may be complementary to 
all other direct dark matter and collider searches~\cite{hailey}.}}
\label{hailey} 
\end{figure}

In figure \ref{scatthb} 
we give a plot of the scalar cross section 
for tan$\beta=10$ and $\mu > 0$. We do observe that the lower rapidly 
falling curve that terminates at $m_\chi^0 = 300$ GeV 
is the branch on which staus co-annihilation occurs. 
The upper curve arises from the low zone of the HB, while the patch 
to the right is the one corresponding to the inversion region of the 
HB~\cite{chatto}. This patch indicates that the scalar cross sections are quite 
significant even though one is in the inversion region 
in the range $10^{-10} - 10^{-7} \;$ pbarns. 
Thus, despite the fact that the direct detection of SUSY in the inversion 
region is difficult, the neutralino-proton scalar cross section are still 
substantial. The sensitivity of the future 
dark matter detectors~\cite{Klapdor} will reach $10^{-9}$ pbarns, 
being capable of probing a significant
portion of the parameter space of fig. \ref{scatthb}.
In this latter respect we also mention novel indirect detectors 
of SUSY dark matter ~\cite{hailey}, 
exploiting possible detector of cosmic antimatter
(antideuteron), as a result of neutralino interactions 
(see figure \ref{hailey}),  
which could detect neutralinos in a mass range from 80-350 GeV, 
much higher than other future 
dark matter experiments~\cite{Klapdor}.

\begin{figure}[htb]
\centering 
\epsfig{file=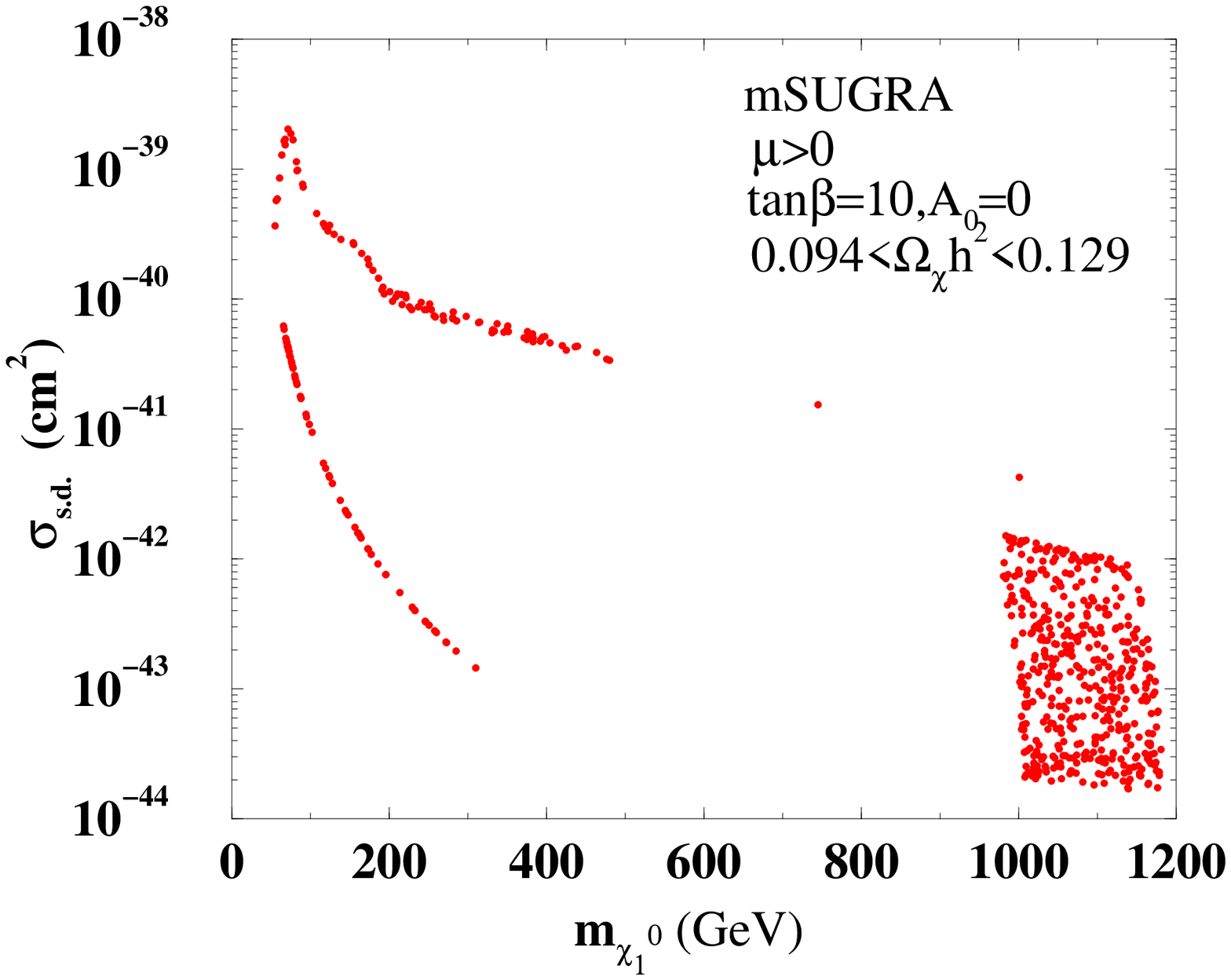,angle=0,width=0.4\linewidth}
\caption{\it Spin dependent part of the 
neutralino-proton cross section vs. $m_{\chi^0}$ including the 
HB~\cite{chatto}.} 
\label{scatthb2}
\end{figure}

Finally, in figure \ref{scatthb2} we plot the spin dependent 
parts of the neutralino-proton cross section, for the same range of 
parameters as in figure \ref{scatthb}. A comparison between these two figures
shows that the spin dependent cross section is much larger, by three or four
orders of magnitude, than the spin independent one.
One can repeat the analysis 
to find that for tan$\beta \le 50$ the neutralino mass range consistent
with the WMAP constraints on the branch corresponding to 
neutralino-stau coannihilation is $m_{\chi_1^0} \le 500$ GeV~\cite{olive,ln},
and $m_{\chi_1^0} \le 1200$ GeV for the high zone of the HB, where the relic density constraints are satisfied due to coannihilation with the next to lightest 
neutralino and light chargino~\cite{chatto}, as discussed previously.
We repeat again that these constraints remain intact under the imposition 
of the Estimate (I) of the $g_\mu -2$, while the constraint arising
from the inversion region of the HB is removed under the
imposition of the Estimate (II) of the $g_\mu -2$.  We re-iterate, therefore,  
once more the importance of having unambiguous measurements of this 
quantity in the immediate future, which will be indispensable in guiding 
supersymmetric searches in future colliders.

\subsection{WMAP, Proton Decay and SUSY Constraints in 
Grand Unified Supersymmetric Models}

In view of the above-described constraints on  
MSSM embedded in mSUGRA, 
it is also natural to ask whether additional constraints are implied
by the lower limit on the proton life time which is 
a prediction of Grand Unified Theories (GUTs). So far we did not consider 
grand unification in the above analysis. 

There are a number of important issues associated with such an extension.
One is the exact value of $\sin^2 \theta_W$, which
acquires important corrections from threshold effects at the electroweak
scale, associated with the spectrum of MSSM particles~\cite{EKNII,ELN},
and at the grand unification scale, associated with the spectrum of GUT
supermultiplets~\cite{EKNII,Heavy}.  Precision measurements indicate a 
small deviation of $\sin^2 \theta_W$ even from the value
predicted in a minimal supersymmetric $SU(5)$ GUT, assuming the range of
$\alpha_s(M_Z)$ now indicated by experiment~\cite{baggeretal}.

The second issue is the lifetime of the proton. 
Minimal supersymmetric
$SU(5)$ avoids the catastrophically rapid $p \to e^+ \pi^0$ decay that
scuppered non-supersymmetric $SU(5)$. However, supersymmetric $SU(5)$
predicts $p \to {\bar \nu} K^+$ decay through $d = 5$ 
operators~\cite{weinbergd5,enrud} at a rate
that may be too fast~\cite{PDK} to satisfy the presently available lower
limit on the lifetime for this decay~\cite{SK,PDG}. The latter requires
the $SU(5)$ colour-triplet Higgs particles to weigh $> 7.6 \times
10^{16}$GeV, whereas conventional $SU(5)$ unification for $\alpha_s(M_Z) =
0.1185 \pm 0.002$, $sin^2 \theta_W = 0.23117 \pm 0.00016$ and
$\alpha_{em}(M_Z) = 1/(127.943 \pm 0.027)$~\cite{PDG} would impose the
upper limit of $3.6 \times 10^{15}$~GeV at the 90\% confidence
level~\cite{PDK}. This problem becomes particularly acute if the sparticle
spectrum is relatively light, as would be indicated if the present
experimental and theoretical central values of $(g_\mu - 2)$~\cite{E821Expt}
remain unchanged as the errors are reduced.

The simplest way to avoid these potential pitfalls is to flip
$SU(5)$~\cite{f5,AEHN}.  As is well known, flipped $SU(5)$ offers the
possibility of decoupling somewhat the scales at which the Standard Model
$SU(3), SU(2)$ and $U(1)$ factors are unified. This would allow the
strength of the $U(1)$ gauge to become smaller than in minimal
supersymmetric $SU(5)$, for the same value of $\alpha_s(M_Z)$~\cite{ELN}.
Moreover, in addition to having a longer $p \to e/\mu^+ \pi^0$ lifetime 
than
non-supersymmetric $SU(5)$, flipped $SU(5)$ also suppresses the $d = 5$
operators that are dangerous in minimal supersymmetric $SU(5)$, by virtue
of its economical missing-partner mechanism~\cite{f5}.

In \cite{enw}, the authors re-analyzed 
the issues of $\sin^2 \theta_W$ and proton
decay in flipped $SU(5)$~\cite{ELN}, in view of the most recent precise
measurements of $\alpha_s(M_Z)$ and $\sin^2\theta_W$, and the latest
(at the time of publication) limits 
on supersymmetric particles (before publication of WMAP data). 
The analysis of such issues was done in the framework of CMSSM,
as above, and consisted of {\it both}
a general analysis in the $(m_{1/2}, m_0)$ plane and also more detailed
specific analyses of benchmark CMSSM parameter choices that respected all
the available experimental constraints at the time~\cite{benchmark}. 
It was found that the $p
\to e/\mu^+ \pi^0$ decay lifetime exceeds the present experimental lower
limit~\cite{SK}, with a significant likelihood that it may be accessible
to the next round of experiments~\cite{UNO}. In the context of the present
article we shall not present the analysis in great detail, in particular 
we shall not discuss 
the ambiguities and
characteristic ratios of proton decay modes in flipped $SU(5)$.
We refer the reader to the existing literature for this~\cite{enw}. 
Instead we shall 
give a comprehensive description of their results in
figures \ref{proton}, \ref{proton2}, which are self explanatory.

\begin{figure}
\begin{center}
\vskip 0.5in
\vspace*{-0.75in}
\begin{minipage}{8in}
\epsfig{file=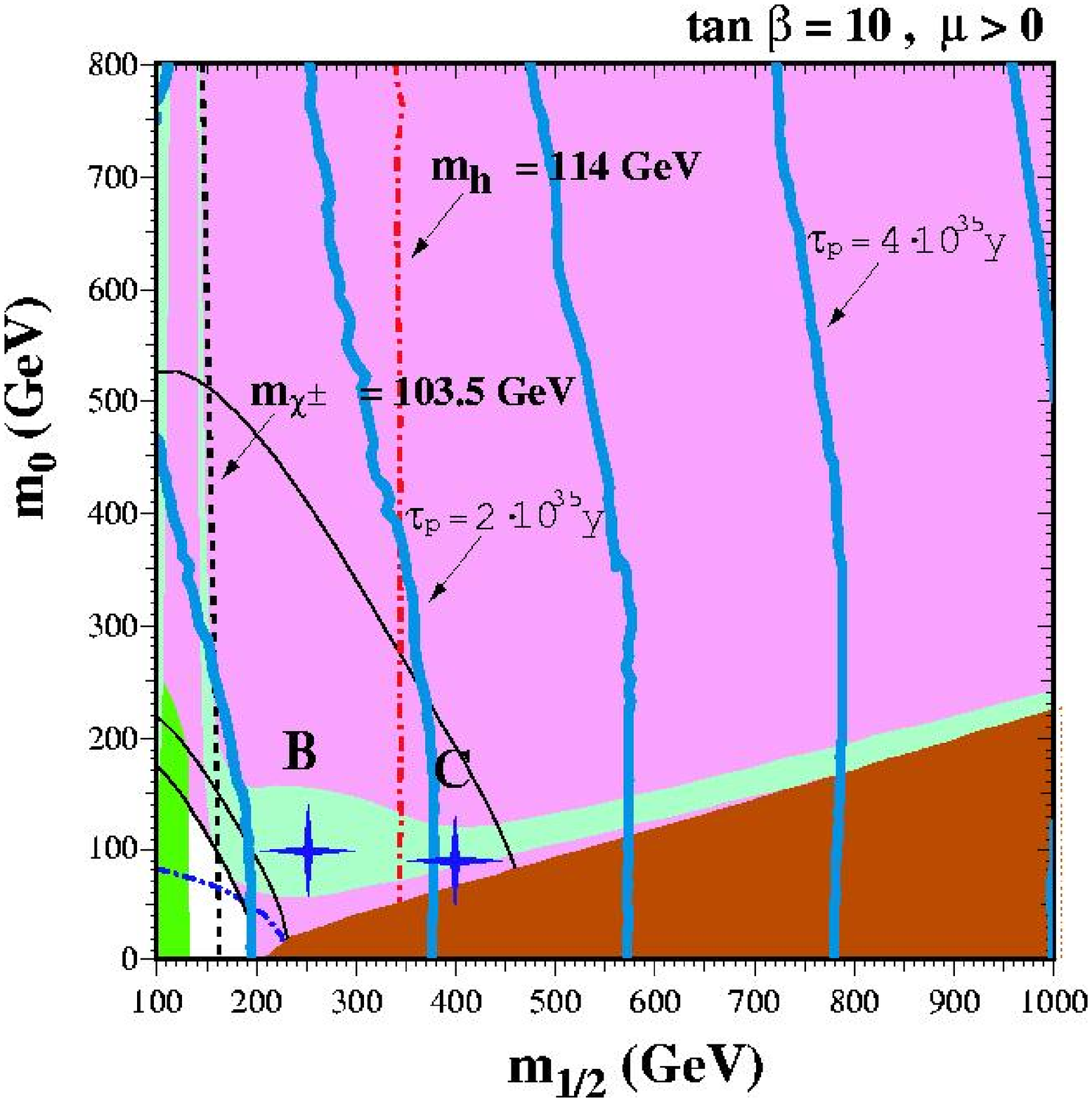,height=2.3in}
\hspace*{-0.17in}
\epsfig{file=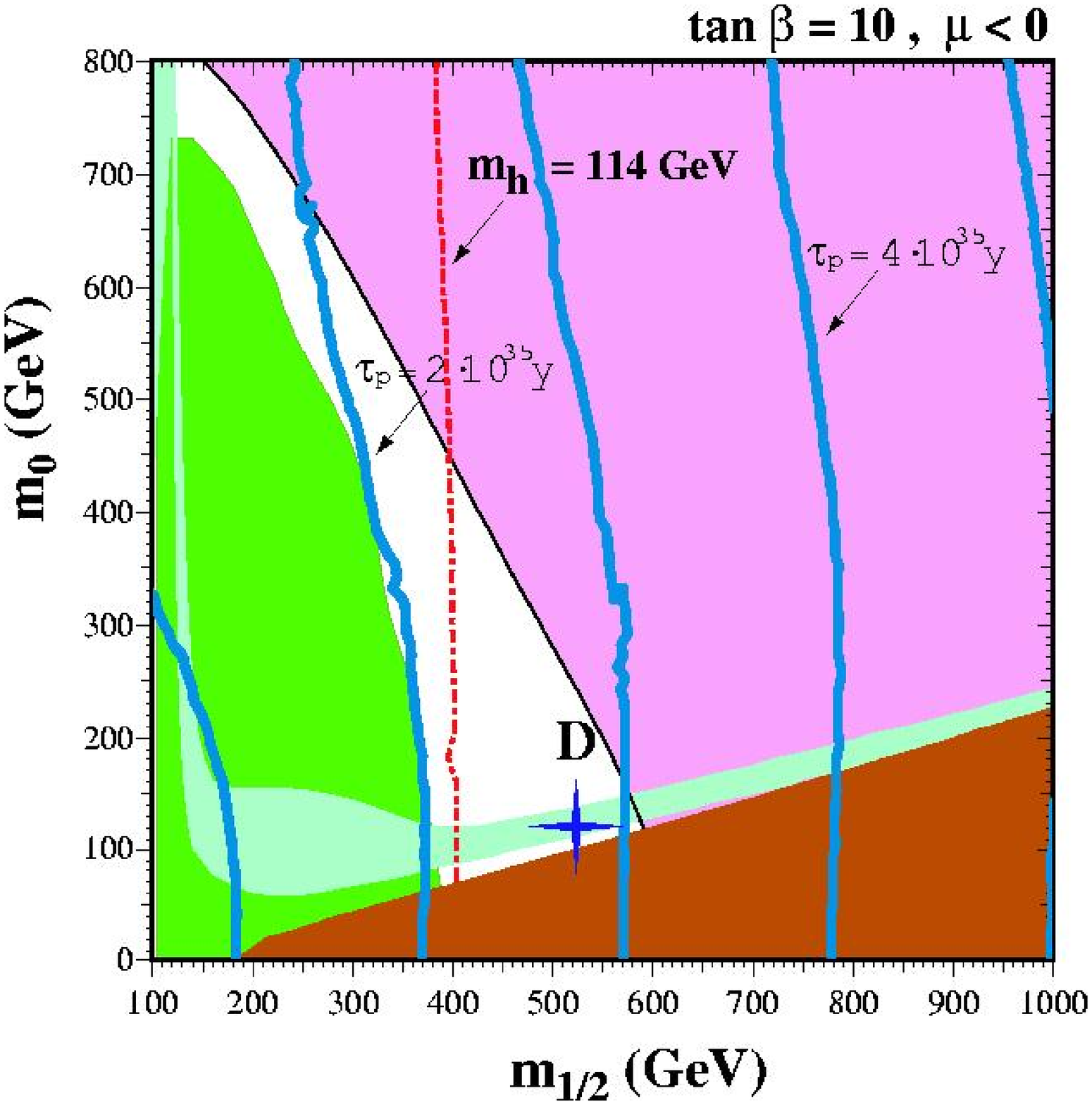,height=2.3in} \hfill
\end{minipage}
\begin{minipage}{8in}
\epsfig{file=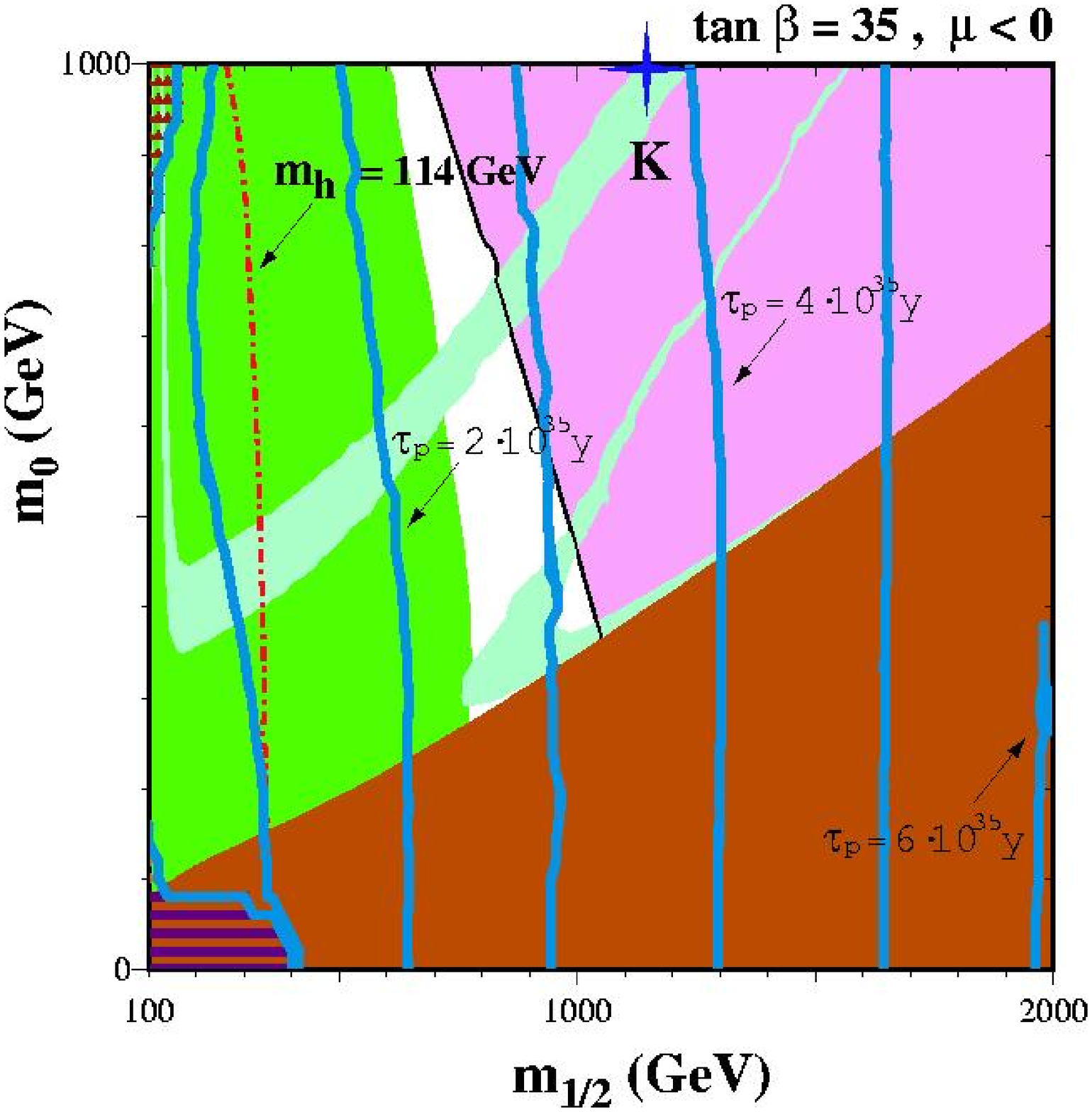,height=2.3in}
\hspace*{-0.2in}
\epsfig{file=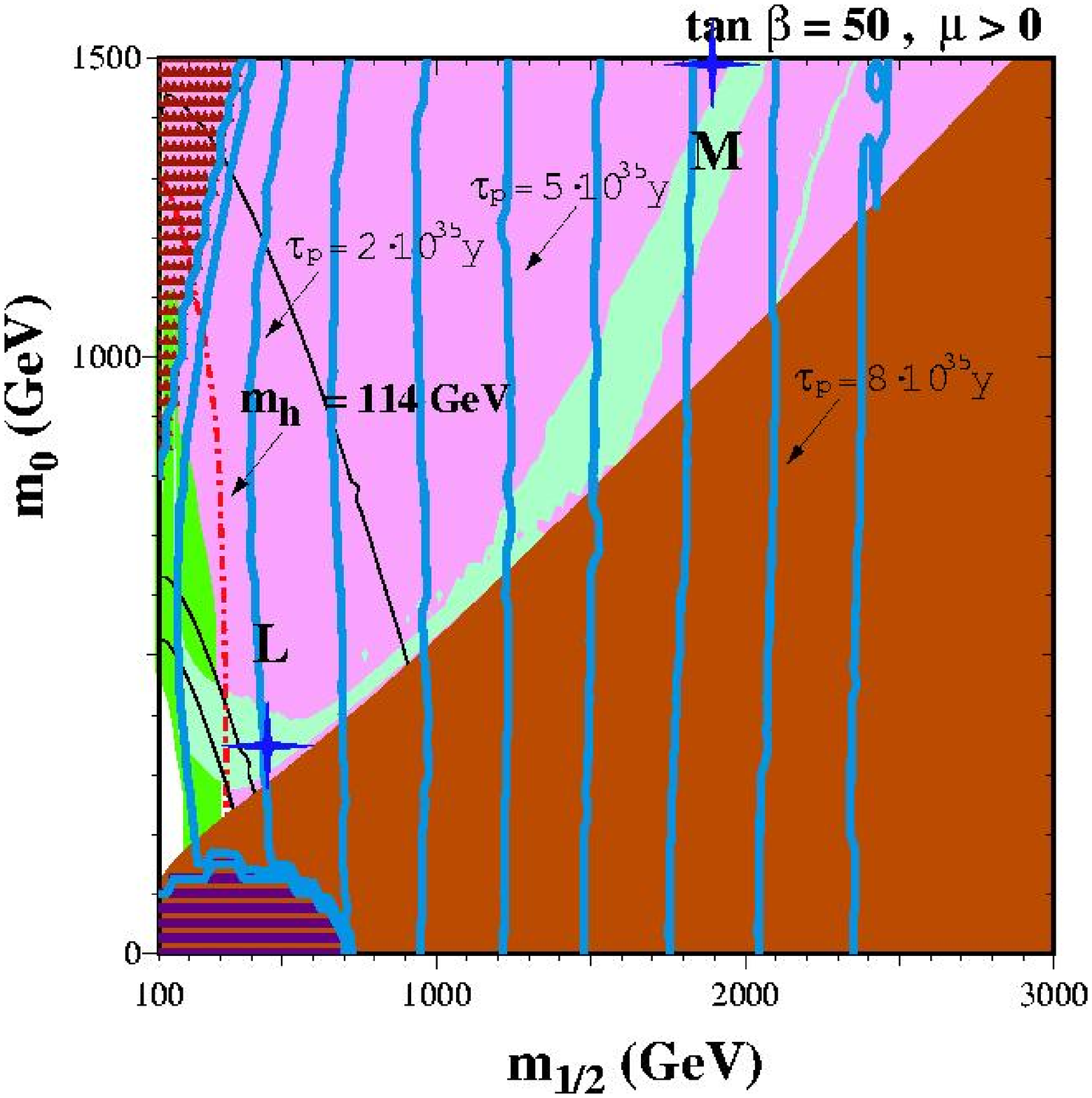,height=2.3in} \hfill
\end{minipage}
\end{center}
\caption{{{\small \it Pre-WMAP constraints on supersymmetric flipped SU(5) 
model~\cite{enw}. 
The solid (blue) lines are contours of $\tau(p \to e/\mu^+ \pi^0)$ in the
$(m_{1/2}, m_0)$ plane for the CMSSM with (a) $\tan \beta = 10, \mu > 0$,
(b) $\tan \beta = 10, \mu < 0$, (c) $\tan \beta = 35, \mu < 0$ and (d)
$\tan \beta = 50, \mu > 0$. The (blue) crosses indicate the CMSSM
benchmark points with the corresponding value of $\tan \beta$ and sign of
$\mu$~\cite{benchmark}. Following~\cite{EOS3}, the dark (red) shaded regions 
are excluded because the 
LSP is
charged, the light (turquoise) shaded regions have $0.1 < \Omega_\chi h^2
< 0.3$, intermediate (green) shaded regions at low $m_{1/2}$ are excluded
by $b \to s \gamma$, shaded (pink) regions at large $(m_{1/2}, m_0)$
are consistent with $g_\mu - 2$ at the 2-$\sigma$ level, and 
electroweak symmetry breaking is not possible in the hatched 
regions. The near-vertical
dashed (black) lines correspond to the LEP lower limit
 on chargino mass, $m_\chi^\pm = 103.5$~GeV, 
the dot-dashed (red) lines to
 the Higgs mass, $m_h = 114$~GeV 
as calculated
using the {\tt FeynHiggs} code~\cite{FeynHiggs}, and the dotted (blue) 
lines at small
$(m_{1/2}, m_0)$ to $m_{\tilde e} = 100$~GeV.
}}} 
\label{proton} 
\end{figure}

\begin{figure}
\begin{center}
\vskip 0.5in
\vspace*{-0.75in}
\begin{minipage}{8in}
\epsfig{file=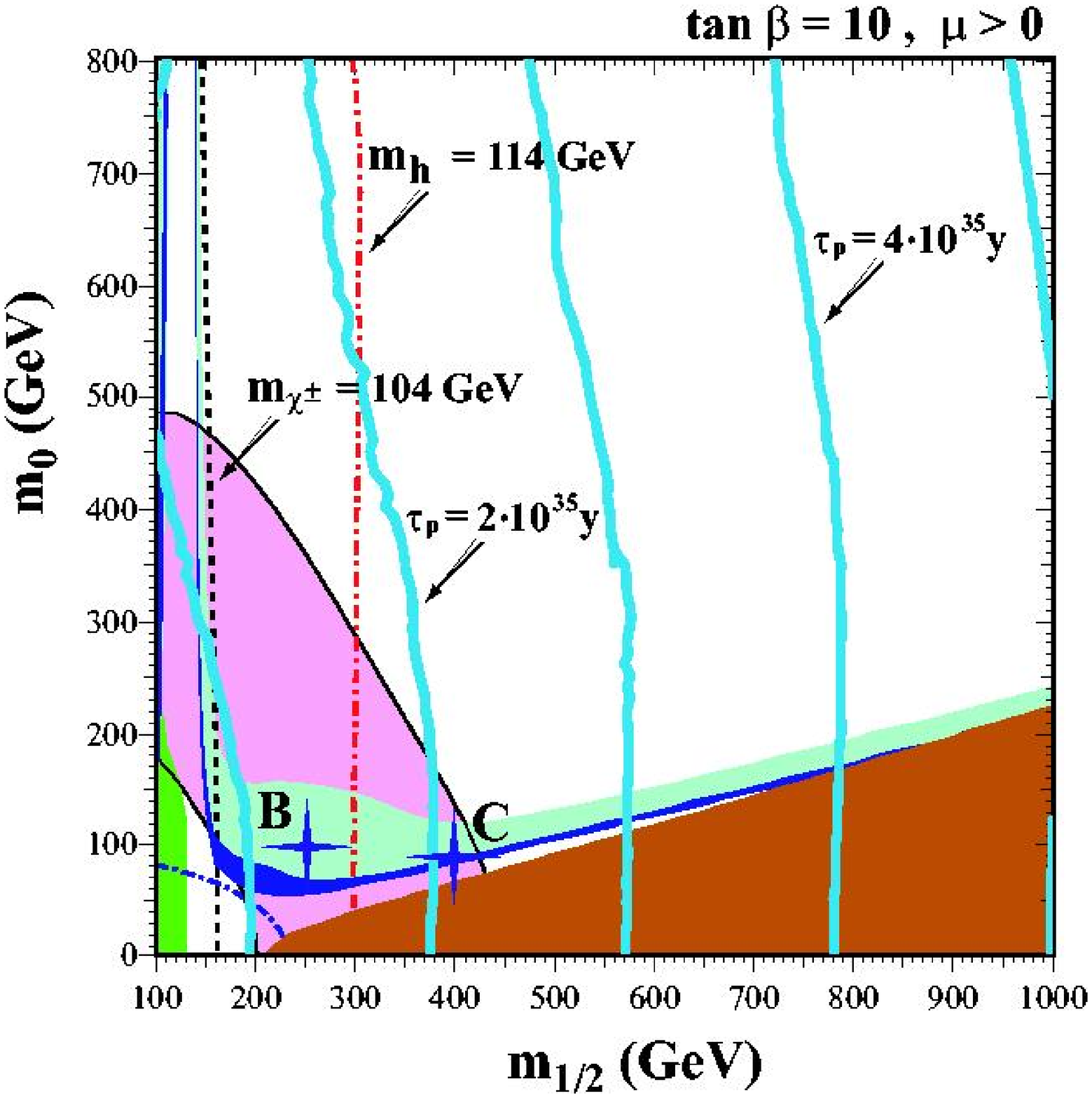,height=2.3in}
\hspace*{-0.17in}
\epsfig{file=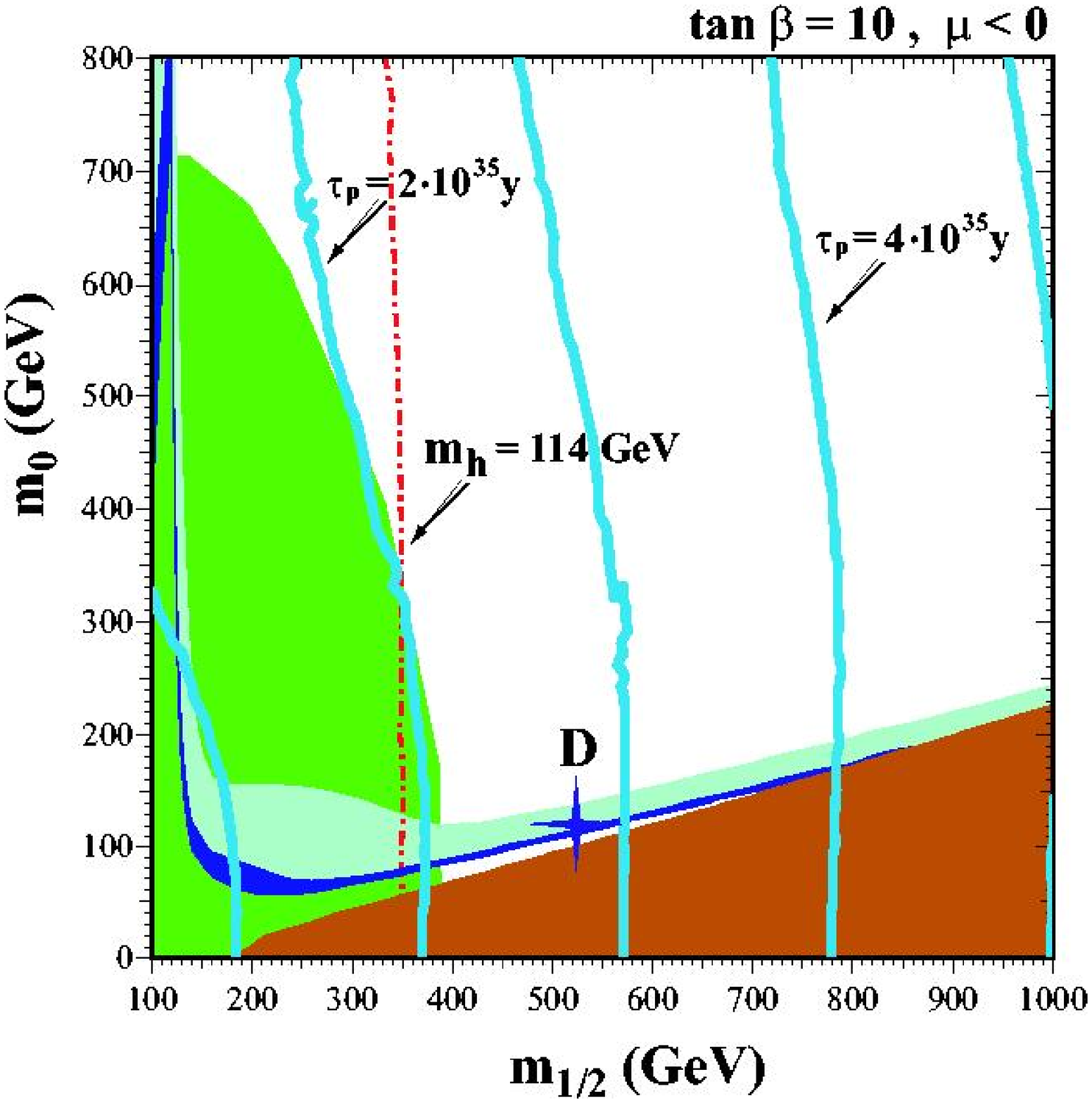,height=2.3in} \hfill
\end{minipage}
\begin{minipage}{8in}
\epsfig{file=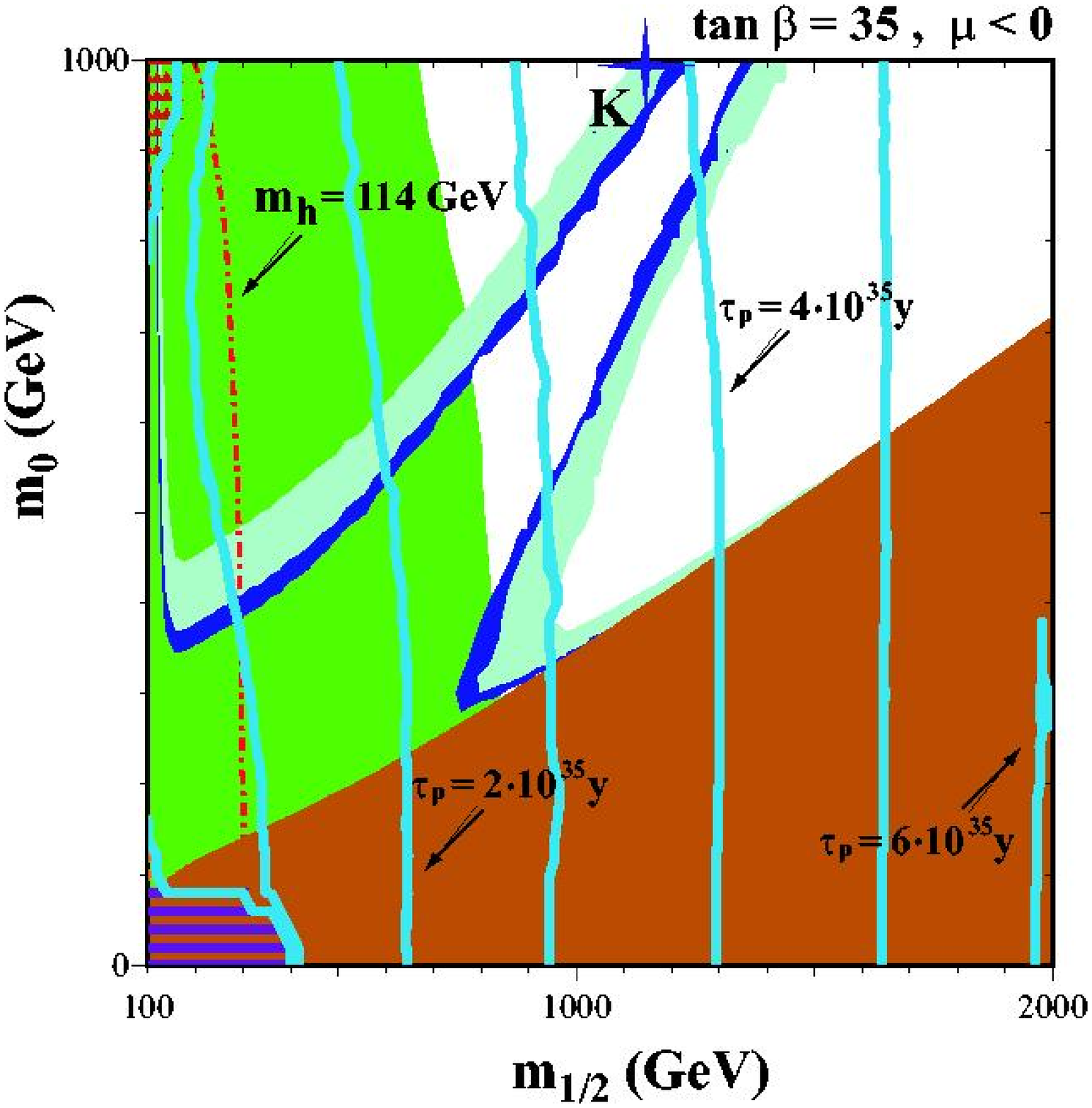,height=2.3in}
\hspace*{-0.2in}
\epsfig{file=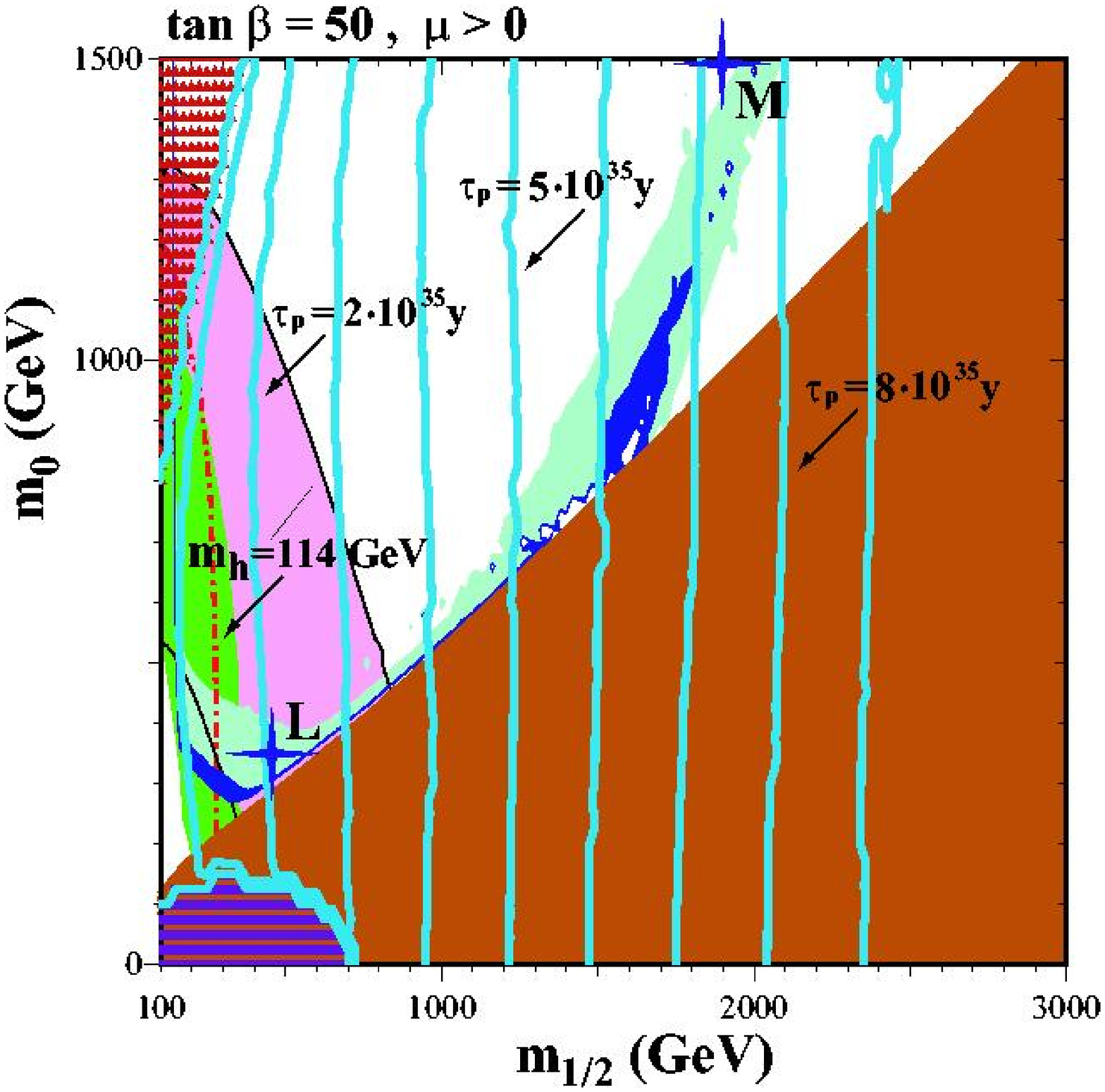,height=2.3in} \hfill
\end{minipage}
\end{center}
\caption{{\it {\small As in figure \ref{proton} but taking into account 
the new cosmological constraints from WMAP on relic densities 
(dark blue regions). The contours of $\tau(p \to e/\mu^+ \pi^0)$
are given here by the light-blue solid  lines (we are grateful to J. Walker 
for providing us with these updated figures).}}} 
\label{proton2} 
\end{figure}

In figure \ref{proton} we give the pre-WMAP results of 
\cite{enw}, while in figure \ref{proton2} we give the updated
post WMAP figures (taking into account 
the updated benchmark points~\cite{benchmarksnew}), 
for comparison. 
From figs.~\ref{proton}, and \ref{proton2} 
it is evident that the `bulk' regions of the parameter
space preferred by astrophysics and cosmology, which occur at relatively
small values of $(m_{1/2}, m_0)$, generally correspond to $\tau (p \to e^+
\pi^0) \sim (1~-~2) \times 10^{35}$~y. However, these `bulk' regions 
are generally disfavoured by the experimental lower limit on
 the Higgs mass, $m_h$, 
and/or by $b \to s \gamma$ decay. Larger values of $\tau (p
\to e^+ \pi^0)$ are found in the `tail' regions of the cosmological
parameter space, which occur at large $m_{1/2}$ where $\chi - {\tilde
\ell}$ coannihilation may be important, and at larger $m_{1/2}$ and $m_0$
where resonant direct-channel annihilation via the heavier Higgs bosons
$A, H$ may be important. 

 Another issue studied in \cite{enw} 
were the possible implications of the GUT threshold effect 
$\delta_{\rm heavy}$~\cite{EKNII,Heavy}.
A general expression for this in 
flipped $SU(5)$ is given in~\cite{EKNII}, where we refer the 
interested reader. 
In fig.~\ref{fig:benchlines} we display~\cite{enw} 
the possible numerical effects of
$\delta_{\rm heavy}$ on $\tau (p \to e/\mu^+ \pi^0)$ in the various 
benchmark
scenarios, assuming the plausible ranges $-0.0016 < \delta_{\rm heavy} < 
0.0005$~\cite{ELN}. The boundary between the different shadings for each 
strip corresponds to the case where $\delta_{\rm heavy} = 0$. The left (red)  
parts of the strips show how much $\tau (p \to e^+ \pi^0)$ could be
reduced by a judicious choice of $\delta_{\rm heavy}$, and the right
(blue) parts of the strips show how much $\tau (p \to e^+ \pi^0)$ 
could be
increased. The inner bars correspond to the uncertainty in $\sin^2 
\theta_W$. On the optimistic side, we see that some models could yield 
$\tau (p \to e^+ \pi^0) < 10^{35}$~y, and all models might have $\tau (p 
\to e^+ \pi^0) < 5 \times 10^{35}$~y. However, on the pessimistic side, in 
no model can we exclude the possibility that $\tau (p \to e^+ \pi^0) > 
10^{36}$~y.

\begin{figure}[ht]
\begin{center}  
\includegraphics[width=0.4\textwidth,angle=0]{lines.eps}
\end{center}
\caption[]{\it {\small
For each of the CMSSM benchmark points, this plot shows,
by the lighter
outer bars, the range of $\tau (p \to e/\mu^+ \pi^0)$ attained by varying
$\delta_{\rm heavy}$ over the range -0.0016 to + 0.0005~\cite{ELN}.  The 
central boundary of
the narrow inner bars (red, blue) corresponds to the effect of
$\delta_{\rm light}$ alone, with $\delta_{\rm heavy} = 0$, while the
narrow bars themselves represent uncertainty in $\sin^2 \theta_W$.
We see that heavy threshold effects could make $\tau (p \to e/\mu^+ 
\pi^0)$    
slightly shorter or considerably longer.
} }
\label{fig:benchlines} \end{figure}

We notice at this stage that a 
new generation of massive water-{\v C}erenkov detectors
weighing up to $10^6$~tonnes is being proposed~\cite{UNO}, that may be
sensitive to $\tau (p \to e^+ \pi^0) < 10^{35}$~y. According to 
the 
calculations of \cite{enw}, 
such an experiment has a chance of detecting proton decay in
flipped $SU(5)$, though nothing can of course be guaranteed.
We also mention~\cite{faspects,ELNO} that flipped $SU(5)$ makes predictions
for ratios of decay rates involving strange particles,
neutrinos and charged leptons that differ characteristically from those of
conventional $SU(5)$. Comparing the rates for $e^+$, $\mu^+$ and neutrino
modes would give novel insights into GUTs as well as mixing patterns.

We conclude from the above analyses~\cite{enw}, therefore,  
that flipped $SU(5)$ evades two of the pitfalls of
conventional supersymmetric $SU(5)$: (i) it
offers the possibility of lowering the prediction for $\alpha_s(M_Z)$ for
any given value of $\sin^2 \theta_W$ and choice of sparticle spectrum, 
and (ii), as far as  
proton decay is concerned, 
flipped $SU(5)$ suppresses $p \to
{\bar \nu} K^+$ decay naturally via its economical missing-partner
mechanism. As in conventional supersymmetric $SU(5)$, the lifetime for $p
\to e/\mu^+ \pi^0$ decay generally exceeds the present experimental lower
limit. However, as shown in \cite{enw}, the flipped $SU(5)$
mechanism for reducing $\alpha_s(M_Z)$ reduces the scale at which
colour $SU(3)$ and electroweak $SU(2)$ are unified, bringing $\tau(p \to
e/\mu^+ \pi^0)$ tantalizingly close to the prospective sensitivity of the
next round of experiments. 
Proton decay has historically been an
embarrassment for minimal $SU(5)$ GUTs, first in their non-supersymmetric
guise and more recently in their minimal supersymmetric version. The
answer may be to flip $SU(5)$ out of trouble.
We remark also that flipped $SU(5)$ grand unified 
supersymmetric theories arise naturally in 
intersecting brane models~\cite{kanti}.

\section{Dark Energy: Cosmological Constant or \\ 
Quintessence or...?}

Before ending this review we would like to devote a few pages 
on an important matter, which so far
 has not been properly  
taken into account.
In our constrained minimal models above, we have taken into consideration
the WMAP results concerning dark matter density constraints, which 
constitutes  only 23 \% of the Uninerse energy density content.
However, as mentioned in section two, a plethora of recent data, including 
those of WMAP, has indicated that  
73 \% of the Universe energy consists of an unknown substance,
termed {\it dark energy} (c.f. figures \ref{budget},\ref{spergel2}).

The origin of this component is as yet unknown, despite enormous theoretical
effort. In the simplest approach, the dark energy is nothing other 
but a cosmological {\it constant}. A negative cosmological 
constant was 
introduced by Einstein in the original
theory of relativity to enforce compatibility 
of Einstein's equation of General Relativity with a static Universe, believed 
(wrongly) to be the case of the observed Universe at the time.
Hubble's observation on the expansion of the Universe (1927) removed the 
necessity for introducing such a constant into the theory. 
Ever since, until five years ago, theoretical physicists believed that 
the vacuum energy density of the Universe was zero, 
and they were trying (alas in vain!) to find a symmetry reason for that. The situation
changed drastically recently, after experimental evidence for a present-era
acceleration of the Universe.   
A FRW Universe with a {\it positive cosmological constant} 
constitutes 
the best fit model of the WMAP and supernovae 
data~\cite{spergel,supernovae}. From a theoretical viewpoint, however,
this 
poses a serious threat in quantization. Indeed, a de Sitter Universe, which 
would be the final state of such 
a positive cosmological constant FRW 
Universe, due to the expansion of the Universe which leads to a dilution
of any matter density, is characterized by the presence of an event horizon,
making the proper definition of asymptotic states, and hence a scattering 
matrix ($S$-matrix), problematic. This presents serious challenges for  
a proper quantization of the theory, which are unresolved at 
present~\cite{cosmolconst}.
It also poses a serious challenge for string theory~\cite{smatrix,emnsmatrix}, given that
the latter, at least in its initial perturbative form, has been formulated
as an S-matrix theory.

Perhaps a more satisfactory, from this point of view, 
approach to the problem of a physical explanation of the 
currently observed acceleration of the Universe~\cite{supernovae},
and the deduced dark energy component~\cite{supernovae,spergel}
is to assume a time varying vacuum energy, $\Lambda (t)$ 
relaxing to zero
asymptotically in the Robertson-Walker time~\cite{timecosmo,lncosmo},
$t \to \infty$. 

One may obtain this time dependent ``vacuum energy'' 
by means of a 
(scalar) field, which has not relaxed yet to the minimum of its 
potential. 
Such a mechanism, known as {\it quintessence}, has attracted
a lot of attention~\cite{quintessence}, 
and represents an elegant {\it non-equilibrium} approach to the issue 
of a positive cosmological constant.
The important advantage of quintessence (or other relaxation) 
models is the possibility of 
defining asymptotic states, and hence an $S$-matrix, 
properly, due to the asymptotic vanishing of the vacuum energy,
when the system reaches its equilibrium. 
Another important advantage is that there exist 
attractor solutions in a cosmological setting for some classes 
of potentials~\cite{quintattr}, which imply that in some cases
the scalar field will join the attractor solution 
before the present epoch for a wide range 
of initial conditions, thereby evading (in principle) 
fine tuning and tracking problems.
There are, however, serious issues 
that remain unresolved, most of them associated
with the nature of the tracking quintessence field, which should be 
important today, but not in the past of our Universe. 
Another important issue 
is the current coincidence in order of magnitude of the
dark energy density with the matter energy density of the Universe, 
which quintessence
models are called to explain in a fine-tuning free way.

A natural relaxation mechanism, compatible 
with the current value of 
$\Lambda(t_{\rm present}) \lsim 10^{-120}M_P^4$ inferred from 
the observations, would be the one in which~\cite{timecosmo,lncosmo}
$\Lambda (t) \sim M_P^4/t^2$, 
for long times $M_Pt \gg 1 $, where $M_P\sim 10^{19}$ GeV is the four-dimensional
Planck energy scale. For $M_Pt_{\rm present} \gsim 10^{60}$, which is the 
Age of the (observed) Universe, one obtains naturally the above limits. 

Such inverse square law relaxation models exist in the literature,
in either field or string/brane quintessence 
theory~\cite{timecosmo,lncosmo}, or even in 
{\it non-equilibrium} approaches to string theory, 
such as non-critical Liouville 
strings~\cite{emninfl,diamandis,gm,mav}, where the vacuum-energy relaxation law is 
obtained from first principles, 
specifically, by means of world-sheet conformal field theory methods. 

In this last respect the quintessence field is actually provided by the 
Liouville mode itself, which is viewed as the time variable. It is the nature
of non-critical strings to associate the Liouville mode with a linear in time
part of the dilaton, and hence non-critical string quintessence may actually
be viewed as a dilaton quintessence, 
which also characterizes some other approaches
within critical string theory~\cite{quintdil}. 
The advantage of dilaton quintessence is that the scalar 
dilaton field exist in the string spectrum,
and there is no need to assume extra fundamental scalars, as 
is the case in quintessence models of local field theory. 

There is also another aspect of non-critical strings, which makes them 
interesting, at least from a theoretical viewpoint. 
Unlike critical strings, which are formulated as a theory of target-space
scattering amplitudes (S-matrix), Liouville strings do not necessarily have 
a well-defined target-space S-matrix~\cite{emnliouv,emnsmatrix}.
Upon the identification of 
target-time with the Liouville mode~\cite{emnliouv},
the world-sheet Liouville correlators among vertex operators may admit the 
interpretation of non-factorizable \$-matrix elements, acting on 
density matrices rather than target-space state vectors~\footnote{The presence 
of density matrices implies an analogy of Liouville strings with open quantum mechanical systems~\cite{emnliouv}. The analogy is exact, given that 
in such models there are modes which are not accessible
by low-energy scattering experiments, such as topological gravitational modes
or back-reaction space-time foam effects, 
or, in colliding brane world scenaria, recoil modes {\it etc.}.
Intgegrating out such modes from a low-energy observer, defines an 
{\it effective} low-energy string theory, which is non critical and 
out of equilibrium. It is such theories that cosmological models are based upon in this context~\cite{emninfl,diamandis,gm,mav}.}, 
$\rho_{\rm out} =$\$$\rho_{\rm in}$, with \$$\ne S~S^\dagger$, 
with $S$ the ordinary S-matrix, which may thus not be 
well defined. However, we stress that 
the operator \$ is well defined 
in Liouville strings, as being associated 
with correlation functions of vertex operators~\cite{emnliouv}.

Such non-critical string models are then capable of describing consistently even quantum cosmological models with a non-zero
cosmological {\it constant}, which, as we mentioned earlier, are hamprered
in a field theory context by the lack of an S-matrix formalism due to the 
existence of cosmic horizons. The lack of a proper $S$-matrix 
may also have implications on violations of CPT symmetry 
in such systems~\cite{wald,emnliouv}, with interesting 
phenomenological consequences in both particle physics and 
cosmology~\cite{lopez}. 
Of course, in Liouville-quintessence relaxation string models 
one may be able of defining an S-matrix asymptotically in time $t$, 
given the possibility 
of defining properly asymptotic 
states at the equilibrium point $t \to \infty$, where critical 
string theory is reached in such cases.

As we have mentioned previously,
in section 3, in some string theory models the dilaton 
has also been viewed as 
the inflaton~\cite{dilaton,aben,emninfl,diamandis,gm,mav}. 
It would be nice to find a consistent 
realistic model in which a single field can explain
both phases of the Universe. From the point of view of the acceleration
of the Universe, implied by the dark energy, 
such an assumption is not unnatural given that currently the Universe
appears to enter an accelerating de Sitter phase which is 
qualitatively similar to that of (the beginning of) inflation.
This may be the result of a particular dilaton potential in a 
string/brane model. However, this issue is still far from being resolved,
given that it is associated with non perturbative effects in 
string theory which are not at hand.  

From the phenomenological point of view quintessence models 
are still compatible with the current observations, including 
those of WMAP~\cite{quintessencetests,spergel}. 
Indeed, as discussed in section 2 (c.f. figure \ref{spergel3}), 
the WMAP best fit model analysis  
has provided  us with the following range of the values of the quintessence 
field ${\cal Q}$ equation of state:
\begin{eqnarray} 
&& w_{\cal Q} = \frac{p_{\cal Q}}{\rho_{\cal Q}} = 
\frac{(\dot {\cal Q})^2/2 - V({\cal Q})}{(\dot {\cal Q})^2/2 + V({\cal Q})} ,
\nonumber \\
&& -1 \le w \le -0.78 
\label{quint} 
\end{eqnarray} 
where $V({\cal Q})$ is the potential 
of the quintessence, and the lower limit is theoretical, corresponding 
to the cosmological 
constant model (if this theoretical assumption is relaxed the upper limit
becomes slightly larger, $\sim -0.67$~\cite{spergel}).

Such values appear naturally in many quintessence 
models~\cite{quintessence,diamandis,gm,mav}.
From our point of view in this work we would like to concentrate and review
approaches that embed such quintessence scenaria in 
supergravity models~\cite{brax,othersusy,CNR}. 
We note that, in this conext, 
the first phenomenological approach was given in \cite{lncosmo}, well before
the recent astrophysical data~\cite{wmap,supernovae}. 
Such models could in principle be 
used, then, to constraint minimal supersymmetric models, such as MSSM,
in the way mSUGRA was used above, but this time taking into account the 
dark energy component of the Universe energy density~\footnote{The cosmological constant problem in mSUGRA models is not solved. Even if one fine tunes 
the tree level contributions to be small, however loop corrections
result in Planck size contributions to vacuum energy.}.

The issue of looking at quintessence in supersymmetric theories
is motivated clearly by the need to reconcile particle physics 
models with cosmology. The presence of cosmological constant 
or dark vacuum energy immediately brings up the issue 
of compatibility with the hierarchy between supersymmetry breaking 
scale and the size of the vacuum energy density of the Universe. 
In supersymmetric quintessence models such issues  may find 
a natural explanation, given that scalar cosmological fields
in cosmology,  
such as quintessence fields 
and inflatons, may arise naturally in a supersymmetric
particle spectrum and it is not unreasonable to expect
that the current height of the relevant potential may become compatible 
with the phenomenologically correct size 
of supersymmetry breaking (of at most a few TeV). 

Scalar field models of quintessence typically require that the 
expectation value of the scalar field today is of order of 
the Planck Mass, in order to explain the observed acceleration
of the Universe. This immediately implies that one should include
supergravity corrections to such models (in the case  
of supersymmetric cosmological models). 
Brax and Martin~\cite{brax} were the first to discuss 
supergravity corrections in the quintessence potential 
in a supergravity model. They considered a (toy) supergravity model 
with a superpotential
\begin{equation}
W({\cal Q}) =\Lambda^{\alpha + 3}{\cal Q}^{-\alpha} 
\label{superpot}
\end{equation}
where $\Lambda$ is the (mass) scale of the particle theory~\footnote{For instance, in supersymmetric SU(N) theories $\Lambda$ is the scale at which the gauge coupling becomes non perturbative~\cite{othersusy}.}, and a flat K\"ahler potential $K= {\cal Q}{\cal Q}^*$, where 
${\cal Q}$ from now on denotes a scalar superfield. 
 
The resulting scalar potential for such a model~\cite{brax}
has an F-term of the general form~\cite{nilles}:
\begin{equation} 
V({\cal Q}) = F^2 - e^{\kappa^2~K} 3\kappa^2 |W|^2
\label{general} 
\end{equation}
which in the case of a flat Kahler potential, gives~\cite{brax}:
\begin{equation}
V({\cal Q})= e^{\frac{\kappa^2}{2}{\cal Q}^2}
\frac{\Lambda^{4 + \beta}}{{\cal Q}^\beta} \left(\frac{(\beta - 2)^2}{4} - 
(\beta + 1)\frac{\kappa^2}{2}{\cal Q}^2 + 
\frac{1}{4}\kappa^4{\cal Q}^4 \right)
\label{scalpote}
\end{equation}
with $\beta = 2(\alpha + 1)$, and 
$\kappa = 8\pi M_P^{-2}$ the gravitational constant.
Notice that, 
for $\alpha > 0$ which is the case of supersymmetric models, the potential
has non-polynomial terms in the quintessence field, which 
when combined with the exponential term turn out, as we shall discuss below, 
to be crucial 
for the correct phenomenology of the model in the present epoch. 

However, as it stands, the scalar potential (\ref{scalpote})  
is ill defined for $<{\cal Q}>=M_P$ due to negative contributions. 
One way out~\cite{brax}
would be the imposition of the constraint 
that the superpotential vanishes $<W>=0$, which eliminates the 
dangerous negative terms in the scalar potential. 

The model has several attractive features then.
The presence of non-polynomial terms in the scalar potential 
(\ref{scalpote})
implies compatibility of the model with a current era acceleration of 
the Universe. 
Moreover, due to the presence of the exponential term, 
the value of the equation 
of state, $\omega _Q$, computed from (\ref{quint})
is pushed towards the value $-1$ in contrast to the usual 
non supersymmetric case 
for which it is difficult to go beyond $\omega _Q\approx -0.7$.  For 
$\Omega _{\rm m}\approx 0.3$, as suggested by the current data, 
the model of~\cite{brax} predicts 
$\omega_Q\approx -0.82$, which is in agreement with the recent WMAP data
implying $-1 < w_{\cal Q} < -0.78$ 
(see figures \ref{spergel3}), as discussed in section 2.

The condition $<W>=0$ is possible to realise in the presence of 
other matter fields but in general is a tight restriction. 
Moreover, such a constraint is not compatible 
with models of supersymmetry breaking. 
Things therefore might have been easier if this restriction is relaxed. 
In ref.~\cite{CNR} a different 
K\"ahler potential has been used than in \cite{brax}: 
\begin{equation} 
K= [{\rm ln}(\kappa {\cal Q} + 
\kappa {\cal Q}^*)]^2/\kappa ^2~.
\label{kahler2}
\end{equation}
The resulting scalar potential has the form:
\begin{equation} 
V = M^4 \left( 2x^2 + (4\alpha -7)x + 2(\alpha -1)^2
\right)~\frac{1}{x}~e^{[(1-x)^2 -2\alpha(1-x)]}
\label{scalarCNR}
\end{equation} 
where $x \equiv (-\frac{3}{2}\kappa {\cal Q})^{2/3}$, and 
$M^4 = \Lambda^{6 + 2\alpha}\kappa^{2+2\alpha}2^{2\alpha}$. 
In this case the minimum of the 
resulting scalar potential is always positive for $\alpha > 1.25$
without the need for the imposition of any constraint on $<W>$.

\begin{figure}[t]
\centering 
\epsfig{file=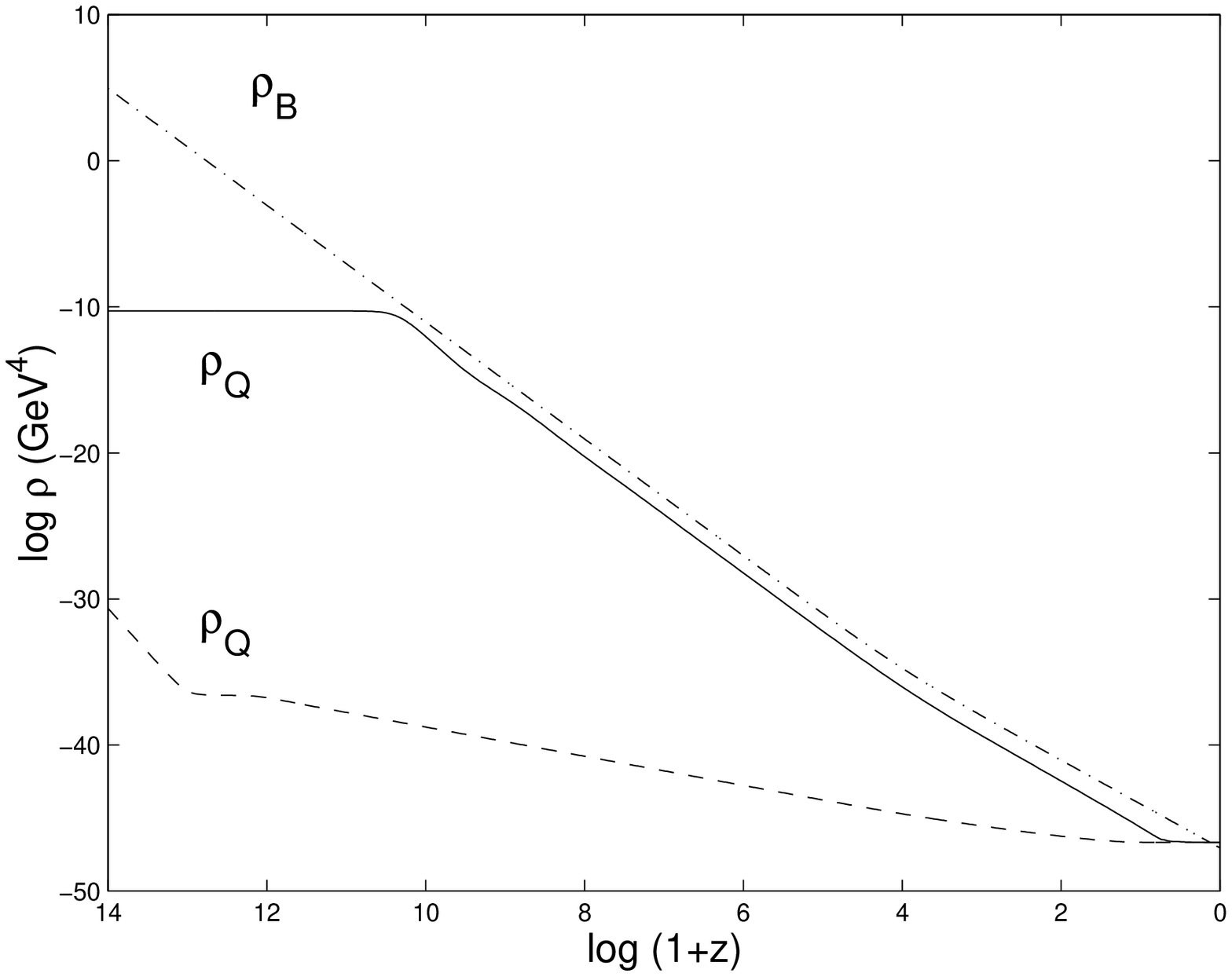,angle=0,width=0.3\linewidth}
\hfill \epsfig{file=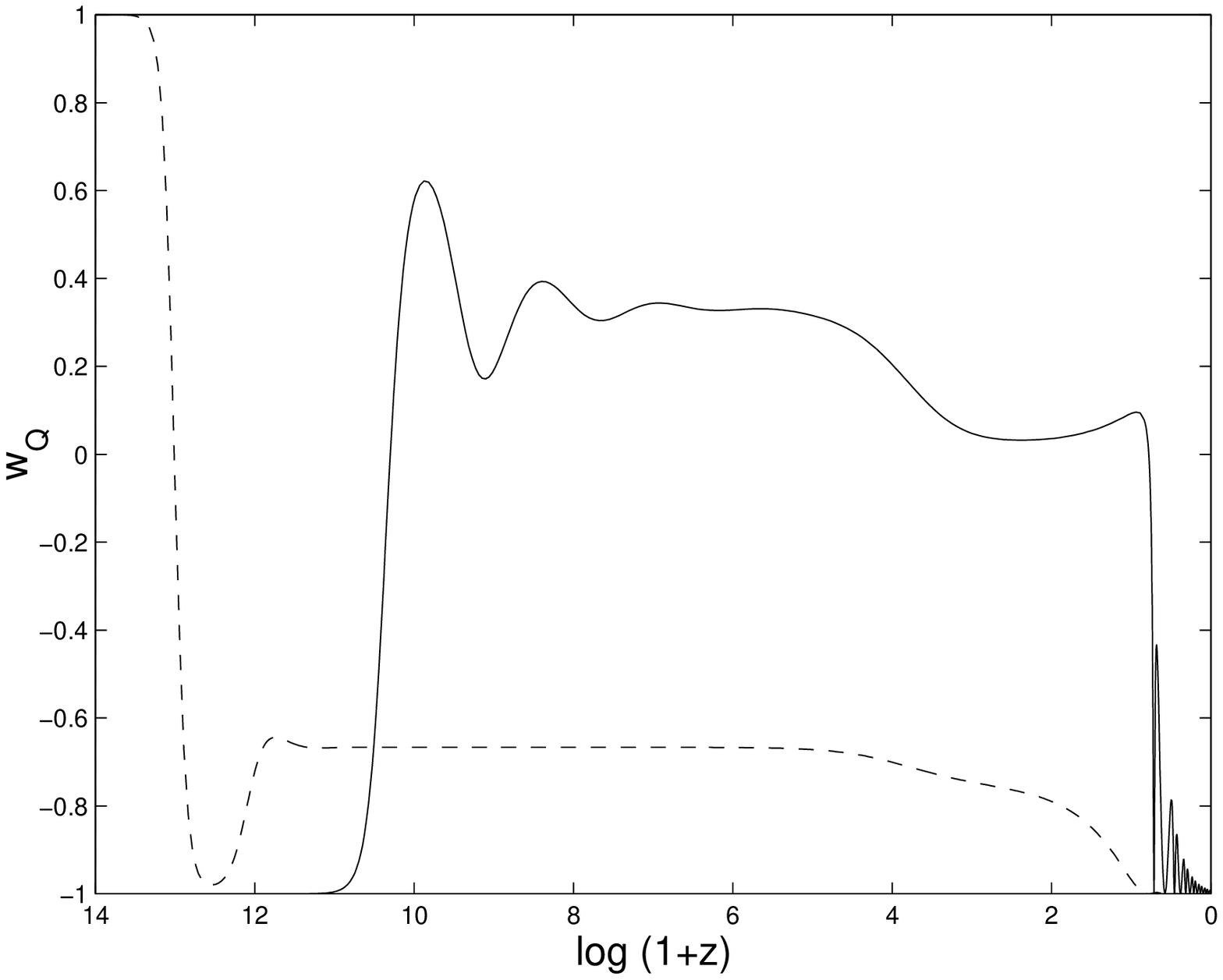,angle=0,width=0.3\linewidth}
\caption{{\it \underline{Left}: The evolution of the 
energy density of the quintessence 
field in the model of ref.~\cite{CNR} 
versus that of the background ($z$ is the redshift). 
\underline{Right}: The evolution
of the equation of state of the quintessence field in that model.
In the figures, the dash-dotted line represents the background energy density, 
the continuous lines correspond to the initial condition 
for the quintessence field $
{\cal Q}_{in}/{\cal Q}_m \gg 1$, while the dashed lines 
correspond to the initial condition 
for the quintessence field $
{\cal Q}_{in}/{\cal Q}_m \ll 1$ (where ${\cal Q}_m$ denotes the value of the 
quintessence field at the minimum of the scalar potential).}} 
\label{quintessfig}
\end{figure} 

The cosmology of that model is given in figure \ref{quintessfig}.
It is important to notice that the presence of field ${\cal Q}$ denominators
in the potential (\ref{scalarCNR}) accompanying the exponential terms
are {\it crucial} for yielding a negative equation of state $w < 0$ 
in the current era, thereby implying acceleration of the Universe today,
as seems to be predicted by observations~\cite{supernovae,wmap}. 
Indeed, a solution for the potential involving only exponential and
polynomial factors
of the quintessence fields  cannot yield a negative $w$
today~\footnote{We note at this stage that the case with only exponential
factors of the quintessence field in the potential  
is discussed in 
\cite{scaling}, and in this case the attractor is a `scaling one'. However
these results apply in field theory where there is 
only a single quintessence field, and may 
not characterise the case of multifield configurations in
more complex situations, such as  
string theory. Indeed, as we shall discuss below, 
a negative 
quintessence equation of state today $w < 0$ can be obtained
for exponential dilaton potentials,
at least in a non-critical string 
cosmology framework~\cite{emninfl,diamandis,gm,mav},
where the vacuum energy scale is set by the central charge deficit 
of the non-critical $\sigma$-model~\cite{ddk}, and expresses
physically a departure from equilibrium, due to cosmically catastrophic 
events in our world.}. After reaching its minimum the scalar
field will mimic the cosmological constant with $w_{\cal Q} = -1$. 
At present, the WMAP data point towards
$w_{\it cal} \to -1$, which is certainly a feature 
that characterizes supergravity
models (c.f. figure \ref{quintessfig}), 
but the big question remains as to whether 
one can distinguish (supersymmetric) 
quintessence with $w \to -1$ 
from a cosmological constant model.
This question is associated clearly with the topic of the present review.
If one constructs an appropriate supergravity model of quintessence
which can incorporate consistently the dark energy component 
of the Universe with a broken supersymmetry at TeV scale 
then one may use such models to constrain particle physics supersymmetry 
searches as done above for the MSSM. From such studies one may eventually
make predictions that could distinguish a quintessence model from a 
naive cosmological constant one. This remains the biggest challenge
for the years to come. The reason is the following. 
 
At this stage it is
 worth mentioning 
that the authors of \cite{CNR} 
have also considered another type of Kahler 
potentials, 
\begin{equation} 
K= -\frac{1}{\kappa ^2} [{\rm ln}(\kappa {\cal Q} + 
\kappa {\cal Q}^*)]~.
\label{noscale}
\end{equation} 
Such logarithmic potentials are characteristic of no-scale supergravity 
models~\cite{noscale}, which are known to be derived at low energies 
from string models~\cite{wittenns}, as well as from M-theory~\cite{horava}.

However, the resulting scalar potential, in terms of canonically 
normalized fields of the form ${\tilde Q} =({\rm ln}\kappa Q)/\sqrt{2}\kappa$,
reads: 
\begin{equation}
V({\tilde {\cal Q}}) = \left(\Lambda^{5+\beta}\kappa^{1+\beta}\frac{(\beta^2 -3)}{2}
\right)e^{-\sqrt{2}\beta \kappa {\tilde {\cal Q}}}
\label{scalingpot}
\end{equation} 
with $\beta = 2\alpha + 1$. 
Positivity of the potential is guaranteed for $\beta > \sqrt{3}$.

The problem with this model, however, is that, as it stands, 
the potential (\ref{scalingpot}) has the form of the above-mentioned  
scaling solution 
of quintessence~\cite{scaling}, according to which 
one cannot have a current era acceleration of the Universe, 
since one cannot obtain a negative equation of state. 
We note, however, that such a conclusion
is valid only if there is one quintessence field. 
As we shall discuss later~\cite{diamandis},  in more complicated
string models, with several field configurations, 
one has the possibility of obtaining a current era acceleration 
and a negative value of the equation of state parameter $w$, by
appropriately arranging couplings and field magnitudes. 
Thus, it may well be that no-scale SUGRA models can still provide realistic, phenomenologically
acceptable, cosmologies to date. This remains to be seen.

A {\bf serious problem} 
for all the supersymmetric quintessence models so far 
is {\it Supersymmetry Breaking (SSB)} at a phenomenologically acceptable 
scale $M_S$, such that 
 $M_S^2 \sim <F>~\gsim (10^{10}~{\rm GeV})^2$ for gravity-mediated cases
of SSB, 
or $M_S^2 \sim <F>~\gsim (10^{4}~{\rm GeV})^2$ for gauge-mediated cases
(with $F$ denoting an F-term contribution, see (\ref{general})).
In the model of~\cite{CNR} 
this requires a value for 
$W \sim <F> \kappa^{-1} \sim m_{3/2}\kappa^{-2}$, 
where $m_{3/2}$ is the gravitino mass,  
in order to cancel the F-term contribution and 
give negligible vacuum energy, in agreement with
the known limits (or recent observations) on the cosmological constant
(we remind the reader that, in this formalism, 
the cosmological constant is given by $<F^2>$, c.f. (\ref{general})).
Although the relaxation of the $<W>=0$ constraint
allows for such solutions, however it is clear that,  
in such models, 
the dynamical cosmological constant provided by the quintessence
field cannot be the dominant source of SUSY breaking. The authors
of \cite{CNR} 
modified their superpotential by adding the term $m_{3/2}\kappa^{-2}$
on the right hand side of (\ref{superpot}) but then the 
scalar potential acquires changes $\delta V$ 
containing this constant 
term,
\begin{equation} 
\delta V \sim \Lambda ^{3 + \alpha}m_{3/2}\kappa^{-\alpha} 
+ m_{3/2}^2 \kappa^{-2}~.
\label{dv}
\end{equation} 
This leads to a serious disruption of the quintessence potential,
due to the constant second term.

\begin{figure}[htb]
\centering 
\epsfig{file=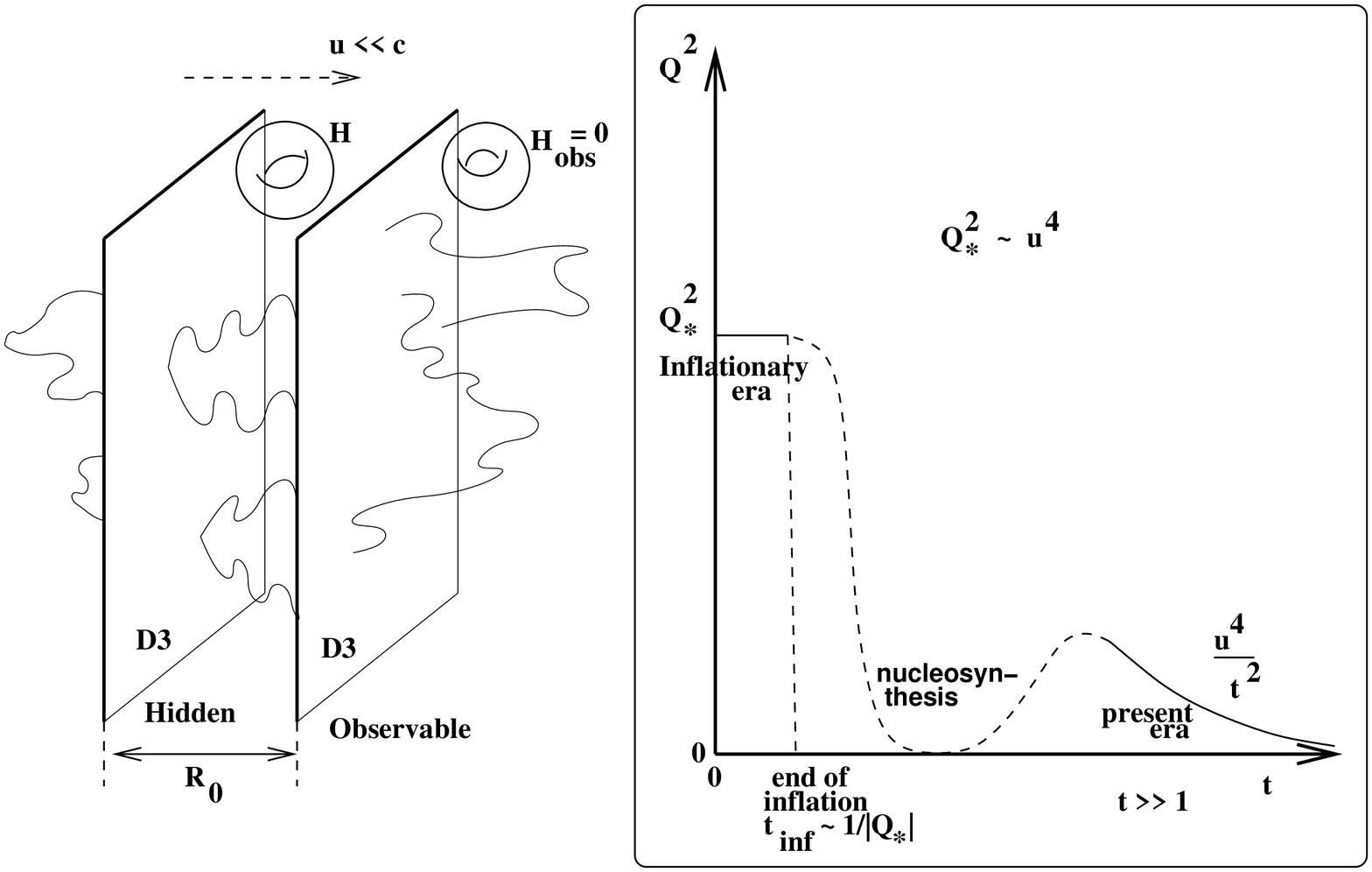,angle=0,width=0.4\linewidth}
\hfill \epsfig{file=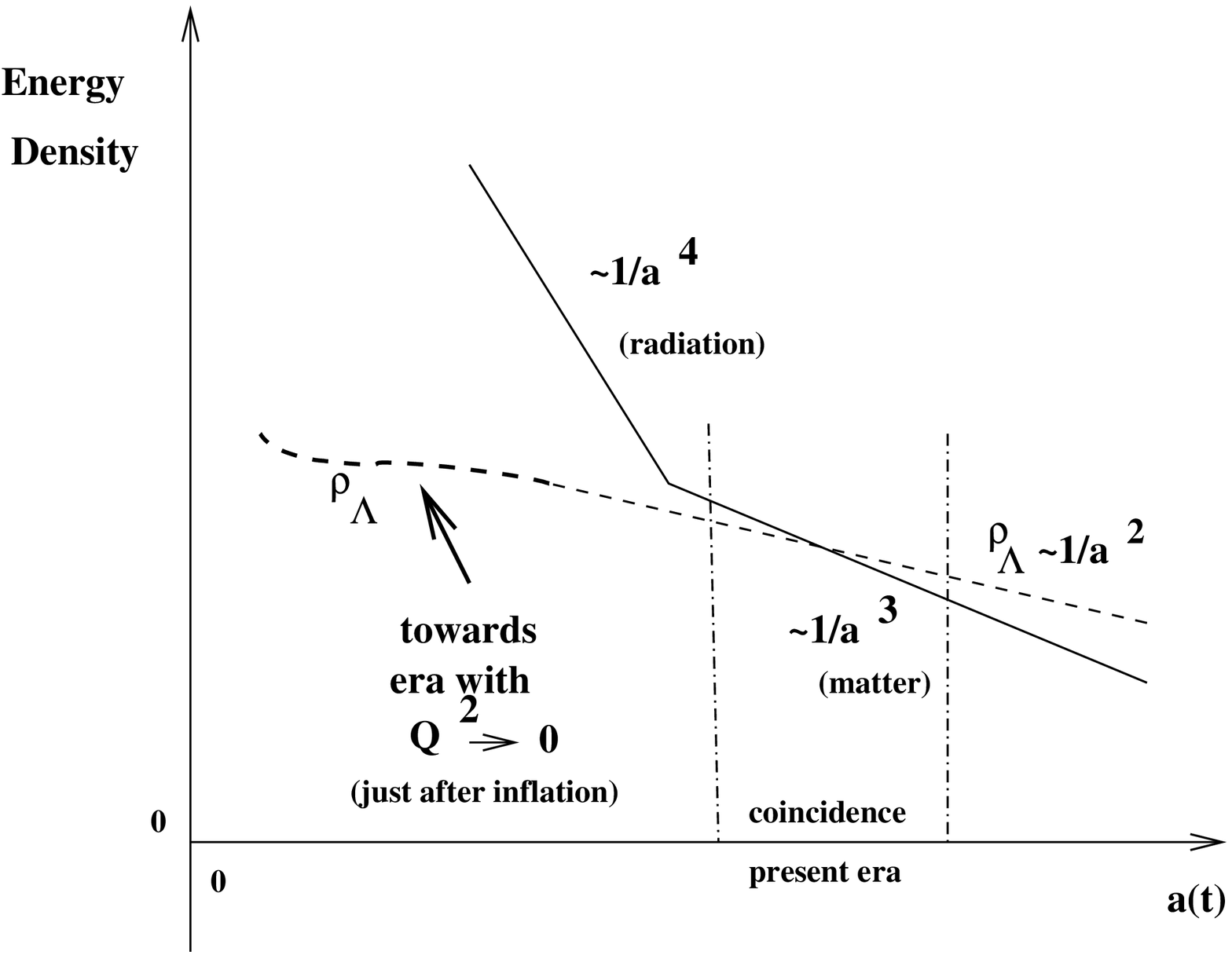,angle=0,width=0.3\linewidth}
\caption{{\small {\it Various non-critical string scenaria of quintessence
\underline{Left}: A non-critical (Liouville) string scenario
for quintessence
involving 
colliding five-brane worlds, compactified on magnetized tori~\cite{gm,mav}. 
In this model the dilaton field today is constant (with vev of order of 
the Planck scale), while it was not constant in inflation era. 
The dashed lines in the evolution of the world-sheet central
charge deficit $Q^2$ are conjectural, in order 
to match the criterion of consistency with 
nucleosynthesis. The continuous lines have been calculated using 
world-sheet (logarithmic) conformal field theory methods.  
\underline{Right}: The evolution of the 
energy density of the quintessence field in the non-critical 
string model of Diamandis {\it et al.}~\cite{diamandis} versus that of the background.
In that model, there is a time dependent dilaton on the observable brane world, whose time dependence today 
is logarithmic, and causes the appearance of a 
relaxing to zero effective vacuum energy  $\Lambda (t) \sim M_P^4 e^{2\Phi}~,
\Phi \sim -{\rm ln}t$.}}}
\label{diamandisfig}
\end{figure}

This is still a serious unresolved problem on field theoretic
quintessence models. It may well be that string theory models
tackle such problems due to unconventional supersymmetry
breaking mechanisms one may devise in that framework, as a 
result of the extra dimensions. 
For instance, it is advocated in \cite{obstr} that one may actually
live in a four-dimensional world with unbroken supersymmetry, and 
that the mass splittings between the superpartners occur as a result 
of the {\it obstruction} of supersymmetry, through some higher dimensional
effects in string/brane theory. 
By `obstruction' we mean here a situation in which the ground 
state of the system is still supersymmetric, characterised by a zero 
cosmological constant, but the excitations of the system 
exhibit supersymmetry breaking mass splittings. 
If this is the case, then 
one does not have to worry about supersymmetry breaking 
terms in the effective SUGRA quintessence potential. However, 
a concrete realisation of these ideas, in the context of realistic
models, is still lacking. 
Another uncnventional idea about SSB 
is advocated in ref.~\cite{banks}, according to which the 
relation between the supersymmetry breaking scale $M_S$ 
and the cosmological 
constant $<F^2>$ (c.f. (\ref{general})) 
may be modified, in such a way that the 
amount of vacuum energy, claimed to be observed today, 
is responsible for the (phenomenologically acceptable) 
supersymmetry breaking. This could be the case if 
one had ther following relation: 
\begin{equation} 
M_S = \kappa^{-1} \left(<F>^2\kappa^4\right)^\beta~, 
\label{ssbcosmo}
\end{equation} 
with $\beta = 1/8$ instead of the usually assumed $1/4$. 
In this case, one would not have any dangerous F-terms in the 
SUGRA potential of the order $\kappa^{-2}M_S^2$, which would disrupt the 
quintessence potential. At present, however, at least in the context 
of conventional field or string theory, such a scenario 
remains only a conjecture. 

We now remark that alternative ways out of these problems 
could be provided by completely unconventional 
ways of looking at stringy cosmology, namely 
the 
approach to quintessence advocated in \cite{diamandis,gm,mav},
by means of non-critical string theory ~\cite{emninfl}.
As we shall discuss below, in such a framework one may also 
obtain in some cases a concrete realisation of (\ref{ssbcosmo}) 
with $\beta =1/8$.

According to such scenaria, the quintessence field is provided by the 
Liouville mode, which is the linear in time part of a string theory dilaton.
The current acceleration of the Universe is due to an excitation of 
our brane world, as a consequence of 
a catastrophic cosmic event. 
For instance, in the model of \cite{gm,mav} 
such a catastrophic event is provided 
by the collision of two brane worlds, moving initially 
with a small relative velocity $u \ll c$ (c.f. figure \ref{diamandisfig}).  
This results 
in a departure from criticality in the resulting world-sheet 
string theory describing excitations on the brane. 
In more general situations, 
such as the models of Diamandis {\it et al.}~\cite{diamandis}, 
such a departure from equilibrium could be due to an initial
quantum fluctuation of a single brane world. 

The vacuum energy is relaxing to zero as $M_P^4/t^2$ for long times
after the collision, or the initial brane  fluctuation, 
and measures essentially the deviation of the excited string system 
from equilibrium. This scaling is obtained by (logarithmic) conformal 
field theory methods on the world sheet of the string describing 
excitations of the (observable) brane world.   
It is
 worth stressing 
that in the model of
\cite{diamandis}, the effective quintessence (dilaton) potential  
is of an exponential form $V(\Phi) \sim M_P^4 e^{2\Phi}$,
with the dilaton scaling with (large) 
time as $\Phi \sim -{\rm ln}t$. Naively, 
one is tempted to think that 
such potentials 
could not 
cause a current-era acceleration of the Universe~\cite{scaling}, 
since this potential characterises the so-called 
`scaling solution' in cosmology~\cite{scaling}. 
However, this conclusion is only true if there are no other fields 
in the problem. As demonstrated in 
\cite{diamandis}, the string models considered there, have a plethora
of other fields (moduli and flux fields), whose magnitudes can be arranged
in such a way so as 
to yield a quintessence equation of state with $w \lsim -0.75$, and thus a   
current-era acceleration of the Universe, 
as claimed by the observations~\cite{supernovae,wmap,spergel}. At present,
however, one could not identify precise dynamical reasons for such arrangements
in the magnitude and the couplings 
of the various field configurations, and hence,
at present, such models may be considered as {\it fine-tuned}.

In the above-described models of non-critical string quintessence  
the basic source of supersymmetry breaking is independent of the 
value of the vacuum energy today. 
For instance, in Diamandis {\it et al.}~\cite{diamandis}, 
one uses type-0 strings which are not 
supersymmetric due to appropriately projecting out superpartners 
in the string spectrum~\cite{type0}. On the other hand, in 
the models of \cite{gm,mav},
involving five-brane worlds, 
the supersymmetry breaking 
is due to 
the compactification 
of the two extra dimensions 
on internal manifolds with magnetic fields.
The supersymmetry breaking in this case is obtained as a 
consequence~\cite{bachas} 
of the mass splittings in the spectrum of string excitations
due to the different coupling of fermionic and bosonic excitations
on the brane world to the internal magnetic fields (the associated
Nielsen-Olesen instabilities of such scenaria are welcome when the model
is considered in a cosmological setting).
In fact in some cases within the 
models of \cite{gm} it is even possible to realise concretely the 
scenario of \cite{banks}, given that one can derive the relation 
(\ref{ssbcosmo}) with $\beta=1/8$, although we stress such a relation
is not the only phenomenologically consistent solution in these models.
As remarked in \cite{mav}, however, the biggest
challenge for such non-critical string theory models of quintessence
is the nucleosynthesis era, which seems to require an almost critical
string setting (c.f. the left picture in figure \ref{diamandisfig}). 

Finally, we wish simply to mention that, in the modern context of 
brane theories, there are new interesting approaches to the cosmological
constant problem, without, however, offering any solution, at present.
These explore either the holographic principle,
by means of a  
conformal-field-theory (CFT)/de-Sitter-space (positive
cosmological constant) correspondence
(in the line of 
the correspondence between CFT and Anti-de-Sitter Space 
(with negative cosmological constant) of  
Maldacena~\cite{maldacena}), or condensation of tachyons in 
(non-supersymmetric) string theory/brane models~\footnote{We note that 
tachyonic instabilities
may be useful in cosmology, as triggering inflationary expansion
of the Universe 
in the past, but being completely harmless today~\cite{diamandis,deboer}.}.
Although we think of such issues as
very interesting and intellectually challenging,
however, we shall not touch upon them here, 
due to lack of space. We refer the reader to the relevant literature
for more details~\cite{deboer}.

This completes our discussion on dark energy and its currently 
unsolved problems. From our brief but hopefully 
comprehensive discussion 
in this section, the reader must have certainly realized that 
the problem is far from being solved, however it presents
many interesting possibilities for future work, both theoretical
and experimental. Theoretically, the biggest challenge is to find 
a correct model that accounts for the currently observed acceleration 
of the Universe, compatible with other constraints from particle 
physics, such as supersymmetry breaking. Such a model can then
be used to constrain supersymmetric particle physics models,
in a similar spirit to mSUGRA discussed here. 
Experimentally, it is desirable to have more direct evidence on 
the current acceleration of the Universe  by having 
more dedicated supernovae (or even other type) experiments.
This will provide the only direct evidence for such a phenomenon,
which when combined with indirect evidence from CMB anisotropies, 
will give a definite result on the important issue of dark energy.

\section{CONCLUSIONS AND OUTLOOK}

In this article we have given a comprehensive 
review of the latest constraints on CMSSM, especially after the 
first data released by the WMAP satellite experiment.   
Specifically, we have combined the new WMAP cosmological data \cite{wmap}  
on Dark Matter with recent 
high energy physics experimental information including measurements 
of the anomalous magnetic moment of the muon, from E821 Brookhaven experiment,  
the $\bsga$ branching ratio  and the light Higgs boson mass bound
from LEP and we studied the imposed constraints on the parameter space of the 
CMSSM. We have assessed the potential of discovering SUSY, if it is based on 
CMSSM, at future colliders and DM direct search experiments. 
The available parameter space seems to be constrained severely
by the combination of the above-mentioned data. 
The use of the 
new WMAP data in conjunction with the $2 \; \sigma$ $( g_{\mu}-2 )$ bound can guarantee 
that in LHC but also in a $e^{+} e^{-}$ collider with center of mass energy 
$\sqrt{s} \approx 1.1 \; \mathrm{TeV}$ CMSSM can be discovered
provided that the high (inversion) region of the HB is not realized. 

However, WMAP data seem to favour HB/focus point region 
(this was based on a $\chi^2$ analysis 
of mSUGRA model). Updated reach studies have indicated  
that, for an integrated luminosity $100 fb^{-1}$, LHC can probe 
$m_{1/2} \sim 1400$ GeV, for small $m_0$ (gluino mass $m_{\tilde g} \sim 
3 $ TeV). For large $m_0$ the HB/focus point region $m_{1/2} \sim 
700$ GeV can be probed (gluino mass $m_{\tilde g} \sim 
1800 $ GeV) via conventional SUSY channels.

However HB/focus point region appears to extent 
indefinitely to large $m_{1/2}, m_0$ values, far beyond LHC reach.
Such regions can be probed by direct dark mater searches.

The effect of these constraints 
is also significant for the direct DM searches. 
For the $\mu>0$ case we found that the minimum
value of the spin-independent $\lsp$-nucleon
cross section attained is of the order of $10^{-9}$ pb~\cite{ln}.

We have also discussed such constraints, combined with those coming 
from 
proton life time current lower limts, in the context of 
grand unified supersymmetric models, specifically the flipped 
SU(5) model, which seems to survive such tight constraints.  

For future directions we point out that one should take 
seriously into account 
the other constraint coming from recent astrophysical data,
that of {\it dark energy}. 
We believe it is 
quintessence, i.e a relaxing to zero (non-equilibrium) field. 
The WMAP data point towards an equation of state 
of quintessence $w=p/\rho \to -1 $ 
which is close to that of a cosmological constant.
This feature is shared by Quintessence SUGRA models~\cite{brax}, 
and should be explored further within Superstring, Brane models
of a relaxing to zero vacuum energy, in conjunction with the 
issue of supersymmetry breaking.

The important theoretical task for the years to come will be 
to determine the (physically) 
correct SUGRA/string/brane  model to constrain SUSY searches
exploiting the non-zero Dark Energy component of Universe.
Experimentally, it is certain that the post-WMAP era, which  now 
begins, will imply that 
Particle Physics 
and Astrophysics will proceed
together for the exciting years to come and provide 
useful and complementary physics input to each other.

\section*{ACKNOWLEDGMENTS}

We wish thank D.Ahluwalia-Khalilova for the invitation to write 
this review for Int. J. Mod. Phys. D, 
which was based on numerous presentations
by the authors, including a plenary talk (NEM) at the 4th ATLAS
Physics Workshop, Athens University (Greece), May 21-25 2003. NEM wishes
to thank Fabiola Gianotti for the invitation in that meeting, and 
acknowledges useful discussions with her and Jim Pinfold. 
It is also a great pleasure to thank J. Walker for providing us with 
material on proton decay prior to publication, which we used in this review. 
The work of A.B.L. is partially supported by the European Union 
(contracts HPRN-CT-2000-00148 and HPRN-CT-2000-00149), and 
by the University of Athens Research Committee. 
That of N.E.M. 
is partially supported by a visiting professorship at the Theoretical 
Physics Department of the University of Valencia (Spain), and by 
the European Union (contract HPRN-CT-2000-00152). 
The work of D.V.N. is supported by D.O.E. grant 
DE-FG03-95-ER-40917.

\end{document}